\documentclass[onecolumn,sort&compress,numbers]{els-mrw} % For numbered references Style 

\usepackage{amsmath,amssymb,amsfonts,amsthm,makeidx,graphicx}
\usepackage{txfonts}
\usepackage{helvet}
\usepackage{wrapfig}
\usepackage{eso-pic}

\newcommand{\Vek}[1]{\mbox{\boldmath$#1$\unboldmath}}
\newcommand{\vek}[1]{\mbox{\footnotesize\boldmath$#1$\unboldmath}}
\newcommand{\fract}[2]{{\textstyle\frac{#1}{#2}}}

\newcommand{\dslash}{\partial \hskip -0.5em /}
\newcommand{\zr}[1]{\mbox{\hspace*{#1em}}}
\newcommand{\ID}{\mbox{{\sf 1}\zr{-0.16}\rule{0.04em}{1.55ex}\zr{0.1}}}
\newcommand{\imu}{{\rm i}}

\begin{document}

%%%%%%%%%%%%%%%%%%%%%%%%%%%%%%%%%%%%%%%%%%%%%%%%%%%%%%%%%%%%%%%%
%% the following items are mandatory: 
%% - title
%% - author names
%% - affiliation details
%% - abstract
%% - keywords

%% Precise, concise, and informative description of the focus of this work. Avoid abbreviations and formulae in the title
\chapter{Chiral Solitons}

%% All author names and affiliations, and email address for corresponding author
\author[1]{Herbert Weigel}%

\address[1]{\orgname{Stellenbosch University}, \orgdiv{Institute for Theoretical Physics, Physics Department}, 
\orgaddress{Stellenbosch 7600, South Africa}}

%\articletag{Chapter Article tagline: update of previous edition, reprint.}

\maketitle

%%%%%%%%%%%%%%%%%%%%%%%%%%%%%%%%%%%%%%%%%%%%%%%%%%%%%%%%%%%%%%%%
%% the following item is mandatory: 
%% 100-150 word summary of the chapter
\begin{abstract}[Abstract]
Generalizing quantum chromodynamics (QCD) from three to arbitrarily many color degrees 
of freedom suggests that baryons can be described as solitons in an effective meson theory 
whose interaction strength decreases with the number of colors. The exact form of that 
theory is unknown, but at low energies chiral symmetry and its breaking are considered 
as the construction recipes for modeling the theory. The Skyrmion is a static, localized 
solution in a non-linear field theory for pions and it is the most prominent version of a 
soliton in a chirally symmetric meson theory. Upon quantization it reproduces the spectrum 
of the low-lying baryons and their static properties reasonably well. Extending that theory 
by vector mesons improves on the agreement between predicted and empirical data. Chiral 
solitons within models for the quark flavor dynamics facilitate the investigation of nucleon 
structure functions. Here we provide a pedagogical overview of these facets.
\end{abstract}

%% 5-10 words that embody the key topics in the chapter. What terms would someone put into a search engine if they were looking for a chapter like this?
\begin{keywords}
soliton\sep baryons\sep chiral symmetry\sep large-$N_C$ QCD \sep Skyrmion\sep baryon properties
\sep flavor symmetry breaking
\end{keywords}

%%%%%%%%%%%%%%%%%%%%%%%%%%%%%%%%%%%%%%%%%%%%%%%%%%%%%%%%%%%%%%%%
%% the following item is optional: 
%% - Single figure visually illustrating the key topic/method/outcome described in the chapter
\begin{figure}[h]
	\centering
	\includegraphics[width=6cm,height=5cm]{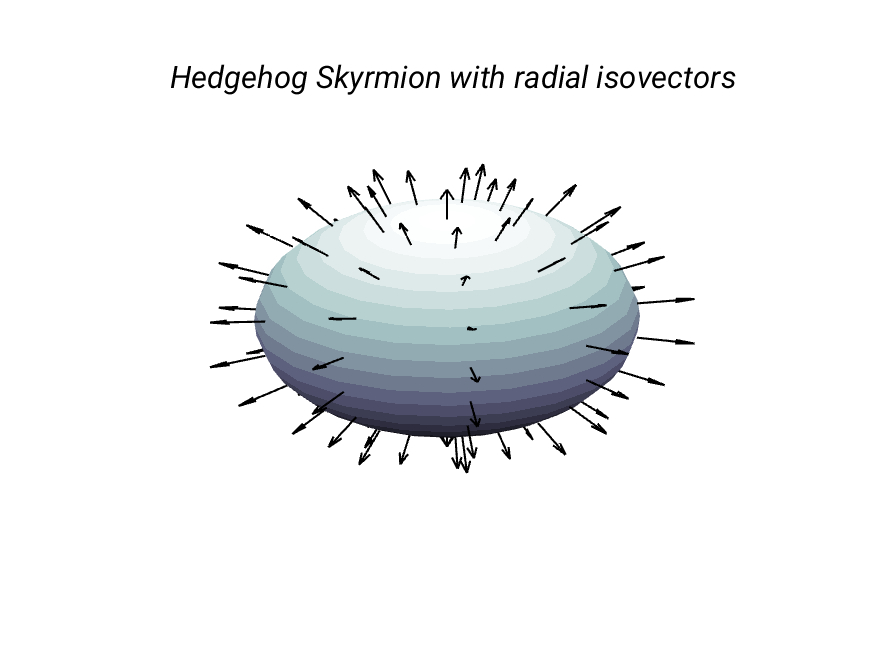}
        \centering
	%\caption{Hedgehog Skymion with radial isovectors.}
	%\label{fig:titlepage}
\end{figure}

%%%%%%%%%%%%%%%%%%%%%%%%%%%%%%%%%%%%%%%%%%%%%%%%%%%%%%%%%%%%%%%%
%% the following item is optional: 
%% - System of abbreviations/terms/symbols used in the specific field of study/community. List and define
\begin{glossary}[Nomenclature]
	\begin{tabular}{@{}lp{34pc}@{}}
		Soliton & Finite energy solution to non-linear field equations\\
                QCD & quantum chromodynamics\\
                $N_C$ & number of colors\\
                VEV & vacuum expectation value\\
                $f_\pi$ & pion decay constant (empirical value: $93\,{\rm MeV}$) \\
                $U=\xi^2$ & chiral field\\
                $F(r)$ & chiral angle (soliton profile)\\
                $L$, $R$ & global chiral transformations\\
                $U_{\rm H}$ & hedgehog configuration\\
                $\Vek{x}\,(\hat{\Vek{x}})$ & coordinate (unit) vector\\
                $\Vek{\tau}$ & vector whose components are the three Pauli matrices\\
                $A(t)$ & matrix for time dependent flavor rotations representing collective coordinates\\
                Hyperons & positive parity baryons with up, down and strange quarks\\
                PCAC & partially conserved axial vector current
	\end{tabular}
\end{glossary}

%%%%%%%%%%%%%%%%%%%%%%%%%%%%%%%%%%%%%%%%%%%%%%%%%%%%%%%%%%%%%%%%
%% the following item is mandatory: 
%% List of the key points and topics a reader can expect to learn from this chapter 
\section*{Key Points}
\begin{itemize}
\item 
Section {\it Introduction} briefly explains the origin of chiral symmetry and its 
breaking in QCD.
\item
Section {\it Large-${\it N_C}$ QCD} reflects on QCD with a large number of color degrees of freedom
and motives the soliton picture.
\item
Section {\it Solitons in field theory} reviews the concept of solitons in non-linear field theories
and connects to the motivation from Section {\it Large-${\it N_C}$ QCD}.
\item
Section {\it Chiral Lagrangian} introduces the concept of effective chiral Lagrangians
which is the point of departure for chiral solitons.
\item
Section {\it The Skyrme model} discusses the Skyrme model for baryons. This section conveys the
main information about the soliton picture. Improvements, in the sense of better
reproduction of empirical data, of this model are (very) technical but do not
alter the soliton picture for baryons substantially. An exception is the 
chiral quark soliton discussed in Section {\it Quarks and Skyrmions}
\item
Section {\it Extension to SU(3)} discusses the extension to three flavors by including kaons in the 
effective Lagrangian. The main objective is the exploration of the hyperon spectrum.
\item
Section {\it Vector mesons} reflects on including light vector mesons. Rather than going into
much of the technical details, the major improvements over the Skyrme model are listed.
\item
Section {\it Quarks and Skyrmions} explains how a chiral soliton can be constructed from a model 
for the quark flavor dynamics. Here it is important to consider both, the energy
from quark levels bound by the chiral soliton and the energy from the distorted Dirac
sea because these two components emerge at the same order in the semi-classical 
expansion. A particular novelty is the possibility to explore nucleon structure 
functions.
\item
Section {\it Further applications} lists further applications of chiral solitons in particle physics 
in compact form. The reader is referred to the quoted literature for more details.
\item
Section {\it Conclusion} briefly summarizes the article.

\end{itemize}

%%%%%%%%%%%%%%%%%%%%%%%%%%%%%%%%%%%%%%%%%%%%%%%%%%%%%%%%%%%%%%%%
%% the following items are mandatory: 
%% - Section: Introduction 
%% - further sections
%% - Section: Conclusion
\section{Introduction}\label{intro}

Ordinary matter is composed of molecules formed from atoms. An atom in turn is 
described as an electron cloud bound to a nucleus through the electromagnetic 
interaction. The nucleus itself is built from protons and neutrons. The latter 
are members of a larger group of (formerly considered elementary) particles: the 
baryons. Together with the mesons (pions, kaons, etc.\@) the baryons constitute
the even larger group of hadrons that comprise all particles which are subject to 
the strong nuclear force. The quark model~ \cite{Gell-Mann:1964nj,Clo79} is
a systematic description of hadrons. Quarks are spin $\frac{1}{2}$ particles
and three quarks comprise baryons, which are known as half-integer spin objects,
while mesons (integer spin) are quark-antiquark compounds. There are different
quark species, called flavor, to accommodate the various quantum numbers of the 
hadrons. In total there are six flavors, though in this context we are mainly 
concerned with the three light ones: {\it up}, {\it down} and {\it strange}, 
{\it cf.\@}, however, Section~{\it Heavy flavor symmetry}. Pauli's exclusion 
principle for fermions dictates an anti-symmetric wave-function and, since the 
ground state baryons have a symmetric spatial wave-function, this requires an additional 
quantum number for the quarks, nowadays known as color. The fundamental field 
theory for the interaction of quarks is the non-Abelian gauge theory for this 
color degree of freedom~\cite{Fritzsch:1973pi,Mu87}: quantum chromodynamics (QCD).
Since three quarks make up the anti-symmetric wave-function, there must be 
equally many color degrees of freedom and therefore QCD is an $SU(3)$ gauge
theory. This theory is kindred to quantum electrodynamics which governs the 
interaction of electrons and photons. The quarks are associated with the
electrons and the photon equivalent is called gluon. The main difference,
however, is that while photons are electrically neutral, the gluons carry color 
charge and there are eight of them. The quarks spinors, $q_f$ and gluon fields 
$G_\mu=\sum_{a=1}^8G_\mu^a\frac{\lambda_a}{2}$ are respectively combined 
in the fundamental and adjoint representations of $SU(3)$. While the label 
''$f$'' stands for flavor, the Dirac and color indexes of the spinors are not 
written out. The matrix character of the gluon field is represented by the 
Gell-Mann matrices, $\lambda_a$. The indexes of these matrices are contracted
with the color index of the spinors. Then the QCD Lagrangian reads
\begin{BoxTypeA}{
\begin{equation}
\mathcal{L}_{\rm QCD}=-\frac{1}{2}{\rm tr}\left(G_{\mu\nu}G^{\mu\nu}\right)
+\sum_f \overline{q}_f\gamma^\mu\left(\partial_\mu-\imu gG_\mu\right)q_f
-m_f\overline{q}_f q_f
+\mbox{gauge fixing terms}\,.
\label{LQCD}\end{equation}}
\end{BoxTypeA}\noindent
The matrix valued field strength tensor,
$G_{\mu\nu}=\partial_\mu G_\nu-\partial_\nu G_\mu-\imu g\left[G_\mu,G_\nu\right]$
induces three- and four-point gluon self-interactions with strengths $g$ 
and $g^2$, respectively. These interactions make the main difference to quantum 
electrodynamics. While in the large energy regime these self-interactions lead to 
asymptotic freedom~\cite{Gross:1973id,Politzer:1973fx}, they hamper the perturbative 
treatment at low energies, which is the standard approach to explore quantum field 
theories. This subtlety is reflected by the confinement hypothesis: {\it Quarks and 
gluons do not exist as isolated states, rather they combine to hadrons}.
In a group theoretical language this translates into the statement that only 
color singlets are asymptotically observable states.

The above discussion indicates that QCD is not too helpful in gaining insight 
into (static) baryon properties like radii or electromagnetic moments\footnote{The 
computationally expensive approach of lattice QCD~\cite{Ro92} might be an 
exception.} and we should instead explore QCD inspired models for that
purpose. Besides the obvious Lorentz-invariance and the above described 
color gauge symmetry, the Lagrangian, Eq.~(\ref{LQCD}) embodies a further
symmetry when $m_f=0$ which is called the {\it chiral} symmetry\footnote{See
also the contribution by Ulf-G.\@ Mei{\ss}ner to the encyclopedia for a 
thorough discussion of chiral symmetry\cite{Meissner:2024ona}.}. It becomes
apparent when introducing left- and right handed components
\begin{equation}
\Psi_{\rm L,R}=\frac{1}{2}\left(1\mp\gamma_5\right)\Psi
\label{ch2:handed}\end{equation}
of the Dirac spinors (for each flavor and color). Without the mass term and with only 
vector interactions, $\overline{\Psi}\gamma^\mu G_\mu\Psi$ as in $\mathcal{L}_{\rm QCD}$, 
these components decouple. The quark mass spectrum separates into two sets: (i)~the light 
quarks (up and down current quark masses are few MeV, the strange quark mass parameter
is about $100\,{\rm MeV}$) and (ii)~charm, bottom and top ($\sim 1.3\,{\rm GeV}, 
4.5\,{\rm GeV}, 170\,{\rm GeV}$, respectively)~\cite{ParticleDataGroup:2024cfk}.
Hence, at a typical energy scale of hadron interactions (several hundred MeV 
and more) chiral symmetry may well be respected by the up and down flavors and 
eventually also by the strange counterpart. In this picture the QCD Lagrangian 
possesses an $U_L(N_f)\times U_R(N_f)$ symmetry. Here $N_f$ is the number of 
quark flavors whose current quark masses may be ignored. Depending on whether 
we consider the strange current quark mass as large or small we have $N_f=2$ 
or $N_f=3$. This symmetry group factorizes according to
\begin{equation}
U_L(N_f)\times U_R(N_f) \cong U_{L+R}(1) \times U_{L-R}(1) \times
SU_L(N_f)\times SU_R(N_f)
\label{symmetrygroup}\end{equation}
and is called the {\it chiral group}. The invariance
under $U_{L+R}(1)$ is responsible for the conservation of baryon number
whereas $U_{L-R}(1)$ is subject to a quantum anomaly~\cite{Ad69,Be69}.
This results in $2N_f-1$ conserved flavor currents. The $2N_f$ flavor
currents are most conveniently presented as linear combination of the
left- and right-handed vector currents that have definite parity: 
the vector current $J_\mu^a$ and the axial vector current $A_\mu^a$,
\begin{equation}
J_\mu^a=\bar{q}_{\rm L}\gamma_\mu\frac{\lambda_a}{2}q_{\rm L}
+\bar{q}_{\rm R}\gamma_\mu\frac{\lambda_a}{2}q_{\rm R}
=\bar{q}\gamma_\mu\frac{\lambda_a}{2}q
\qquad {\rm and}\qquad
A_\mu^a=-\bar{q}_{\rm L}\gamma_\mu\frac{\lambda_a}{2}q_{\rm L}
+\bar{q}_{\rm R}\gamma_\mu\frac{\lambda_a}{2}q_{\rm R}
=\bar{q}\gamma_\mu\gamma_5\frac{\lambda_a}{2}q\,.
\label{conscurr}\end{equation}
Here $\lambda_a$ ($a=1,\ldots,N_f^2-1$) are the Gell-Mann matrices of 
$SU(N_f)$ and $\lambda_0=\sqrt{2/N_f}\,\ID_{N_f\times N_f}$ is proportional 
to the unit matrix in flavor space. In the above construction the spinors
are column vectors with $N_f$ entries of the quark flavors
\begin{equation}
q=\begin{pmatrix}\Psi_1\cr \Psi_2 \cr\vdots\cr \Psi_{N_f}\end{pmatrix}=
\begin{pmatrix}\Psi_u\cr \Psi_d \cr\vdots\cr \Psi_{N_f}\end{pmatrix}\,.
\label{flavorspinor}\end{equation}
In the construction of the currents, Eq.~(\ref{conscurr}) the color label of the
spinors is simply summed over. The vector current, $J^a_\mu$ is conserved when the 
quark masses are equal while the conservation of the axial current, $A^a_\mu$ 
emerges only when all masses are fully ignored, while $\partial^\mu A_\mu^0\ne0$ due
to the above mentioned anomaly.

The symmetry structure, Eq.~(\ref{symmetrygroup}) suggests that QCD predicts
two sets of matter that, up to minor corrections from $m_f\ne0$, have the
same mass spectrum. Hadrons in one set are composed of left-handed quarks,
the other set has the right-handed partners. Hence, the elements of the two sets 
differ by their parity quantum number. This scenario would result from the 
Wigner-Weyl realization of chiral symmetry when the lowest energy configuration 
(vacuum) is invariant under the symmetry and the generators of the symmetry, 
i.e., the Noether charges constructed from the currents in Eq.~(\ref{conscurr}), 
transform degenerate physical states into one another. However, Nature is 
different: There are low-lying pseudoscalar mesons like the pions 
($m_\pi\approx135\,{\rm MeV}$) and the kaons ($m_K\approx495\,{\rm MeV}$) but no 
scalar mesons with similar masses exist~\cite{ParticleDataGroup:2024cfk}. 
Hence we conclude that chiral symmetry is subject to the Nambu-Goldstone
realization: the vacuum is not invariant under the symmetry, rather massless
bosons (called Goldstone bosons) are excited. This scenario is called spontaneous
chiral symmetry breaking. Furthermore non-singlet operators possess non-zero VEVs.
Vector transformations do not mix left- and right-handed 
spinors while axial transformations do. Observing that
\begin{equation}
\bar{\Psi}\Psi=\bar{\Psi}_{\rm L}\Psi_{\rm R}
+\bar{\Psi}_{\rm R}\Psi_{\rm L}\,,
\label{LRdecomp}
\end{equation}
it is perspicuous that the simplest non-singlet operator is $\bar{q}q$ and
that the dynamics of QCD implies a non-zero quark condensate,
\begin{equation}
\langle\bar{q}q\rangle\ne0\,.
\label{nonzerovev}
\end{equation}
This non-zero VEV is called the quark condensate.
Model building therefore requires to (i)~ find a simple mechanism that yields 
such a VEV or (ii)~start from a formulation that has Eq.~(\ref{nonzerovev}) built in.

\section{Large-${\bf N_C}$ QCD}\label{QCD}
\begin{figure}[h]
\centering
\includegraphics[width=12cm,height=2cm]{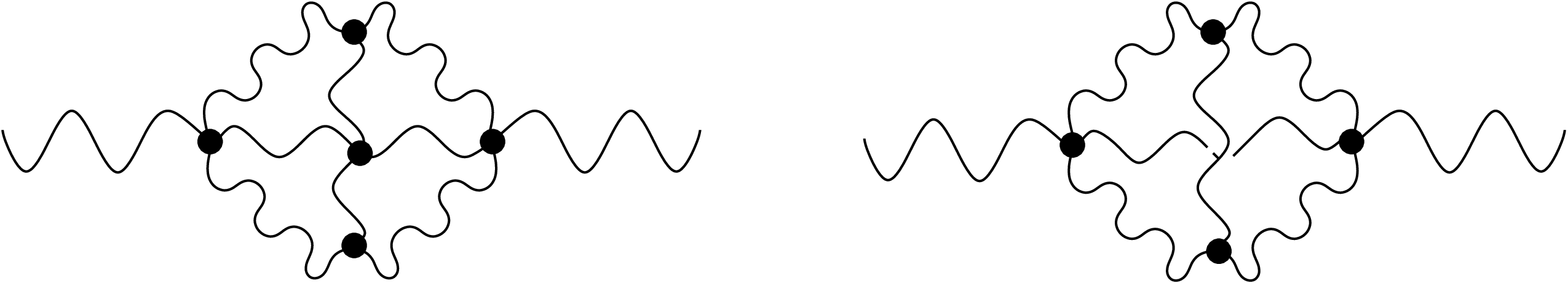}
\caption{A planar (left) and a non-planar (right) diagram that contribute to the
Feynman diagrams for the gluon propagator. The dots indicate three- and
four-point gluon self-interaction vertices.}
\label{fig:g44}
\end{figure}

The number of color degrees of freedom can be generalized from three to $N_C$ so that
$\frac{1}{N_C}$ becomes a hidden expansion parameter for QCD~\cite{tHooft:1973alw}. 
This expansion is mainly a matter of combinatorics. The leading large-$N_C$ components 
of the combinatoric factors are obtained by treating gluons as quark-antiquark compounds 
where each may have an arbitrary color quantum number. In Feynman diagrams (anti)quark 
lines maintain their color quantum number. Demanding that the gluon propagator has a 
smooth large-$N_C$ limit, stipulates the scaling $g\sim\frac{1}{\sqrt{N_C}}$ and leads 
to three important observations for the Feynman diagrams that are relevant in the 
large-$N_C$ limit~\cite{Witten:1979kh}: (i)~internal lines are made from gluons, (ii)~all 
lines are in a single plane, and (iii)~the edges only have quark lines. These diagrams 
scale with $N_C$. Fig.~\ref{fig:g44} shows two typical higher order interaction diagrams 
that contribute to the gluon propagator. The left diagram is planar and $\mathcal{O}(N_C)$,
but the right diagram is non-planar and $\mathcal{O}(1/N_C)$.
The fact that for large-$N_C$ neither non-planar nor diagrams with internal quark 
loops contribute causes the color indices along any Cutkosky-cut of a Feynman diagram 
for a correlation function to be combined in a single trace, rather than products 
of traces. This translates into the statement that in the large-$N_C$ limit all 
intermediate states are quark bilinear color singlets objects, i.e., single 
mesons, taking for granted that QCD is a confining theory. The absence of 
multiple-meson intermediate states implies that the quadratic correlation function 
of a quark bilinear operator $J(x)$ that is inserted into the quark line at
the edge has the simple spectral decomposition
\begin{equation}
\langle J(x) J(y)\rangle=\int \frac{d^4k}{(2\pi)^4}\,
{\rm e}^{ik(x-y)} \sum_i \frac{a_i(k)a^\dagger_i(k)}{k^2-m_i^2}\,.
\label{spectral}
\end{equation}
The above sum goes over all meson states that can couple to $J(x)$
and $a_i\sim\langle0|J|i\rangle$ is the amplitude for $J$
to create the meson $i$ (with mass $m_i$) from the vacuum. The
left-hand-side of Eq.~(\ref{spectral}) is linear in $N_C$ so 
that $a_i\sim\sqrt{N_C}$ and $\lim_{N_C\to\infty}m_i={\rm const.}$.
On the other hand, infinitely many mesons must contribute in the 
sum, Eq.~(\ref{spectral}), in order to reproduce the ${\rm ln}(k^2)$
behavior in the asymptotically free regime of QCD. 

A diagram with
$n$ insertions of $J(x)$ scales with $N_C$ as well and we can write
\begin{equation}
\underbrace{\langle0|J|i_1\rangle\ldots\langle0|J|i_n\rangle}_
{n\,\, {\rm terms}}\Gamma^{(n)}_{i_1\ldots i_n}=
{\mathcal O}\left(N_C^{\frac{n}{2}}\right)\times
\Gamma^{(n)}_{i_1\ldots i_n}={\mathcal O}\left(N_C\right)\,.
\label{npointfctn}\end{equation}
Thus the coupling constant for a vertex with $n$ mesons scales as
\begin{BoxTypeA}{
\begin{equation}
\Gamma^{(n)}_{i_1\ldots i_n}
={\mathcal O}\left(N_C^{1-\frac{n}{2}}\right)\,.
\label{nvertex}\end{equation}}
\end{BoxTypeA}
\noindent
In particular the four-point function vanishes like $\Gamma^{(4)}=g_{\rm eff}\sim\frac{1}{N_C}$
in the large $N_C$-limit. This leads to the conjecture~\cite{Witten:1979kh} that 
QCD is equivalent to an effective weakly (i.e., $g_{\rm eff}\,\to\,0$) 
interacting meson theory. The caveat of having infinitely many mesons should not 
be a problem in the low energy regime where only a few light mesons are involved.

\begin{figure}[t]
\centering
\includegraphics[height=1.8cm,width=9.0cm]{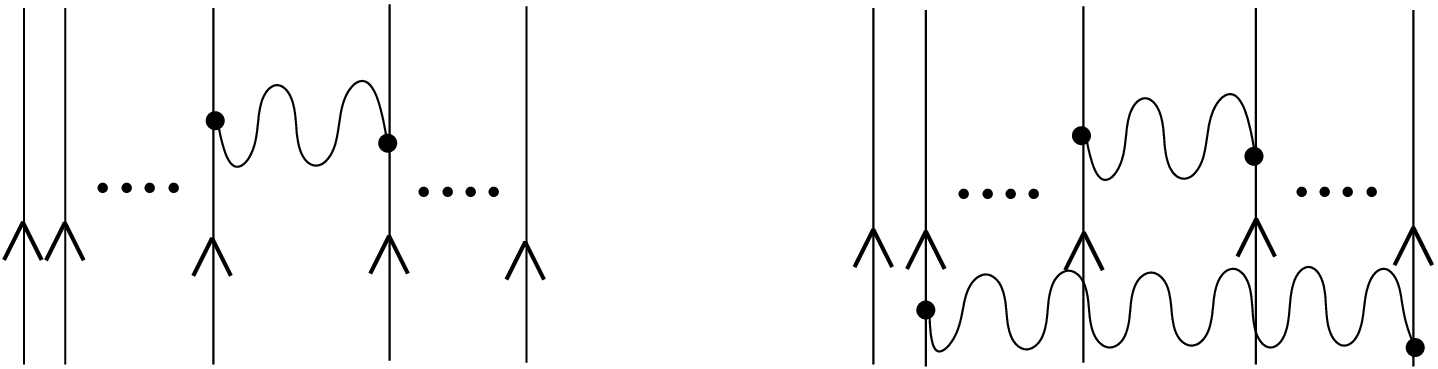}
\caption{\label{fig:baryonlargeN}Feynman diagram for a propagating baryon built from 
$N_C$ quarks with one (left) and two (right) gluon exchanges. Interaction vertices are 
shown as big dots.}
\end{figure}

The large-$N_C$ description of baryons is more complicated because we need (at least)
$N_C$ quark constituents to form a color singlet. In turn the relevant Feynman diagrams 
like those in Fig.~\ref{fig:baryonlargeN} do not have a finite limit as $N_C$ approaches 
infinity.
More specifically we expect a diagram with $m$ single gluon exchanges to scale as
\begin{equation}
g^{2m}\frac{1}{m!}\left[N_C(N_C-1)\right]^{m}
={\mathcal O}\left(N_C^m\right)\,.
\label{gluonexchange} \end{equation}
The proper way to sum all contributions uses many body techniques as
explained in  Ref.~\cite{Witten:1979kh}. Since we are solely interested
in the large-$N_C$ combinatorics it suffices to consider the many body problem 
in a non-relativistic Hartree approach. In that framework
a single quark reacts on an average potential generated
by the $N_C-1$ remaining quarks. In the non-relativistic case only
the two-body forces contribute and spin dependent
forces may been omitted. Then the Hamilton operator reads
\begin{equation}
H=N_CM+\sum_{i=1}^{N_C}\frac{-\Vek{\partial}_i^2}{2M}
-\frac{g_s^2}{N_C}\sum_{i<j}^{N_C}\frac{1}{|\Vek{r}_i-\Vek{r}_j|}\,,
\label{hartree1}
\end{equation}
where $M$ is the mass of a single quark and $g_s=\sqrt{N_C}g$ is ${\mathcal O}(N_C^0)$. 
In the ground state all quarks should be in the $S$-wave channel. 
This motivates the {\it ansatz} for the (scalar) many body wave-function
(anti-symmetrization for color labels not explicitly written):
\begin{equation}
\Psi(\Vek{r}_1,\ldots,\Vek{r}_{N_C})=
\prod_{i=1}^{N_C}\phi(\Vek{r}_i)\,.
\label{hartree2}
\end{equation}
We write $E=N_C\epsilon$ and attempt to apply the variational principle to
\begin{equation}
\langle \Psi |H-E|\Psi\rangle=
-N_C\epsilon+N_CM+\frac{N_C}{2M}\int d^3r\,
\Vek{\partial}\phi^*(\Vek{r})\cdot\Vek{\partial}\phi(\Vek{r})
-\frac{N_C(N_C-1)}{2}\,\frac{g_s^2}{N_C}\int d^3r_1 \int d^3r_2
\frac{|\phi(\Vek{r}_1)|^2|\phi(\Vek{r}_2)|^2}
{|\Vek{r}_1-\Vek{r}_2|}\,.\qquad
\label{hartree3}
\end{equation}
As $N_C\gg1$ we pull out an overall factor $N_C$ and the remaining
variational principle then has a solution with both $\phi$ and 
$\epsilon$ being ${\mathcal O}(N_C^0)$. This implies
that baryon masses are ${\mathcal O}(N_C)$. The fact that $\phi$ has
a smooth large-$N_C$ limit causes the typical extension of a baryon
$\langle r^2\rangle=\langle\Psi|\sum_i\Vek{r}_i^2|\Psi\rangle/N_C$
to be ${\mathcal O}(N_C^0)$. The gluon self-interactions may 
indeed give rise to three and four body forces. In leading order
the respective combinatoric factors are  $N_C^3$ and $N_C^4$.
The coupling constants in the Hartree Hamiltonian are
$g^4=g_s^4/N_C^2$ and $g^6=g_s^6/N_C^3$. Hence these interactions 
add terms $\mathcal{O}\left(N_C\right)$ to Eq.~(\ref{hartree3}) but
do not lead to higher powers of $N_C$. In a similar fashion,
Ref.~\cite{Witten:1979kh} shows that the baryon-baryon scattering
amplitude is also $\mathcal{O}\left(N_C\right)$. The situation is 
different for meson-baryon scattering. Since we can only pick a single 
quark from the meson, the one gluon exchange contribution to the 
energy functional is ${\mathcal O}(N_C^0)$, i.e., of the same 
order as the meson mass but suppressed compared to the baryon mass. 
Hence only the meson reacts in the scattering process while the 
(infinitely heavy) baryon is essentially unaffected. Thus the baryon 
piece ($\phi$) in the $N_C+2$ body Hartree wave-function
\begin{equation}
\Psi(\Vek{r}_1,\ldots\Vek{r}_{N_C};\Vek{x},\Vek{y},t)=
\sum_{P}(-1)^P\Big[\prod_{i=1}^{N_C}\phi(\Vek{r}_i)\Big]
u(\Vek{x},\Vek{y},t)
\label{hartreemb}
\end{equation}
is the same as in Eq.~(\ref{hartree2}) up to corrections of
${\mathcal O}(1/N_C)$. The anti-symmetrization is with respect to
the color labels of the quarks in the baryon and the quark in
the meson. Then the wave-function is symmetric under the exchange of
the quarks' spatial coordinates, $\Vek{r}_i$ and $\Vek{x}$. The meson part
of the wave-function $u(\Vek{x},\Vek{y},t)$ satisfies a {\it linear}
integro-differential equation~\cite{Witten:1979kh},
\begin{equation}
\imu\frac{\partial}{\partial t}u(\Vek{x},\Vek{y},t)=
\frac{-1}{2M}\left(\Vek{\partial}_x^2+\Vek{\partial}_y^2\right)
u(\Vek{x},\Vek{y},t)
-g_s^2\frac{u(\Vek{x},\Vek{y},t)}{|\Vek{x}-\Vek{y}|}
-g_s^2\phi(\Vek{x})\int d^3z\, \phi^\ast(\Vek{z})\,u(\Vek{z},\Vek{y},t)
\left[\frac{1}{|\Vek{z}-\Vek{x}|}+\frac{1}{|\Vek{z}-\Vek{y}|}\right]\,.
\label{hartreeU}\end{equation}
The non-symmetric appearance of the coordinates $\Vek{x}$ and $\Vek{y}$
under the integral stems from the fact the anti-quark coordinate
$\Vek{y}$ is not subject to anti-symmetrization in Eq.~(\ref{hartreemb}).
Eq.~(\ref{hartreeU}) describes the scattering
of a meson in a background potential parameterized by $\phi(\Vek{r})$,
i.e., generated by the baryon field. Obviously $u(\Vek{x},\Vek{y},t)$
is $\mathcal{O}\left(N_C^0\right)$ and so are the meson-baryon scattering
data that are extracted thereof.

Combining these results for the large-$N_C$ scaling behavior of baryon 
properties with the discussion that QCD is equivalent to an effective meson
theory with coupling constant $g_{\rm eff}\sim\frac{1}{N_C}$, we find that 
baryon masses are ${\mathcal O}(1/g_{\rm eff})$ while radii and meson-baryon
scattering amplitudes approach constants at large-$N_C$.

\section{Solitons in field theory}\label{soliton}

In this section we will exemplify that soliton solutions in meson 
theories match the above derived coupling constant dependences for 
baryon properties. For this purpose we briefly review the so-called 
{\it kink} as a simple example for a classical soliton in $D=1+1$ 
dimensions. Ref.~\cite{Ra82} provides a more thorough discussion and
additional examples. The model Lagrangian contains a fourth order
self-interaction 
\begin{equation}
{\mathcal L}_{\rm kink}=
\frac{1}{2}\left[(\partial_t\Phi)^2-(\partial_x\Phi)^2\right]
-\frac{\lambda}{4}\left(\Phi^2-\frac{m^2}{\lambda}\right)^2
\label{kink1}
\end{equation}
for the scalar field $\Phi$. Obviously we identify $\lambda\sim g_{\rm eff}$
as the effective meson coupling constant, $\Gamma^{(4)}$ for two particle 
scattering. To complete the analogy with the large-$N_C$ discussion we demand 
that the mass parameter $m$ remains constant when $\lambda$ is changed.

There are two distinct vacuum configurations
$\Phi=\pm\frac{m}{\sqrt{\lambda}}$. It is straightforward to verify that
the Lagrangian, Eq.~(\ref{kink1}) has static solutions
\begin{equation}
\Phi_\pm(x)=\pm\frac{m}{\sqrt{\lambda}}
{\rm tanh}\,\left(\frac{m}{\sqrt{2}}\,x\right)\,,
\label{kink2}\end{equation}
that mediate between these vacua. The profile function with the positive slope is 
called the kink soliton, the other antikink. The energy density (equal to the 
negative Lagrange density for static fields) is
$\epsilon(x)=\frac{m^4}{2\lambda}{\rm sech}^4\left(\frac{m}{\sqrt{2}}\,x\right)$.
The classical energy and the average extension are computed as spatial integrals 
involving this density:
\begin{equation}
E_{\rm cl}=\int_{-\infty}^\infty dx\,\epsilon(x)=\frac{2\sqrt{2}}{3}\, \frac{m^3}{\lambda}
\qquad{\rm and}\qquad
\langle x^2 \rangle=\frac{1}{E_{\rm cl}}\int_{-\infty}^\infty dx\,x^2\epsilon(x)
=\frac{1}{m^2}\left(\frac{\pi^2}{6}-1\right)\,.
\label{kink3}\end{equation}
Scattering is described by expanding around the kink,
$\Phi(x,t)=\Phi_+(x)+\eta(x,t)$ to harmonic order so that the field equation
for the fluctuation $\eta(x,t)$ reads
\begin{equation}
\left[\partial_t^2-\partial_x^2
+m^2+3m^2{\rm tanh}^2\,\left(\frac{m}{\sqrt{2}}\,x\right)\right]\eta(x,t)=0\,.
\label{kink5}\end{equation}
This wave-equation does not contain $\lambda$ and thus the scattering data 
extracted from $\eta(x,t)$ are $\mathcal{O}(\lambda^0)$.

When we identify the soliton as a baryon with mass $E_{\rm cl}$ and its
fluctuations as scattering mesons, we recognize that this construction 
matches all the results from large-$N_C$ QCD.
The program is now set out: construct a chirally invariant theory for the 
low-mass mesons and find its soliton solutions.

\section{Chiral Lagrangian}\label{chiral}

In this section we will sketch the construction of a chirally invariant
Lagrangian for the low mass pions that would be Goldstone bosons in the limit
of $m_u=m_d=0$. Nevertheless we start from non-zero quark masses and write 
the mass term using Eq.~(\ref{LRdecomp}) as
\begin{equation}
\sum_f m_f \overline{q}_fq_f=
\sum_{i,j=1}^2 \overline{\Psi}_{L,i}\mathcal{M}_{ij}\Psi_{R,i}
+\overline{\Psi}_{R,i}\mathcal{M}^\ast_{ji}\Psi_{L,j}
\qquad {\rm with}\qquad
\mathcal{M}={\rm diag}\left(m_u,m_d\right)\,.
\label{massmatrix}\end{equation}
The lower-case Roman letters are the flavor labels for the up and down quarks
($1\to u$, $2\to d$), {\it cf.} Eq.~(\ref{flavorspinor}). Eventually we 
will also include strange flavors with $\mathcal{M}={\rm diag}\left(m_u,m_d,m_s\right)$.
With these expressions for the mass matrix $\mathcal{M}$, its conjugation in 
Eq.~(\ref{massmatrix}) is redundant. But when we consider $\mathcal{M}$ as a more 
general quantity (called the spurion field) that transforms as 
$\mathcal{M}\,\to\,L\mathcal{M}R^\dagger$, where $L$ and $R$ are 
constant unitary matrices that respectively mix left- and right-handed fermion 
flavors, the mass term is also chirally invariant. We use the same language to
write the transformed quark condensate, Eq.~(\ref{nonzerovev}) as
\begin{equation}
\langle \overline{\Psi}_{R,j}\Psi_{L,i}\rangle=-\sigma \delta_{ij}
\,\rightarrow\, -\sigma\left(LR^\dagger\right)_{ij}
=-\sigma U_{ij}\,,
\label{transfromQbarQ}\end{equation}
where $\sigma$ is a (positive) constant with cubic mass dimension. Note that $U$ is 
also unitary.  From Eq.~(\ref{conscurr}) we observe that left- and right-handed 
fermions transform in the same way for vector transformations, i.e., for $L=R$ 
and the condensate remains unchanged. In case $L\ne R$ the transformed condensate 
describes a different vacuum that would be degenerate for $\mathcal{M}=0$. Applying 
space-time dependent axial transformations to this VEV generates the fields of 
the Goldstone bosons $\phi_a(x)$. Hence we should introduce these bosons via
\begin{BoxTypeA}{
\begin{equation}
U(x)={\rm exp}\left[\imu \phi_a(x)\tau_a/f_\pi\right]\,,
\qquad\mbox{where the $\tau_a$ are the Pauli matrices.}
\label{chiralfield}
\end{equation}}
\end{BoxTypeA}\noindent
The constant $f_\pi$ will be linked to an observable shortly. The next step is to 
establish a Lagrangian for the Goldstone boson fields $\phi_a(x)$. From the 
above discussions we deduce that under constant chiral rotations the so-called 
{\it chiral field} transforms as $U\,\to\,LUR^\dagger$. For $\mathcal{M}=0$, that 
should be an invariance, and since $UU^\dagger=\ID_{2\times2}$, there cannot be a 
potential term in that case. From Lorentz symmetry the leading term must have 
two derivatives and the candidates are
$$
\left[{\rm tr}\left(U\partial_\mu U^\dagger\right)\right]^2\,,\qquad
{\rm tr}\left(\partial_\mu U \partial^\mu U^\dagger\right)\quad {\rm and}\quad
{\rm tr}\left(U\partial_\mu U^\dagger U\partial^\mu U^\dagger\right)\,.
$$
The first term is actually zero and the two others are proportional to each
other because $U\partial^\mu U^\dagger=-\partial^\mu U U^\dagger$. For 
$\mathcal{M}\propto\ID_{2\times2}$ the vector transformation should still be an invariance 
and we can have terms like ${\rm tr}U$. Demanding the canonical normalization
of the fields which we want to identify with the physical pions, the 
first guess effective meson Lagrangian reads
\begin{equation}
\mathcal{L}_\sigma
=\frac{f_\pi^2}{4}{\rm tr}\left(\partial_\mu U \partial^\mu U^\dagger\right)
+\frac{f_\pi^2m_\pi^2}{4}{\rm tr}\left(U+U^\dagger-2\ID_{2\times2}\right)\,.
\label{Lsigma}\end{equation}
In the last term we have subtracted a constant so that the 
energy density of the to-be-constructed soliton vanishes at spatial infinity.
Expanding this Lagrangian yields terms which are quartic in $\phi_a$ or 
$\partial_\mu\phi_a$ with coefficients proportional to $\frac{m_\pi^2}{f_\pi^2}$ 
or $\frac{1}{f_\pi^2}$. The consideration in the first part of 
Section {\it Large $N_C$ QCD} implies that $f_\pi=\mathcal{O}\left(\sqrt{N_C}\right)$. 

To get a number for $f_\pi$ we use empirical information 
from the decay of the charged pions into muons ($\mu$) and their (anti)neutrinos 
($\nu$). The relevant electroweak interaction Lagrangian at energies way below 
the $W$-mass is of a
current-current form
\begin{equation}
\mathcal{L}_{\pi\to\mu\nu_\mu}=\frac{G_F}{2}
L^{\rm (hadr)}_\alpha L^{\rm (lept),\alpha}\,,
\label{Lpimu}
\end{equation}
where $G_F$ is the Fermi constant and $L_\alpha$ are the left-handed currents. The 
leptonic matrix element $\langle0|L^{\rm (lept),\alpha}|\mu\nu\rangle$ is computed 
straightforwardly using Feynman rules. We get its hadronic counterpart from the currents 
in Eq.~(\ref{conscurr}) as $L^{\rm (hadr)}_\alpha=J^{\pm}_\alpha-A^{\pm}_\alpha$, 
where the superscript denotes the pion charge $a=1\pm\imu2$. By symmetry we have
$\langle0|J^a(x)|\pi(p)\rangle=0$, while the axial current matrix element 
defines the pion decay constant\footnote{In general it is momentum dependent, but 
for $q^2\sim m_\pi^2$ that dependence can be ignored.} $f_\pi$ via
$$
\langle 0 |A_\mu^a(x)|\pi^b(q)\rangle=\imu f_\pi \delta_{ab}\,q_\mu\,{\rm e}^{-iqx}\,. 
$$
Hence one can extract this constant from the pion decay width as 
$f_\pi\approx92.2\,{\rm MeV}$~\cite{ParticleDataGroup:2024cfk}\footnote{Unless otherwise stated, 
numerical results in this chapter imply the historic value of $93\,{\rm MeV}$.}.
It remains to verify that this constant is indeed the one in the Lagrangian, Eq.~(\ref{Lsigma}).
To do so we need to get the axial current as the (would-be) Noether current of the transformation
$U\,\to\,U+\delta U$ with $\delta U=\frac{\imu}{2}\left\{\tau_a,U\right\}\epsilon_a$. The result
is
\begin{equation}
\Vek{A}^a_\mu(x)
=\imu\frac{f_\pi^2}{2}{\rm tr} \left[\frac{\tau^a}{2}
\left(U^\dagger\partial_\mu U-U\partial_\mu U^\dagger\right)\right]
=-f_\pi\partial_\mu\phi^a(x)+\mathcal{O}\left(\phi^2\right)\,.
\label{axcurexp}\end{equation}
Taking its matrix element between the vacuum and a one-pion state with momentum $q_\mu$
confirms the stated equality.

Here we used that we can relate the symmetry transformations of QCD to those in the effective 
theory. This allows us to compute relevant matrix elements of hadrons, even though we are not
able to relate the hadron fields to the quark and gluon fields of QCD. This principal concept 
will recur when computing baryon properties.

\section{The Skyrme model}\label{Skyrme}

Before introducing the Skyrme model and discussing (some of) its predictions for baryons
we note that there are numerous review articles
\cite{Holzwarth:1985rb,Zahed:1986qz,Meissner:1987ge,Schwesinger:1988af,Meier:1996ng}
and textbooks~\cite{Makhankov:1993ti,Weigel:2008zz,BrownRho:2010}
that provide detailed background information and references to the original
research publications.

\subsection{The soliton}

The next step is to find a soliton solution from the chiral Lagrangian, Eq.~(\ref{Lsigma}).
However, Derrick's theorem~\cite{Derrick:1964ww} quickly shows that there is no such static
solution. Assume that $U_1(\Vek{r})$ would be such as solution. Then the energy of
$U_s(\Vek{r})=U_1(s\Vek{r})$ must be minimal for $s=1$. Straightforward calculation
yields
\begin{equation}
E[U_s]=\frac{f\pi^2}{4s}\int d^3r\,{\rm tr}
\Big(\Vek{\partial}U_1\Big)\cdot\left(\Vek{\partial}U_1^\dagger\right)
+\frac{f\pi^2m_\pi^2}{4s^3}\int d^3r\,{\rm tr}\left(2\ID_{2\times2}-U_1-U_1^\dagger\right)\,.
\label{Derrick}\end{equation}
The two integrals are non-negative and therefore $\frac{\partial E[U_s]}{\partial s}\ne0$ for
any value of $s$. To remedy the problem, Skyrme added a chirally invariant term
with four derivatives, but only two time derivatives~\cite{Skyrme:1961vq,Skyrme:1962vh}.
Introducing $\alpha_\mu=U^\dagger\partial_\mu U$ the Skyrme model Lagrangian reads
\begin{BoxTypeA}{
\begin{equation}
\mathcal{L}_{\rm Sk}=-\frac{f_\pi^2}{4}\, {\rm tr}\left(\alpha_\mu\alpha^\mu\right)
+\frac{1}{32e^2}\, {\rm tr} \left(\left[\alpha_\mu,\alpha_\nu\right]
\Big[\alpha^\mu,\alpha^\nu\right]\Big)
+\frac{f_\pi^2m_\pi^2}{4}\,{\rm tr}\left(U+U^\dagger-2\ID_{2\times2}\right)\,.
\label{LSkyrme}\end{equation}}
\end{BoxTypeA}\noindent
The expansion of the additional, so-called Skyrme term in powers of $\phi_a$ has the 
leading contribution proportional to $\frac{1}{e^2f_\pi^4}\left(\partial \phi_a\right)^4$. Hence
the Skyrme parameter scales as $e=\mathcal{O}(1/\sqrt{N_C})$.

To construct the soliton solution we consider a spherically 
symmetric {\it ansatz} which is called {\it hedgehog}
configuration and was already explored by Pauli~\cite{Pa46}
\begin{BoxTypeA}{
\begin{equation}
U(\Vek{x})=U_{\rm H}(\Vek{x})\equiv{\rm exp} 
\left(\imu\Vek{\tau}\cdot\hat{\Vek{x}}\,F(r)\right)
\qquad {\rm with} \qquad r=|\Vek{x}|\,. 
\label{hedgehog1}
\end{equation}}
\end{BoxTypeA}\noindent
The radial function $F(r)$ is called the chiral angle. The chiral field is in flavor space, 
i.e., isospin for two flavors. The expression {\it hedgehog} refers to isovectors 
pointing away from the origin, similar to the spines of a hedgehog. This is indicated by 
the picture on the front page of this article. Substituting the {\it ansatz} from 
Eq.~(\ref{hedgehog1}) yields the classical energy functional
\begin{equation}
E_{\rm cl}[F]=\frac{2\pi f_\pi}{e}\int_0^\infty dx\left\{
x^2F^{\prime2}+2{\rm sin}^2F+2\mu_\pi^2x^2\left(1-{\rm cos}F\right)
+{\rm sin}^2F\left(2F^{\prime2}+\frac{{\rm sin}^2F}{x^2}\right)\right\}\,.
\label{Skeng}
\end{equation}
Dimensionless quantities $x=ef_\pi r$ and  $\mu_\pi=m_\pi/(ef_\pi)$ have been 
introduced and the prime denotes the derivative with respect to $x$. Since
$\mu_\pi=\mathcal{O}(N_C^0)$, the $N_C$ scaling is fully determined by the 
coefficient of the spatial integral and we immediately deduce that indeed
$E_{\rm cl}=\mathcal{O}(N_C)$. Upon variation we find the stationary condition,
\begin{equation}
\left(x^2+2{\rm sin}^2F\right)F^{\prime\prime}
+2xF^\prime-{\rm sin}2F-{\rm sin}2F\left(\frac{{\rm sin}^2F}{x^2}
+F^{\prime2}\right)=\mu_\pi^2x^2{\rm sin}F\,.
\label{eqmotionF}
\end{equation}
Asymptotically the chiral field should approach the vacuum value $U\,\to\,\ID_{2\times2}$. 
Linearizing the stationary condition leads to
\setlength{\unitlength}{1cm}
\begin{equation}
F(x)\quad
\begin{array}{c}
\begin{picture}(0,0)
\put(-0.7,-0.12){\vector(1,0){1.5}}
\end{picture}\cr
\mbox{\footnotesize $r\to\infty$}
\end{array}\quad\quad
c\left(1+\mu_\pi x\right)\frac{{\rm e}^{-\mu_\pi x}}{x^2}\,,
\label{asympF}\end{equation}
with some constant $c$. Though the pion mass term originates from the small explicit 
breaking of chiral symmetry and typically contributes only marginally to $E_{\rm cl}$,
it induces an exponential decay of the chiral angle without which some of the
integrals to be computed later would not converge. 

We have determined the boundary condition for $x\,\to\,\infty$ but still need to find 
$F(0)$. To this end we note that the Lagrangian, Eq.~(\ref{LSkyrme}) does not reflect 
the pseudoscalar nature of the pion which relates to the simultaneous transformation
of the fields and the spatial coordinates
$$
\phi_a(\Vek{x},t)\,\to\, -\phi_a(\Vek{-x},t)
\qquad\Longleftrightarrow\qquad U(\Vek{x},t)\,\to\, U^\dagger(\Vek{-x},t)\,.
$$
However, the action associated with the Lagrangian, Eq.~(\ref{LSkyrme}) is invariant 
under $\Vek{x}\,\to\,-\Vek{x}$ and $U(\Vek{x},t)\,\to\, U^\dagger(\Vek{x},t)$
individually. To break that, a contribution with odd powers in $\alpha_\mu$
would be needed, which is not possible in four space-time dimensions. On the level 
of the field equation the pseudoscalar nature can however, be implemented. This is 
achieved by the additional $\lambda$ term in
\begin{equation}
\frac{f_\pi^2}{2}\,\partial_\mu\alpha^\mu
+\frac{1}{8e^2}\partial_\mu\Big(\alpha_\nu\left[\alpha^\mu,\alpha^\nu\right]\Big)
-\frac{f_\pi^2m_\pi^2}{2}\left(U-U^\dagger\right)
+5i\lambda\epsilon_{\mu\nu\rho\sigma} 
\alpha^\mu\alpha^\nu\alpha^\rho\alpha^\sigma=0\,.
\label{variation4}
\end{equation}
The additional term does not effect the field equations in the static case. 
Even though it cannot be derived from the variation of a local Lagrangian, 
Witten showed in Refs.~\cite{Witten:1983tw,Witten:1983tx} that it can 
be obtained when applying the variational principle to the so-called
Wess-Zumino term
\begin{equation}
\Gamma_{\rm WZ}=\imu\lambda\,
\epsilon_{\mu\nu\rho\sigma\tau} \int_{M_5} d^5x\, {\rm tr}\,
\left[\alpha^\mu\alpha^\nu\alpha^\rho\alpha^\sigma\alpha^\tau\right]\,,
\label{WZterm0}\end{equation}
where the boundary of the five-dimensional manifold, $M_5$ is Minkowski space. 
The physics requirement that gauging $\Gamma_{\rm WZ}$ with respect to the 
electromagnetic interaction reproduces the QCD result for the neutral pion 
decay $\pi^0\to\gamma\gamma$ determines the so-far unknown 
coefficient\footnote{One could equally well choose the complement of
$M_5$. This should not alter the physics in the sense that the 
path integral does not depend on that choice. Then 
$\imu\lambda \int_{\mathbb{R}^5}d^5x\,\ldots =2\pi n$. That is, the 
numerator in Eq.~(\ref{WZterm1}), must be an integer.}
\begin{BoxTypeA}{
\begin{equation}
\Gamma_{\rm WZ}=\imu\frac{N_C}{240\pi^2}\,\epsilon_{\mu\nu\rho\sigma\tau}
\int_{M_5} d^5x\, {\rm tr}\,
\left[\alpha^\mu\alpha^\nu\alpha^\rho\alpha^\sigma\alpha^\tau\right]\,.
\label{WZterm1}\end{equation}}
\end{BoxTypeA}\noindent
We recognize that each term in Eq.~(\ref{variation4}) is linear in $N_C$. As in the 
electromagnetic case, the interaction with a (background) baryon gauge field $b^\mu$ 
is incorporated by the gauge principle and we need to 
gauge the $U_V(1)$ symmetry. Since $U$ by itself is not altered by this transformation
no local Lagrangian will contribute to that interaction. However, the non-local
Wess-Zumino term produces a term that is linear in the baryon gauge field\footnote{A 
detailed calculation is given in App.~C of Ref.~\cite{Weigel:2008zz}
based on the techniques from Ref.~\cite{Kaymakcalan:1983qq}.}:\\[-7mm]
\begin{equation}
\mathcal{L}_{b.-int.}=b^\mu B_\mu +\mathcal{O}(b_\mu^2)
\quad{\rm with}\quad
B_\mu=\frac{1}{24\pi^2}\,\epsilon_{\mu\nu\rho\sigma}\,
{\rm tr} \left[\alpha^\nu\alpha^\rho\alpha^\sigma\right]\,. 
\label{barcurWZ}\end{equation}
In electrodynamics the electromagnetic field is contracted with the 
electromagnetic current. By analogy, $B_\mu$ is the baryon number current.
In particular $B=\int d^3r\,B_0$ is the baryon number. For the hedgehog
configuration, Eq.~(\ref{hedgehog1}) we get
\begin{equation}
B[F]=-\int d^3r\,F^\prime\,\frac{\sin^2F}{2\pi^2}=
\frac{1}{\pi}\left[F(0)-F(\infty)\right]\,.
\label{Bnumber}\end{equation}
\begin{wrapfigure}{r}{5.5cm}
\begin{center}
~\vskip-8mm
\includegraphics[width=5.0cm,height=3.5cm]{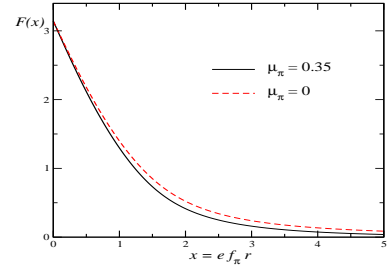}
\end{center}
~\vskip-6mm
\caption{\label{fig:Xangle}Chiral angle solving the   
differential equation ~(\ref{eqmotionF}).}
\end{wrapfigure}
This finally determines $F(0)=\pi$ for a unit baryon number configuration.
The solutions for both a non-zero pion mass ($\mu_\pi=0.35$) and the 
chiral limit ($\mu_\pi=0$) are displayed in Fig.~\ref{fig:Xangle}.

In the chiral limit a few general statements can be made. With $\mu_\pi=0$
the integrand in Eq.~(\ref{Skeng}) is free of any parameters and the classical
energy of the soliton becomes $E_{\rm cl}\approx72.9\frac{f_\pi}{e}$. 
Furthermore, from 
\begin{equation}
{\rm tr}\left(\frac{f_\pi}{2}\alpha_i
\pm\frac{1}{4e}\epsilon_{ijk}\alpha_j\alpha_k\right)^2
={\rm tr}\left(\frac{f_\pi^2}{4}\alpha_i\alpha_i
+\frac{1}{32e^2}\left[\alpha_i,\alpha_j\right]^2
\pm6\frac{\pi^2}{e}f_\pi B_0\right)
\label{Bogobound1}
\end{equation}
we obtain the bound $E_{\rm cl}\ge 6\frac{\pi^2}{e}f_\pi|B_0|$.
This relation is often called the Bogomol'ny~\cite{Bogomolny:1975de} bound 
because of its similarity to the energy bound for the 't Hooft-Polyakov 
monopole~\cite{'tHooft:1974qc,Polyakov:1974ek}. Yet, this bound was already 
known to Skyrme~\cite{Skyrme:1961vq}. The unit baryon number hedgehog
solution, Eq.~(\ref{hedgehog1}) exceeds that bound by about 20\%.

\subsection{Quantization}\label{quant}

Except for the baryon number itself, the hedgehog soliton does not have appropriate
quantum numbers, like spin or isospin. These will be generated in the context 
of canonical quantization by elevating the Noether charges for (iso)rotations 
to generators. To find these Noether charges we require time dependent 
solutions to the non-linear field equation~(\ref{variation4}) which, 
unfortunately, are not known. To approximate them we observe that global
(iso)rotations of the chiral field are zero modes that do not cost energy. 
It is thus suggestive that taking those rotations to be time dependent will 
be a suitable approximation. For this (adiabatic) approximation we introduce 
collective $SU(2)$ rotations as 
\begin{equation}
U(\Vek{x},t)=A(t)U_H(\Vek{x})A^\dagger(t)
\qquad {\rm with}\qquad A(t)\in SU(2)\,,
\label{coll1}\end{equation}
where $U_H(\Vek{x})$ is the hedgehog configuration from Eq.~(\ref{hedgehog1}).
At first sight, this seems as we had collective coordinates only for the 
iso-rotations. However, we observe 
\begin{equation}
A(t)\hat{\Vek{x}}\cdot\Vek{\tau}A^\dagger(t)=\hat{x}_aD_{ab}(A)\tau_b
\qquad \mbox{with the $SO(3)$ matrix}\qquad
D_{ab}(A)=\frac{1}{2}{\rm tr}\left[\tau_aA\tau_bA^\dagger\right]\,.
\label{coll2}\end{equation}
Hence, the collective coordinates for iso-rotations are simultaneously
collective coordinates for spatial rotations. The equivalence of spatial
and iso-rotations is a key feature of the hedgehog configuration.
The Noether charges are associated with the (infinitesimal) transformations
\begin{equation}
U\to U-\Vek{\epsilon}\cdot\left[U,\frac{\Vek{\tau}}{2}\right]+\ldots
\quad {\rm and}\quad
U\to U-\Vek{\epsilon}^\prime\cdot
\left[\imu\Vek{x}\times\Vek{\partial},U\right]+\ldots\,.
\label{inftrans} \end{equation}
for iso- and spatial rotations, respectively. At this point it is advantageous
to write the time derivative (denoted by a dot) of the collective coordinates in 
terms of angular velocities 
\begin{equation}
\Vek{\Omega}=-\imu{\rm tr}\left(A^\dagger(t)\dot{A}(t)\Vek{\tau}\right)
\label{angvel}\end{equation}
so that
\begin{equation}
\dot{U}=A\left[\frac{\imu}{2}\Vek{\tau}\cdot\Vek{\Omega},U_H\right]A^\dagger
=\left[\frac{\imu}{2}\Vek{\tau}\cdot\Vek{\Omega}^\prime,U\right]
\qquad {\rm with}\qquad \Omega_a=\Omega_b^\prime D_{ba}\,.
\label{dotU}\end{equation}
We immediately recognize the commutator from the iso-rotation which gives
easy access to the iso-spin operator (generator):
\begin{equation}
\Vek{I}=-\int d^3x\, {\rm tr}
\left\{\frac{\partial \mathcal{L}(U)}{\partial\dot{U}}
\frac{\partial \dot{U}}{\partial\Vek{\Omega}^\prime}+{\rm h.c.}\right\}
=-\frac{\partial L[U]}{\partial \Vek{\Omega}^\prime}
=-D\cdot \frac{\partial L[U]}{\partial \Vek{\Omega}}\,.
\label{isospin3}\end{equation}
Similarly the hedgehog identity, Eq.~(\ref{coll2}) yields
\begin{equation}
\left[i\Vek{x}\times\Vek{\partial},U\right]=
-\left[U,A\frac{\Vek{\tau}}{2}A^\dagger\right]=
-D^\dagger\cdot\left[U,\frac{\Vek{\tau}}{2}\right]
\label{spinisospin1} \end{equation}
and therefore the spin operator (generator) becomes
\begin{equation}
\Vek{J}=-D^\dagger\cdot\Vek{I}=\frac{\partial L[U]}{\partial \Vek{\Omega}}\,.
\label{spin3}\end{equation}
We immediately observe that $\Vek{I}^2=\Vek{J}^2$ implying that the hedgehog soliton 
describes baryons with equal spin and iso-spin; like the nucleon ($I=J=\frac{1}{2}$) 
or the $\Delta$-resonance ($I=J=\frac{3}{2}$). At this point it is not yet clear 
why spin and and iso-spin should be restricted to half-integer quantum 
numbers. Obtaining that result needs a non-trivial contribution from the 
Wess-Zumino term, Eq.~(\ref{WZterm1}) which only occurs for three 
flavors, {\it cf.} Section {\it Extension to SU(3)}.

Global iso-rotations are parameterized by a left-multiplication of $A$ 
with a constant $SU(2)$ matrix. The hedgehog structure not only leads to 
Eq.~(\ref{spinisospin1}) but also has global spatial rotations parameterized 
by a right-multiplication of $A$ with a constant $SU(2)$ matrix. Elevating
$\Vek{I}$ and $\Vek{J}$ to the corresponding generators therefore implies
(The relative sign originates from Eq.~(\ref{spinisospin1}).)
\begin{equation}
\left[\Vek{I},A\right]=-\frac{\Vek{\tau}}{2}A
\qquad {\rm and}\qquad 
\left[\Vek{J},A\right]=A\frac{\Vek{\tau}}{2}\,.
\label{generator1}\end{equation}
Introducing Euler angles\footnote{The authors of Ref.~\cite{Adkins:1983ya} 
originally parameterized $A(t)=a_4(t)+\imu\Vek{a}(t)\cdot\Vek{\tau}$. This 
complicates matters because of the constraint $\sum_{i=1}^4a_i^2(t)=1$ and 
evades a straightforward generalization to $SU(3)$.} $\Phi$, $\Theta$ and 
$\Psi$ via
\begin{equation}
A(t)={\rm exp}\left(-\imu\Phi(t)\frac{\tau_3}{2}\right)
{\rm exp}\left(-\imu\Theta(t)\frac{\tau_2}{2}\right)
{\rm exp}\left(-\imu\Psi(t)\frac{\tau_3}{2}\right)
\label{generator2}\end{equation}
allows one to express the generators in terms of differential operators
with respect to these Euler angles. These are given in the literature,
{\it cf.\@} the textbook~\cite{Var88}. Here it suffices to list
\begin{equation}
I_z=-\imu\frac{\partial}{\partial \Phi}
\qquad {\rm and}\qquad 
J_z=\imu\frac{\partial}{\partial \Psi}\,.
\label{generator3}\end{equation}
Either set of generators is subject to an $SU(2)$ algebra
$\left[I_a,I_b\right]=\imu\epsilon_{abc}I_c$ and 
$\left[J_a,J_b\right]=\imu\epsilon_{abc}J_c$.

Lastly, it remains to find the $\Vek{\Omega}$ dependence of the 
Lagrange function. This is straightforward: we substitute the 
rotating hedgehog configuration, Eq.~(\ref{coll1}) into the 
Lagrangian, Eq.~(\ref{LSkyrme}) and integrate over space:
\begin{equation}
L[U]=\frac{1}{2}\alpha^2[F]\,\Vek{\Omega}\cdot\Vek{\Omega}-E_{\rm cl}[F]
\qquad {\rm with}\qquad
\alpha^2[F]=\frac{8\pi}{3f_\pi e^3}\int_0^\infty dx\, x^2 {\rm sin}^2F
\left[1+\left(F^{\prime2}+\frac{{\rm sin}^2F}{x^2}\right)\right]\,.
\label{lagrigid1}\end{equation}
Again, we can easily read off that the moment of inertia scales like
$\alpha^2=\mathcal{O}(N_C)$ and in the chiral limit ($\mu_\pi=0$) we
have the universal result $\alpha^2=51.2/(e^3f_\pi)$. Inverting 
$\Vek{\Omega}=\frac{1}{\alpha^2}\Vek{J}$
we are in the position to write down a baryon mass formula 
\begin{equation}
E_{I=J}=\frac{J(J+1)}{2\alpha^2}+E_{\rm cl}
\qquad \Longrightarrow\qquad
E_N=\frac{3}{8\alpha^2}+E_{\rm cl}
\quad{\rm and}\quad
E_{\Delta}=\frac{15}{8\alpha^2}+E_{\rm cl}\,,
\label{massformula1}\end{equation}
where we used that nucleon and $\Delta$-resonance have (iso)spin $\frac{1}{2}$ and 
$\frac{3}{2}$, respectively. Obviously, the correction from generating spin and 
isospin quantum numbers is $\mathcal{O}(1/N_C)$.  Using the above universal 
results for $E_{\rm cl}$ and $\alpha^2$ when $\mu_\pi=0$ would then predict 
$f_\pi\approx64.5\,{\rm MeV}$ and $e\approx5.4$. Stated conversely, when using
the empirical value for the pion decay constant from Section {\it chiral Lagrangian}, 
the Skyrme model overestimates the nucleon mass by 30\% or more. Taking the actual 
pion mass does not alter that conclusion significantly~\cite{Adkins:1983hy}. 
Adjusting the pion decay constant was the original approach in 
Ref.~\cite{Adkins:1983ya}. Nowadays one considers quantum corrects to the classical 
mass that are $\mathcal{O}(N_C^0)$, and neither depend on spin nor isospin, to give a 
substantial negative contribution to the energy~\cite{Meier:1996ng}. It is
therefore customary to keep $f_\pi=93\,{\rm MeV}$ as well as $m_\pi=135\,{\rm MeV}$
and tune $e\approx4.25$ to reproduce $E_\Delta-E_N=293\,{\rm MeV}$; even though
these corrections are not fully under control since the model is not renormalizable.
Here one may also remark that the chiral-limit decay constant is smaller than 
the physical value~\cite{Langacker:1973hh}.

\subsection{Static properties}\label{static}

To compute static properties we need to determine the symmetry currents from vector and 
axial-vector transformations. That is, we seek the Skyrme model analog of Eq.~(\ref{conscurr}).
A convenient method (already mentioned above) is to upgrade these global symmetries to local 
ones by extending the Skyrme model action with appropriate external gauge fields ({\it e.g.\@} 
the gauge fields of the electroweak interactions). The Noether currents are subsequently 
read off as the objects which couple linearly to these gauge fields. This procedure 
results in Lorentz covariant expressions for the currents and is 
especially appropriate for the Wess-Zumino term (\ref{WZterm0}) because this non-local 
term can only be made gauge invariant by a trial and error type  
procedure~\cite{Witten:1983tw,Kaymakcalan:1983qq,Kaymakcalan:1984bz}. Substituting
the rotating hedgehog, Eq.~(\ref{coll1}) into these expressions yields the densities
\begin{equation}
\frac{1}{3}V_0^0=\frac{1}{2}b(r)\,, \qquad
\frac{1}{3}V_i^0=\frac{1}{2}b(r)\epsilon_{ijk}\Omega_jx_k\,, \qquad
V_0^a=-\frac{2}{3}v(r)D_{ai}\Omega_i\qquad {\rm and}\qquad
V_i^a=\frac{v(r)}{r^2}\epsilon_{ijk}x_jD_{ak}\,,
\label{densities1}\end{equation}
with (here primes are derivatives with respect to $r$)
\begin{equation}
b(r)=-F^\prime\,\frac{{\rm sin}^2F}{2\pi^2}
\qquad {\rm and}\qquad
v(r)={\rm sin}^2F\left[f_\pi^2+\frac{1}{e^2}
\left(F^{\prime2}+\frac{{\rm sin}^2F}{r^2}\right)\right]\,,
\label{densities2} 
\end{equation}
for the vector currents. Subscripts are Lorentz labels while the superscripts
refer to the flavor, i.e., isospin quantum number. Similarly the non-zero
elements of the axial vector current are
\begin{equation}
A_i^a=\left[a_1(r)\delta_{ik}+a_2(r)\hat{x}_i\hat{x_k}\right]D_{ak}\,,
\label{axialcurr}
\end{equation}
with the densities
\begin{equation}
a_1(r)=\frac{{\rm sin}2F}{2r}\left[f_\pi^2+\frac{1}{e^2}
\left(F^{\prime2}+\frac{{\rm sin}^2F}{r^2}\right)\right]
\qquad {\rm and}\qquad
a_2(r)=-a_1(r)+F^\prime\left[f_\pi^2+\frac{2}{e^2}
\frac{{\rm sin}^2F}{r^2}\right]\,.
\label{acurradial}
\end{equation}
Here, and in Eq.~(\ref{densities2}), primes denote derivatives with respect to $r$.
While the conservation of the vector current $\partial^\mu V^a_\mu=0$ follows from 
the structure of the currents in Eq.~(\ref{densities1}), the axial vector current analog
$\partial^\mu A^a_\mu=m^2_\pi\hat{x}_a\sin F$ results from Eq.~(\ref{eqmotionF}). 
The fact that it vanishes in the chiral limit is known as PCAC (partially conserved
axial vector current). 

We have already discussed how to relate some of the functions
of the collective coordinates to spin and isospin in the preceding subsection. To 
compute matrix elements of $D_{ai}$ we either use Wigner-$D$ functions as wave-functions,
$D^{(I=J)\ast}_{I_3,-J_3}\left(\Phi,\Theta,\Psi\right)$ and integrate over the Euler 
angles, or note that the left index ($a$) is isospin, the right one ($i$) is spin, hence 
$\langle D_{ai}\rangle\sim I_aJ_i$. The remaining reduced matrix element is easily 
obtained from the expectation value $\langle D_{ai}J_i\rangle=-\langle I_a\rangle$. 
We summarize 
\begin{equation}
\Omega_i=\frac{1}{\alpha^2}J_i\,,\qquad
D_{ai}\Omega_i=-\frac{1}{\alpha^2}I_a\quad {\rm and}\quad
\langle D_{ai}\rangle=-\frac{1}{J(J+1)}I_aJ_i\,.
\label{matrixelements}\end{equation}
With $\int d^3r\,\, v(r)=\frac{3}{2}\alpha^2$, we immediately get the electromagnetic 
charges\footnote{The relative factors are inherited from the quark charge matrix
${\rm diag}\left(\frac{2}{3},-\frac{1}{3}\right)=\frac{1}{6}\ID_{2\times2}+\frac{1}{2}\tau_3$.}
$\langle Q\rangle=\int d^3r\, V_0^{\rm e.m.}=
\int d^3r \left(\frac{1}{3}V_0^0+V_0^3\right)=\frac{1}{2}+I_3$.
We are now prepared to compute nucleon static properties. As an explicit example we 
present the magnetic moment
\begin{equation}
\frac{\mu_N}{\mu_{\rm n.m.}}
=\frac{1}{\mu_{\rm n.m.}}\left\langle N J_3=\frac{1}{2}\left|\int d^3x
\left(\Vek{x}\times\Vek{V}^{\rm e.m.}\right)_3\right|N J_3=\frac{1}{2}\right\rangle
=\frac{2\pi}{3}M_N\int_0^\infty drr^2\left[
\pm\frac{2}{3}v(r)+\frac{r^2}{2\alpha^2}b(r)\right]
\label{magmom1}
\end{equation}
typically measured in nucleon magnetons $\mu_{\rm n.m.}=\frac{e\hbar}{2M_N}$. Here
$M_N=939\,{\rm MeV}$ is the physical nucleon mass not the model prediction because it
enters solely via $\mu_{\rm n.m.}$. The two signs in Eq.~(\ref{magmom1}) refer to 
proton ($+$) and neutron ($+$). Numerical results for two sets of model parameters are 
compared to data in table~\ref{table:nuclstatskyrme}, which also contains the axial 
vector charge $g_a=2\langle p|A_3^3|p\rangle$.
\begin{table}
\centerline{
\begin{tabular}{c|cccccccc}
& $\mu_p$ & $\mu_n$ & $\frac{\mu_p}{\mu_n}$ & $r_p^2$ 
& $r_n^2$ & $r_{I=0}^2$ & $r_{I=1}^2$ & $g_A$\\
\hline
A &$1.78$&$-1.42$&$-1.26$&$0.48$&$-0.23$&$0.25$&$0.71$&$0.90$\\
B &$1.97$&$-1.24$&$-1.59$&$0.77$&$-0.31$&$0.46$&$1.08$&$0.86$\\
Expt. &$2.79$&$-1.91$&$-1.46$&$0.84$&$-0.12$&$0.72$&$0.96$&$1.28$
\end{tabular}}
\caption{\label{table:nuclstatskyrme}Predictions for nucleon static properties 
(magnetic moments, electric radii and axial vector charge) in the Skyrme model 
compared to experimental data~\cite{ParticleDataGroup:2024cfk}.
Two parameter sets are considered: (A: $f_\pi=93\,{\rm MeV}$ $e=4.25$)
and (B:$f_\pi=54\,{\rm MeV}$ $e=4.84$~\cite{Adkins:1983hy}). Either case 
has the physical pion mass $m_\pi=135\,{\rm MeV}$.}
\end{table}
Both sets reproduce the empirical $\Delta$-nucleon mass difference. While set A uses 
the physical pion decay constant, set B tunes it to match the nucleon mass. We see 
that, in either case, the model predictions are too low in magnitude by up to 30\%. 
The main cause is the too narrow chiral angle, needed to get $\alpha^2$ small enough 
at the expense of small radii. 

Magnetic radii (not listed in table~\ref{table:nuclstatskyrme}) have an additional 
factor $r^2$ under the integral, Eq.~(\ref{magmom1}) and in view of Eq.~(\ref{asympF}) 
require $m_\pi\ne0$ to be finite.

An important result is $A_0^0\equiv0$, i.e., the axial singlet current
vanishes identically in the Skyrme model. This has been viewed as an
elegant explanation of the {\it proton spin puzzle} \cite{Brodsky:1988ip}
and ignited a kind of a revival (after most of the static properties had 
been calculated) of chiral soliton models. Among other topics we will get 
back to this one in the Section {\it Vector mesons}.

The computation of momentum dependent form factors is also possible. It requires 
an additional time-dependent collective coordinate for the position of 
the soliton, $\Vek{R}(t)$:
\begin{equation}
U(\Vek{x},t)=A(t)U_H(\Vek{x}-\Vek{R}(t))A^\dagger(t)\,.
\label{momentum1}\end{equation}
Its canonical quantization leads to a plane wave wave-function with
momentum $\Vek{P}$. The form factors are then obtained as Fourier transforms
of the above listed densities. In fact, the above calculation just corresponds 
to the $\Vek{P}=0$ case. In Ref.~\cite{Braaten:1986md} the calculational
techniques are detailed and the Skyrme model predictions are compared to
the standard dipole parameterization of experimental data based on
a vector meson dominance picture.

As an alternative to stabilize the soliton, a sixth order term 
has been considered~\cite{Jackson:1985yz}: 
$\mathcal{L}_6=-\frac{e^2_6}{4}B_\mu B^\mu$. No substantial differences
have been encountered.

\subsection{Pion-nucleon scattering}\label{piN}

As for the simple kink model in Section {\it Solitons in field theory}, meson-baryon scattering is 
analyzed in terms of small amplitude fluctuations about the soliton. A suggestive
parameterization  is~\cite{Hayashi:1984bc,Walliser:1984wn}
\begin{equation}
U(\Vek{x},t)=A\, {\rm exp}\left[\imu\Vek{\tau}\cdot
\hat{\Vek{x}}F(r)+\imu\Vek{\tau}\cdot\Vek{\eta}(\Vek{x},t)\right]\,A^\dagger\,.
\label{fluctuation1}\end{equation}
In the adiabatic approximation the collective rotation $A$ is treated as
time-independent and the linearized field equations 
\begin{equation}
M^{\mu\nu}_{ij}(\Vek{x})\partial_\mu\partial_\nu\eta_j(\Vek{x},t)
+V_{ij}(\Vek{x})\eta_j(\Vek{x},t)=0
\label{fluctuation2}\end{equation}
reduce to Klein-Gordon equations asymptotically, i.e., $r\to\infty$:
$M^{\mu\nu}_{ij}(\Vek{x})\to M^{(0)\mu\nu}_{ij}=g^{\mu\nu}\delta_{ij}$ and
$V_{ij}(\Vek{x})\to V^{(0)}_{ij}=m_\pi^2\delta_{ij}$.
The space dependence of the matrices $M^{\mu\nu}_{ij}(\Vek{x})$ and 
$V_{ij}(\Vek{x})$ is generated by the soliton, $U_H(\Vek{x})$.
The fluctuations have the partial wave decomposition
\begin{equation}
\Vek{\eta}(\Vek{x},t)={\rm e}^{-\imu \omega t}\sum_{GLM}\eta_{GL}(r,\omega)
\Vek{Y}_{GLM}(\hat{\Vek{x}})\,,
\label{granddecomp}\end{equation}
where $\Vek{Y}_{GLM}$ are vector spherical harmonics for the conserved 
grand spin, $\Vek{G}=\Vek{L}+\Vek{I}$ which is the operator sum of the pion's orbital
angular momentum and its isospin. The conservation of grand spin results
from the hedgehog being invariant under a combination of spatial and 
iso-rotations, {\it cf.\@} Eq.~(\ref{spinisospin1}). The time dependence factorizes
because the soliton is static, at least in the adiabatic approximation.

The scattering amplitudes are extracted from the asymptotic behavior 
of the regular radial functions in Eq.~(\ref{granddecomp}). The resulting,
so-called intrinsic scattering matrix elements are labeled as 
$\widetilde{S}^{G}_{\ell,\ell^\prime}$, where 
$\ell,\ell^\prime\in\left[G-1,G,G+1\right]$
are the possible pion angular momenta for prescribed grand spin $G$.
The physical $S$-matrix elements are finally obtained from the 
recoupling scheme~\cite{Hayashi:1984bc}\footnote{See 
Refs.~\cite{Mattis:1986wc,Schwesinger:1988af} for the recoupling scheme
for general meson-baryon scattering.}
\begin{equation}
S_{\ell(2I)(2J)}=
\sum_{G}\zeta\zeta^\prime \widetilde{S}^{G}_{\ell,\ell^\prime}
\qquad \mbox{with the recoupling coefficient}\qquad
\zeta=(-1)^{\ell+\frac{1}{2}+J}
\left[2(2G+1)\right]^{\frac{1}{2}}
\begin{Bmatrix}
I & 1 & \frac{1}{2}\cr
\ell & J & G
\end{Bmatrix}\,.
\label{recoupling2} \end{equation}
Here, $J$ and $I$ are total spin and isospin. The entry '$1$' arises from the 
pion isospin while '$\fract{1}{2}$' reflects the nucleon spin. The interesting
observation is that there are more elements of the physical $S$-matrix than 
there are for the intrinsic one, $\widetilde{S}$. The latter can thus be 
eliminated in favor of linear relations among the physical $S$-matrix 
elements. For the elastic pion-nucleon scattering this predicts linear
dependences of the $I=\frac{3}{2}$ and $I=\frac{1}{2}$ scattering
amplitudes~\cite{Hayashi:1984bc},
\begin{equation}
2\left(2\ell+1\right)S_{\ell(3)(2\ell+1)}=3\ell S_{\ell(1)(2\ell+1)}
+\left(\ell+2\right)S_{\ell(1)(2\ell+1)}
\qquad {\rm and}\qquad
2\left(2\ell+1\right)S_{\ell(3)(2\ell-1)}=\left(\ell-1\right)S_{\ell(1)(2L-1)}+
3\left(\ell+2\right)S_{\ell(1)(2\ell-1)}\,.
\label{ch8:recouprel}
\end{equation}
These relations are model independent in the sense that they are valid
irrespective of the details of the Lagrangian, as long as scattering is
explored in the adiabatic approximation. They are well satisfied by data 
for $\ell\ge3$ but for lower angular momenta the approximation is not 
adequate. This short-coming is closely related to the {\it Yukawa} 
problem of soliton models: by definition, there is no term linear in the 
fluctuations about the soliton and thus there is no Yukawa coupling to a
resonance. A linear term nevertheless emerges in the context
of the collective coordinate quantization because Eq.~(\ref{coll1}) is
not an exact solution. Numerous attempts have been made to solve this 
problem, {\it cf.\@} Ref.~\cite{Holzwarth:1990eh} and references therein.
Another problem arises from the fact that in the Skyrme model the metric 
tensor, Eq.~(\ref{fluctuation2}) has $M_{ij}^{00}\ne M_{ij}^{rr}$.
As a result, the phase shifts do not saturate at large energies and the 
identification of genuine resonances is indistinct. That problem, however,
has a solution and we will reflect on it in Section {\it Vector mesons}.

Another important result arises from Eq.~(\ref{fluctuation2}): for the
parameterization, Eq.~(\ref{granddecomp}) all energy eigenvalues $\omega$
are found to be real. In fact the lowest eigenvalues are the zero modes
associated with translations and (iso)rotations~\cite{Walliser:1984wn}. Hence 
the solution arising from the {\it ansatz}, Eq.~(\ref{hedgehog1}) is stable, 
at least locally.

\section{Extension to $SU(3)$}\label{SU3}

Empirically chiral symmetry can be considered to be approximately realized
when pseudoscalar and vector mesons with the same quark content differ
significantly in mass. For the up and down flavors this is obviously
the case: $m_\pi=135\,{\rm MeV}\ll770\,{\rm MeV}\approx m_\rho$. For mesons 
containing an up or a down and a strange (anti)quark, it is approximately true: 
$m_K=495\,{\rm MeV}<890\,{\rm MeV}\approx m_{K^\ast}$~\cite{ParticleDataGroup:2024cfk}.
Therefore it is legitimate to extend the Skyrme model to flavor $SU(3)$.
Without flavor symmetry breaking, any embedding of the hedgehog,
Eq.~(\ref{hedgehog1}) would be a solution to the field equations. Including
(some) explicit breaking as measured by $m_K>m_\pi$ has a larger classical
energy when the hedgehog has non-zero kaon components. Whence the initial 
configuration in the three flavor model 
is~\cite{Witten:1983tx,Guadagnini:1983uv,Mazur:1984yf}
\renewcommand{\arraystretch}{1.6}
\begin{BoxTypeA}{
\begin{equation}
U_H(\Vek{x})=\begin{pmatrix}
\begin{array}{c|c}
\hspace{0.1cm}{\rm exp}\left(i\hat{\Vek{x}} \cdot \Vek{\tau\,} F(r)\right)
\hspace{0.1cm} & \hspace{0.3cm}
\begin{picture}(0,0)
\put(-0.1,-0.15){\mbox{\footnotesize $0$}}
\put(-0.1,0.15){\mbox{\footnotesize $0$}}
\end{picture}\hspace{3.0mm} \\[1mm]
\hline
0\hspace{1.0cm}0 & 1
\end{array}\end{pmatrix}\,,
\label{embedding}
\end{equation}}
\end{BoxTypeA}
\renewcommand{\arraystretch}{1.0}\noindent
where, for the Skyrme model case, $F(r)$ is again the solution to the equation of 
motion, Eq.~(\ref{eqmotionF}) which is displayed in Fig.~\ref{fig:Xangle}. We 
introduce collective coordinates for all zero modes in the $SU(3)$ symmetric model
\begin{equation}
U(\Vek{x},t)=A(t)U_H(\Vek{x})A^\dagger(t)
\label{rotsol}
\end{equation}
to parameterize time dependent configurations as in Eq.~(\ref{rotsol}). Here the 
soliton rotates without deformation (i.e., rigidly) in flavor space. 
Therefore treatments based on Eq.~(\ref{rotsol}) are commonly summarized as 
the {\it rigid rotator approach} (RRA). When we substitute this {\it ansatz} 
into the sum $\int d^4x\,{\mathcal L}_{\rm Sk}+\Gamma_{\rm WZ}$, we obtain the 
Lagrange function for the collective coordinates
\begin{equation}
L(A,\Omega_a)=-E_{\rm cl}+\frac{1}{2}\alpha^2\sum_{i=1}^3\Omega_i^2
+\frac{1}{2}\beta^2\sum_{\alpha=4}^7\Omega_\alpha^2
-\frac{N_CB}{2\sqrt3}\Omega_8\,,
\label{lagrigid2}
\end{equation}
with the three flavor generalization of Eq.~(\ref{angvel}):
$\Omega_a=-\imu{\rm tr}\left[A^\dagger(t)\dot{A}(t)\lambda_a\right]$.
A second moment of inertia 
\begin{equation}
\beta^2=\frac{\pi}{e^3 f_\pi}\int_0^\infty dx\, x^2\, {\rm sin}^2\frac{F}{2}
\left[4+\left(F^{\prime2}+\frac{2}{x^2}\,{\rm sin}^2F\right)\right]
\label{beta2}
\end{equation}
emerges because the embedding, Eq.~(\ref{embedding}) breaks the symmetry for rotations
into strangeness direction. It is $\mathcal{O}(N_C)$ and typical model results for this inertia 
parameter are about $4\,{\rm GeV}^{-1}$. In the chiral limit we have the universal result 
$\beta^2=56.5/(e^3f_\pi)$. Even though $\left[U_H,\lambda_8\right]=0$, $\Omega_8$ appears in 
Eq.~(\ref{lagrigid2}) because of the non-local nature of the Wess-Zumino term. Quantization 
of this system generalizes Eq.~(\ref{spin3}) by specifying (intrinsic) $SU(3)$ generators
\begin{equation}
R_a=-\frac{\partial L}{\partial\Omega_a}=\begin{cases}
-\alpha^2\Omega_a=-J_a,&a=1,2,3\cr
-\beta^2\Omega_a,&a=4,..,7\cr
\frac{N_CB}{2\sqrt3},&a=8\,.
\end{cases}
\label{Rgen}\end{equation}
Quantization imposes the commutation relations $\left[R_a,R_b\right]=-\imu f_{abc}R_c$,
with $SU(3)$ structure constants $f_{abc}$. The fact
that the first three generators are identified with the spin operator
is a mere consequence of the hedgehog structure that identifies spin
and isospin rotations. We easily find the Hamiltonian from the Legendre
transformation 
\begin{BoxTypeA}{
\begin{equation}
H=E_{\rm cl}+\frac{1}{2\alpha^2}\Vek{J}^2
+\frac{1}{2\beta^2}\sum_{\alpha=4}^7R_\alpha^2
=E_{\rm cl}
+\frac{1}{2}\left(\frac{1}{\alpha^2}-\frac{1}{\beta^2}\right)\Vek{J}^2
+\frac{1}{2\beta^2}C_2[SU(3)]-\frac{N_C^2B^2}{24\beta^2}\,,
\label{hamrigid1}\end{equation}}
\end{BoxTypeA}\noindent
where $C_2[SU(2)]=\sum_{a=1}^8R_a^2$ is the quadratic Casimir
operator of $SU(3)$. The last term in Eq.~(\ref{hamrigid1}) does
not affect mass differences as long as only unit baryon number states
are considered. However, we need to find the eigenstates subject to the 
constraint $Y_R=\frac{2}{\sqrt{3}}R_8=\frac{N_CB}{3}$. This can be
done by a detailed analysis of $SU(3)$ representations. A more intuitive
derivation notes that the well-known Gell-Mann-Nishijima relation for the 
electric charges has the analog $Q_R=R_3+\frac{1}{2}Y_R=-J_3+\frac{1}{2}Y_R$. 
The quark model tells us that possible eigenvalues are
$Q_R=0,\pm\frac{1}{3},\pm\frac{2}{3},\pm1,\pm\frac{4}{3},\ldots$, implying that
half-integer charges are not allowed. For $Y_R=1$ the constraint can thus only 
be consistently accommodated with $J_3$ being half-integer. Hence for 
$N_C=3$ baryons indeed have half-integer spin~\cite{Witten:1983tx}.  Furthermore, 
for $Y_R=1$ the allowed low dimensional $SU(3)$ representations are the octet 
and the decuplet, containing the established spin $\frac{1}{2}$ and $\frac{3}{2}$ 
baryons, respectively. The corresponding $C_2[SU(2)]$ eigenvalues are three and six, 
so that the mass difference between
the spin $\frac{1}{2}$ and $\frac{3}{2}$ baryons is
$M_{\mathbf{10}}-M_{\mathbf{8}}=\frac{3}{2\alpha^2}$, as in the two
flavor version.

Flavor symmetry breaking terms that transform like $(LR^\dagger+{\rm h.c.})$ are 
added to the chiral Lagrangian to account for different masses and decay 
constants $f_K\approx1.22f_\pi$:
\begin{equation}
{\mathcal L}_{\rm sb}=\frac{f_\pi^2 m_\pi^2+2f_K^2 m_K^2}{12}
{\rm tr} \left[ U + U^\dagger - 2 \right]
+ \sqrt{3}\frac{f_\pi^2 m_\pi^2-f_K^2 m_K^2}{6}
{\rm tr}\left[\lambda_8\left(U+U^\dagger\right)\right]
+ \frac{f_K^2-f_\pi^2}{12}
{\rm tr}\left[(1-\sqrt{3}\lambda_8) \left(
U \partial_\mu U^\dagger \partial^\mu U  
+ U^\dagger \partial_\mu U \partial^\mu U^\dagger \right)\right]\,.
\label{lagsb}\end{equation}
Generalizing the definition in Eq.~(\ref{coll2}) to
$D_{ab}(A)=\frac{1}{2}{\rm tr}\left[\lambda_aA\lambda_bA^\dagger\right]$,
${\mathcal L}_{\rm sb}$ adds
\begin{equation}
H_{\rm sb}=\frac{1}{2}\gamma\left(1-D_{88}(A)\right)
\qquad {\rm with}\qquad
\gamma=\frac{4}{3}\int d^3r \left\{
\left(m^2_Kf^2_K-m^2_\pi f^2_\pi\right)\left(1-{\rm \cos}F\right)
+\frac{1}{2}\left(f^2_K-f^2_\pi\right){\rm cos}F
\left(F^{\prime2}+2\frac{{\rm sin}^2F}{r^2}\right)\right\}
\label{Hsb}\end{equation}
to the Hamilton operator, Eq.~(\ref{hamrigid1}). The constraint for $R_8$ is not 
effected. Using the $SU(3)$ Clebsch-Gordan coefficients \cite{deSwart:1963gc} $H_{\rm sb}$
can be treated in stationary perturbation theory~\cite{Guadagnini:1983uv,Park:1989wz}. 
At leading order this yields the model independent predictions
\begin{equation}
2(M_N+M_\Xi)=3M_\Lambda+M_\Sigma\,,\quad
M_\Delta-M_{\Sigma^\ast}=M_{\Sigma^\ast}-M_{\Xi^\ast}
=M_{\Xi^\ast}-M_\Omega
\quad{\rm and}\quad
M_{\Xi^\ast}-M_{\Sigma^\ast}+M_N=\frac{1}{8}\left(11M_\Lambda-3M_\Sigma\right)\,.
\label{Guadagnini}\end{equation}
The first two are the well-established Gell--Mann-Okubo relations but the third one is 
specific to $H_{\rm sb}$ and its treatment in first order perturbation theory. Empirically
it is satisfied within a fraction of a percent 
($1088\,{\rm MeV}$ versus \@ $1087\,{\rm MeV}$). On the other hand, this first order treatment
also predicts the ratios $\left(M_\Lambda-M_N\right): 
\left(M_\Sigma-M_\Lambda\right):\left(M_\Xi-M_\Sigma\right)=1:1:\frac{1}{2}$,
while the data have a reversed order: $1:0.43:0.69$. The inclusion of higher
orders improves on that when the product $\beta^2\gamma$ is large 
enough~\cite{Park:1989wz}. More elegantly, it is possible to diagonalize
the Hamiltonian $H+H_{\rm sb}$ exactly~\cite{Yabu:1987hm}:
introducing eight Euler angles via 
$A=A_2(\alpha,\beta,\gamma){\rm e}^{-\imu\nu\lambda_4}
A_2(\alpha^\prime,\beta^\prime,\gamma^\prime){\rm e}^{-\imu\rho\lambda_8/\sqrt{3}}$,
where $A_2$ is a two flavor matrix like in Eq.~(\ref{generator2}), embedded as in 
Eq.~(\ref{embedding}), and expressing $C_2$ in terms of differential operators with 
respect to these angles, the eigenvalue problem is cast into sets of ordinary 
differential equations with respect to the strangeness changing angle $\nu$ since 
$1-D_{88}=\frac{3}{2}\sin^2\nu$. A typical result for the mass differences is 
listed in Tab.~\ref{SU3Skyrme}. 
\begin{table}
\centerline{
\begin{tabular}{c|ccccccc}
Baryon & $\Lambda$ & $\Sigma$ & $\Xi$ &
$\Delta$ & $\Sigma^*$ & $\Xi^*$ & $\Omega$\cr
\hline
$e=4.0$ &$163$&$264$&$388$&$268$&$406$&$545$&$680$\cr
VM  &$159$&$270$&$398$&$311$&$448$&$592$&$718$\cr
Expt.  &$177$&$254$&$379$&$293$&$446$&$591$&$733$
\end{tabular}}
\caption{\label{SU3Skyrme}Skyrme model prediction for the mass differences with 
respect to the nucleon using $e=4.0$~\cite{Weigel:2008zz}. For later reference we
include the results from the vector meson model (VM) of Section {\it Vector mesons}.}
\end{table}
The above mentioned ratios improve to $1:0.61:0.72$; though it not a perfect match, the 
empirical order is reproduced. This is achieved by assuming a small value for the 
Skyrme parameter $e$, which has the disadvantage that the spin $\frac{3}{2}$ baryon 
masses are predicted too low. Yet, refined models also improve on that. 

Static properties can be computed in parallel to the procedure of the two-flavor model, 
Section {\it Static properties}. But it becomes more involved because additional collective coordinate 
operators emerge for the symmetry currents. As an example we list the spatial components 
of the octet axial-vector current (Roman letters run from one to three, except $a=1,\ldots,8$;
Greek letters from four to seven)
\begin{equation}
A_i^a=\left[a_1(r)\delta_{ik}+a_2(r)\hat x_i\hat x_k\right]D_{ak}
+\left[a_3(r)\delta_{ik}+a_4(r)\hat x_i\hat x_k\right]d_{k\alpha\beta}
D_{a\alpha}\Omega_\beta+\left[a_5(r)\delta_{ik}+a_6(r)\hat x_i\hat x_k\right]D_{a8}\Omega_k
+\,\mbox{symmetry breaking pieces}\,,
\label{axialcurr1}
\end{equation}
where $d_{abc}$ are the symmetric $SU(3)$ structure constants. Though tedious, all 
collective coordinate matrix elements can be computed in the exact diagonalization 
program mentioned above.  The radial functions $a_1(r)$ and $a_2(r)$ are those from 
Eq.~(\ref{acurradial}) while the others originate from the Wess-Zumino term. They 
actually cause a conceptual problem, as they violate $\Vek{\partial}\cdot\Vek{A}^a=0$ 
in the chiral limit. The reason being that the Wess-Zumino term does not contribute 
to the classical equations of motion~\cite{Weigel:2007yx}. Hence the flavor rotating 
hedgehog violates PCAC! Induced kaon fields~\cite{Weigel:1989iw} are only a partial cure, 
as they are not orthogonal to the zero mode of the infinitesimal flavor transformation into 
strangeness direction, i.e., $m_K>m_\pi$ is compulsory since otherwise the field equation 
for the induced components has no solution. Regardless of that problem, together with the 
vector current pendant, numerous static properties can be computed, for example the hyperon 
magnetic moments. Results and further references can be found in the review~\cite{Weigel:2008zz}. 
In particular the sensitivity on flavor symmetry breaking can be tested by varying 
$\gamma$ which is the crucial entry of the diagonalization program. For example, the 
axial vector transition matrix elements that change the strangeness of hyperons by one 
unit have a moderate dependence on $\gamma$ while the nucleon matrix element of the 
strange axial current, $\Delta s =\langle N |A_i^{(s)}|N\rangle$ significantly decreases 
as $\gamma$ grows.  Stated otherwise, the historic flavor symmetric treatment, {\it cf.\@} 
Ref.~\cite{Ge64} or chapter IV in Ref.~\cite{Garcia:1985xz}, may (to some extend) be a 
legitimate tool to describe semi-leptonic hyperon decays, but it fails to determine the 
strangeness content of the proton.

This above described rigid rotator approach assumes that flavor symmetry is still
approximately valid. There is the alternative view that it is strongly violated which
motivates the so-called {\it bound state approach} \cite{Callan:1985hy,Callan:1987xt}.
Similar to the approach in Section {\it Pion-nucleon scattering}, strangeness degrees of 
freedom, $K(\Vek{x},t)$ are incorporated as small amplitude fluctuations about the hedgehog. 
This can, {\it e.g.} be realized by the parameterization
\begin{equation}
U(\Vek{x},t)=
\begin{pmatrix}
\begin{array}{c|c}
\hspace{0.1cm}A_2(t)
\hspace{0.1cm} & \hspace{0.3cm}
\begin{picture}(0,0)
\put(-0.1,-0.15){\mbox{\footnotesize $0$}}
\put(-0.1,0.15){\mbox{\footnotesize $0$}}
\end{picture}\hspace{0.3cm} \\[1mm]
\hline
0\hspace{0.4cm}0 & 1
\end{array}\end{pmatrix}
\xi_H\, {\rm e}^{iZ}\,\xi_H
\begin{pmatrix}
\begin{array}{c|c}
\hspace{0.1cm}A^\dagger_2(t)
\hspace{0.1cm} & \hspace{0.3cm}
\begin{picture}(0,0)
\put(-0.1,-0.15){\mbox{\footnotesize $0$}}
\put(-0.1,0.15){\mbox{\footnotesize $0$}}
\end{picture}\hspace{0.3cm} \\[1mm]
\hline
0\hspace{0.4cm}0 & 1
\end{array}\end{pmatrix} 
\quad{\rm with}\quad
Z=Z(\Vek{x},t)=\frac{\sqrt{2}}{f_K}\,
\begin{pmatrix}
\begin{array}{c|c}
\mbox{\large $0$} & K(\Vek{x},t)\cr
\hline
K^\dagger(\Vek{x},t)\, & 0
\end{array}
\end{pmatrix}
\quad{\rm and}\quad \xi_H^2=U_H\,.
\end{equation}
The kaon fluctuations $K(\Vek{x},t)$ are treated to harmonic order. Then 
the $P$-wave zero mode for rotations into strangeness direction, that 
emerges for $m_K=m_\pi$, turns into a bound state with energy eigenvalue 
$\omega_P$ when $m_K>m_\pi$. The Wess-Zumino term contributes a term
that is linear in the energy eigenvalue and thus lifts the degeneracy
of positive and negative strangeness and only the latter has a bound 
state.  Hyperons states are then constructed by
(multiple) occupation of this bound state. The quantization of this 
system, which is discussed at length in the literature, {\it cf.\@}
Ref.~\cite{Callan:1987xt}, yields the mass formula
\begin{equation}
M_B=E_{\rm cl}+|S|\omega_P+\frac{1}{2\alpha^2}
\left[\chi J(J+1)+(1-\chi)I(I+1)\right]\,,
\label{bsmass}
\end{equation}
for the low-lying $P$-wave baryons with strangeness $S=0,-1,-2,-3$.
and total spin (isospin) $J$($I$). The hyperfine splitting parameter 
$\chi$ is computed as a spatial integral of the chiral angle and the 
bound state wave-function. Numerical results for the mass differences
are shown in Tab.~\ref{bsmasses}.
\begin{table}[b]
~\vskip-5mm
\begin{minipage}{0.44\linewidth}\centering
\begin{tabular}{l|ccccccc}
& $\Lambda$ & $\Sigma$ & $\Xi$ & $\Delta$ &
$\Sigma^*$ & $\Xi^*$ & $\Omega$\cr
\hline
model &$205$&$334$&$505$&$293$&$431$&$602$&$805$\cr
Expt. &$177$&$254$&$379$&$293$&$446$&$591$&$733$
\end{tabular}
\caption{\label{bsmasses}Mass differences of the low-lying baryons with 
respect to the nucleon in the bound state approach to the Skyrme model 
with $f_\pi=93\,{\rm MeV}$. The Skyrme model parameter, $e=4.25$ properly
reproduces $M_\Delta-M_N$. All data are in MeV. \\ \\ \\ }
\end{minipage}\hspace{0.02\linewidth}
\begin{minipage}{0.53\linewidth}\centering
\begin{tabular}{l|ccccccc}
& $n$ & $\Lambda$ & $\Sigma^{+}$ & $\Sigma^{-}$ & $\Xi^{0}$ & 
 $\Xi^{-}$ & $\Sigma^{0}\to\Lambda$ \cr
\hline
BSA &$-0.80$&$-0.31$&$1.10$&$-0.60$&$-0.74$& $-0.18$&$-0.86$\cr
SK  &$-0.78$&$-0.35$&$0.98$& $-0.39$&$-0.76$&$-0.32$&$-0.68$\cr
VM  &$-0.79$&$-0.25$&$1.02$& $-0.47$&$-0.83$&$-0.36$&$-0.74$\cr
Expt. &$-0.68$&$-0.22$&$0.87$&$-0.42$&$-0.45$& $-0.25$&$-0.56$
\end{tabular}
\caption{\label{bsmagmom}Predications for the bound state approach (BSA)
for baryon magnetic moments relative to the proton magnetic moment
which is predicted too low: 1.78, versus  2.79 (Expt.). Parameters 
are as in Tab.~\ref{bsmasses}. Also listed are predictions from the
rigid rotator approach from the Skyrme model (SK) and the vector meson
model (VM) of Section {\it Vector mesons}.}
\end{minipage}
\end{table}

The hyperon state/wave-function is a combination of products
$D^{(I=J)\ast}_{I_3,-J^\prime_3}\left(\Phi,\Theta,\Psi\right)
a^\dagger_{J^\prime_3}(\omega_P)|0\rangle$ with the Wigner-$D$ function 
for the collective rotation, $A_2$ and $a^\dagger$ is the creation operator 
for the bound state. The combination goes over $J^\prime_3$ and appropriate 
Clebsch-Gordan coefficients. The isospin is carried by $U_H$ so that $A_2K$ 
has zero isospin. Hence the rotator part must be quantized with integer 
(half-integer) spin for odd (even) occupation numbers of the bound state. It 
is then possible to compute static properties like magnetic moments~\cite{Kunz:1989zc}.
A set of typical results is presented in Tab.~\ref{bsmagmom}.

We conclude this section with the remark that also the $S$-wave channel
contains a bound state solution with negative strangeness. This state has 
been employed to describe the $\Lambda(1405)$ resonance~\cite{Schat:1994gm}.

\section{Vector mesons}\label{VM}

In Section {\it The Skyrme Model} we have seen deficits of the Skyrme model predictions.
Some of them can be cured by extending the model such that it contains vector meson 
fields. Unfortunately matters become very technical quickly. Therefore we will 
only discuss the concepts here and refer the interested reader to the review 
articles of Refs.~\cite{Meissner:1987ge,Schwesinger:1988af} for more details.

The goal is to construct a chiral Lagrangian that contains the vector mesons
$\omega$ and $\rho$ besides the chiral field. These vector mesons have masses 
$m_V\approx 770\,{\rm MeV}$. (We ignore the difference $m_\omega-m_\rho\approx12\,{\rm MeV}$.)
Chiral symmetry requires to also consider the fields for their chiral partners, like the 
axial-vector meson $a_1$, which is much heavier, $m_{a_1}\approx1260\,{\rm MeV}$, and which 
one would not want to include, at least for simplicity. The starting point is to write a 
chirally invariant Lagrangian for left- and right-handed vector fields, $A_L^\mu$ and 
$A_R^\mu$, which are sums and differences of vector and axial-vector meson fields. 
Among standard terms like ${\rm tr}\left[F_{L,\mu\nu}F_{L}^{\mu\nu}\right]$ this
Lagrangian contains unconventional, chirally invariant terms like~\cite{Kaymakcalan:1983qq}
\begin{equation}
{\rm tr}\left[F_{L,\mu\nu}UF_{R}^{\mu\nu}U^\dagger\right]\,,\quad
{\rm tr}\left[A_{L,\mu}UA_{R}^{\mu}U^\dagger\right]\quad {\rm or}\quad
\epsilon_{\mu\nu\rho\sigma}{\rm tr}\left[\partial^\mu A_L^\nu\partial^\rho 
U A_R^\sigma U^\dagger\right]\,.
\label{VM1}\end{equation}
To eliminate the $a_1$ field without violating chiral symmetry the vector mesons are
parameterized as
\begin{equation}
A_L^\mu=\xi\left(V^\mu+\frac{\imu}{g}\partial^\mu\right)\xi^\dagger\,,\quad
A_R^\mu=\xi^\dagger\left(V^\mu+\frac{\imu}{g}\partial^\mu\right)\xi
\quad {\rm and}\quad U=\xi\ID_{2\times2}\xi\,.
\label{VM2}\end{equation}
The $\omega$ and $\rho$ fields are then identified as the isoscalar and -vector
components of the new field, $V^\mu=\omega^\mu\ID_{2\times2}+\Vek{\rho}^\mu\cdot\Vek{\tau}$. 
Possible model parameters, i.e., the coupling constant $g$ and the 
coefficients of the terms in Eq.~(\ref{VM1}) and others, are determined from vector 
meson properties like their masses and the decay widths for  $\rho\,\to\,\pi\,\pi$ or 
$\omega\,\to\,\pi\,\pi\,\pi$~\cite{Jain:1987sz}. Stabilizing terms, like the Skyrme 
term or $\mathcal{L}_6$ discussed in Section {\it The Skyrme model} are not included. 
On the contrary, the local approximation in which $m_V^2 V_\mu \gg \partial^2  V_\mu$
yields $F_{\mu\nu}\left(\Vek{\rho}\right)\sim \left[\alpha_\mu,\alpha_\nu\right]$
and $\omega_\mu\sim B_\mu$. That is, in this limit the vector meson coupling 
generates those stabilizing terms, but now with coefficients known from meson 
properties. Of course, the ongoing calculations will not adopt the local 
approximation. The so-called hidden symmetry approach~\cite{Bando:1987br} obtains 
the same model Lagrangian by writing $U=\xi_L^\dagger\xi_R$ and elevating
the transformation $\xi_L\,\to\,h\xi_L L$, $\xi_R\,\to\,h\xi_R R$ to a
local symmetry with $h=h(x)$ by means of introducing $V_\mu$ as a gauge 
field.

The spherically symmetric soliton configuration has three profile functions
$F(r)$, $\omega(r)$ and $G(r)$ in
\begin{BoxTypeA}{
\begin{equation}
\xi_H={\rm exp}\left[\frac{\imu}{2}\hat{\Vek{x}}\cdot\Vek{\tau}F(r)\right]\,,\quad
\omega^0=\frac{\omega(r)}{g}\quad {\rm and}\quad
\Vek{\rho}^i=\frac{G(r)}{gr} \left(\hat{\Vek{x}}\times\Vek{\tau}\right)_i\,.
\label{VMprofiles}\end{equation}}
\end{BoxTypeA}\noindent
The space components of $\omega^\mu$ and the time components $\Vek{\rho}^0$
are zero. 
\begin{figure}
\begin{minipage}{0.48\linewidth}\centering
\includegraphics[width=7.0cm,height=4.5cm]{vmsol.eps}
\caption{\label{fig:vmsol}Profile functions for the vector meson 
soliton. The vector meson profile $\omega$ is measured in
units of the vector meson mass~$m_V$.}
\end{minipage}\hspace{0.04\linewidth}
\begin{minipage}{0.48\linewidth}\centering
\includegraphics[width=7.0cm,height=4.5cm]{vmind.eps}
\caption{\label{fig:vmind}Profile functions for the induced components
of the vector meson soliton. \\ }
\end{minipage}
\end{figure}
A set of profiles that solves the classical equations is displayed in 
Fig.~\ref{fig:vmsol}. The result $G(0)=-2$ is not a singularity 
because the coupling to the chiral angle is 
$\left(G+1-\cos F\right)^2$~\cite{Jain:1987sz}.

A major difference to the Skyrme model is that even for two flavors the 
quantization is not as straightforward as in Section {\it Quantization}. 
On top of the time-dependent rotations, Eq.~(\ref{coll1}), field components 
that vanish classically get induced~\cite{Meissner:1986js,Meissner:1988iv}. 
These components are incorporated by writing
\begin{BoxTypeA}{
\begin{equation}
\xi={\rm exp}\left[\frac{\imu}{2f_\pi}\eta(r)
\hat{\Vek{x}}\cdot\Vek{\Omega}\right]A(t)\xi_HA^\dagger(t)\,,\quad
\Vek{\rho}^0=\frac{1}{2g}A(t)\left[\xi_1(r)\Vek{\Omega}
+\xi_2(r)(\hat{\Vek{x}}\cdot\Vek{\Omega})\hat{\Vek{x}}\right]A^\dagger(t)
\quad {\rm and}\quad
\omega^i=\frac{\phi(r)}{2g}\epsilon_{ijk}\Omega_j\hat{x}_k\,.\qquad
\label{vmind1}\end{equation}}
\end{BoxTypeA}\noindent
The four profile functions $\eta$, $\xi_{1,2}$ and $\phi$ obey inhomogeneous linear 
differential equations that arise from extremizing the moment of inertia,~$\alpha^2$. 
The classical soliton provides the source terms in these equations. 
Typical solutions are shown in Fig.~\ref{fig:vmind}.
The calculations of baryon properties are quite technical and we merely summarize 
some of the substantial improvements compared to the Skyrme model:
\begin{itemize}
\item[a)]
Due to the induced components the iso-singlet axial vector current is now 
non-zero~\cite{Johnson:1990kr}
$A_0^i=a_0(r)\Omega_i+\overline{a}_0(r)\hat{x}_i\hat{\Vek{x}}\cdot\Vek{\Omega}$
and its nucleon matrix element $\Sigma_N=\langle N|3A_0^{3}|N\rangle\approx0.3$
agrees reasonably well with data extracted from deep-inelastic nucleon 
scattering~\cite{Ellis:1995de}. 
\item[b)]
(Approximate) vector meson dominance relates the $\omega$ field to the 
iso-scalar density, $V_0^0$ which is thus subject to an equation of the form,
$\left(\Vek{\partial}^2-m_V^2\right)V_0^0\sim -m_V^2 B_0(r)$. Multiplication  
by $r^2$ and integration over space 
yields~\cite{Meissner:1986js}\footnote{That publication also considers 
momentum dependent form factors as does Ref.~\cite{Meissner:1988iv}.}
$r^2_{I=0}
=r^2_{B}+\frac{6}{m_V^2}$\,,
where $\langle r^2\rangle_{B}$ is the radius of the baryon density and equals 
the isoscalar radius in the Skyrme model. The additional $0.4{\rm fm}$ bring the too 
small predictions in Tab.~\ref{table:nuclstatskyrme} into better agreement with data.
\item[c)]
The $\eta$ profile is induced via its coupling to the vector meson fields. Though it 
is small, {\it cf.\@} Fig.~\ref{fig:vmind}, it yields the important result 
${\rm tr}\left[\tau_3\left(U-U^\dagger\right)\right] \sim 
D_{3i}\hat{x}_i(\hat{\Vek{x}}\cdot\Vek{\Omega})\eta(r)\sin F(r)\ne0$. As a consequence 
the two flavor vector meson model, in contrast to the Skyrme model, predicts a non-zero 
strong interaction contribution to the neutron-proton mass difference: 
$\left(m_n-m_p\right)_{\rm str.}\approx1.2\,{\rm MeV}$~\cite{Jain:1989kn}. 
\item[d)]
It was already mentioned that the vector mesons replace higher order stabilizing
terms in the Skyrme model. In the context of the adiabatic approach to meson-baryon
scattering the dynamical vector mesons replace contact interactions by a propagator 
(like the transition from the Fermi model to the Standard Model of particle physics 
with massive vector mesons): $\frac{G^2}{m_V^2}\,\to\,\frac{G^2}{m_V^2-k^2}$. In 
turn this leads to $M_{ij}^{00}=M_{ij}^{rr}$ and solves the ever-rising phase shift 
problem~\cite{Schwesinger:1988af}. We also refer to that article for detailed
comparison with data and the application of the distorted wave-functions that 
emerge from this scattering problem to meson-photon production processes.
\item[e)]
The $SU(3)$ vector meson model calculations are as cumbersome as the one for the 
scattering in d)~\cite{Park:1991fb}. It brings into the game further induced components as 
in Eq.~(\ref{vmind1}) but now related to the angular velocities $\Omega_4,\ldots\Omega_7$.
These components are obtained by extremizing the second moment of inertia $\beta^2$. 
Moreover, this model comes with additional flavor symmetry breaking structures like
\begin{equation}
\sum_{i=1}^3D_{8i}J_i\,,\quad
\sum_{\alpha=1}^7 D_{8\alpha}R_\alpha\,,\quad
\sum_{i=1}^3D_{8i}D_{8i}\quad{\rm and}\quad
\sum_{\alpha=1}^7 D_{8\alpha}D_{8\alpha}\,.
\label{addSB}\end{equation}
In particular the first of these operators is very welcome to improve on the
predictions for the ratios $\left(M_\Lambda-M_N\right): 
\left(M_\Sigma-M_\Lambda\right):\left(M_\Xi-M_\Sigma\right)$.
\end{itemize}
Some of the many prediction of this model have already been listed in Tabs.~\ref{SU3Skyrme} 
and~\ref{bsmagmom}. For additional ones we refer to the many references quoted in the review 
articles~\cite{Meissner:1987ge,Weigel:2008zz} on top of those listed under the above items.

\section{Quarks and Skyrmions}\label{NJL}

So far we have considered chiral solitons that were directly written as functions of the 
chiral field $U$. However, there is an alternative, and essentially equivalent approach 
that starts from a chirally invariant formulation of the quark flavor dynamics. The 
number of publications on this topic is legion and thus we just mention the two main 
review articles~\cite{Alkofer:1994ph,Christov:1995vm}. Most of the many publications repeat 
the studies under items a), b), c) and e) above. There are improvements when it comes 
to agreement with empirical data, but there is not much of a conceptual difference. 
An exception may be the incorporation of sub-leading $1/N_C$ terms for the axial vector
charge, $g_A$ and the iso-vector magnetic moment, $\mu_p-\mu_n$~\cite{Christov:1993ny},
though there are inconsistencies with PCAC in these approaches~\cite{Alkofer:1993pv}.
In any event, we will focus here on a particular application of quark models that (at 
least without further approximations) is not accessible in purely meson theories: nucleon 
structure functions.

It has been long known~\cite{Friedberg:1976eg,Kahana:1984dx,Jain:1988ix} that when the 
Dirac equation $(\imu\dslash -m)q=0$ is augmented with a chiral interaction of the form 
$\overline{q}_LU_Hq$, a strongly bound quark {\it valence} quark solution emerges with 
energy eigenvalue $|\epsilon_{\rm v}|<m$ for wide enough chiral angles, $F(r)$. However, 
considering only this single level is merely the quasi-classical approximation \cite{Friedberg:1976eg}
which does not have a well defined expansion parameter. Rather, one needs to consider the
semi-classical approximation which includes the contribution from the distorted 
Dirac sea: $-\frac{1}{2}\sum_n\left[|\epsilon_n|-|\epsilon^{(0)}_n|\right]$. Here 
$\epsilon_n$ and $\epsilon^{(0)}_n$ are quark energy eigenvalues with and without
coupling to $U_H$, respectively. The leading order of the semi-classical expansion
is determined from the one-loop effective action
\begin{BoxTypeA}{
\begin{equation}
\mathcal{A}_F=-\imu {\rm Log}\,{\rm Det}_{\Lambda}
\left[\imu\dslash-m\left(U_H\right)^{\gamma_5}\right]
\qquad {\rm with}\qquad
\left(U_H\right)^{\gamma_5}={\rm exp}
\left[\imu \hat{\Vek{x}}\cdot\Vek{\tau}\gamma_5F(r)\right]\,.
\label{Aeff}\end{equation}}\end{BoxTypeA}\noindent
This part of the action can be motivated from QCD by a number of 
successive steps: (i) approximate the gluon exchange by an effective four quark 
interaction $\frac{1}{4}(\overline{q}q)^2$, (ii) provide an auxiliary meson field
$\Phi$ to write this interaction as $\Phi\overline{q}q-\Phi^2$, (iii) only quark 
bilinears are left and the functional integral can be performed, (iv) for 
sufficiently large coupling a non-zero VEV, $\langle\Phi\rangle=m$ emerges dynamically
as the constituent quark mass, (v) introduce fluctuations about the VEV in form 
of the chiral field via $\Phi=mU$, (vi) the fluctuating modes are the pseudoscalar 
mesons, Eq.~(\ref{chiralfield})~\cite{Alkofer:1995mv}. These steps are rather 
sketchy and the full approach is quite sophisticated. In particular it is crucial 
to maintain chiral symmetry throughout the calculation. Unfortunately, confinement 
(if it ever was there) and renormalizability are lost. The latter is indicated 
by the cut-off $\Lambda$ in Eq.~(\ref{Aeff}) which acquires a physical meaning 
in the model. In fact, it is related to the pion mass and decay constant. In many 
applications Schwinger's proper-time regularization scheme~\cite{Schwinger:1951nm}
has been used. Here we will, however, consider a variant of the Pauli-Villars
scheme, which, as an important feature is implemented on the level of Eq.~(\ref{Aeff}). 
The final classical energy functional then reads
\begin{equation}
E_{\rm tot}[F]=
\frac{N_C}{2}\left[1+{\rm sign}(\epsilon_{\rm v})\right]\epsilon_{\rm v}
-\frac{N_C}{2}\sum_n\left\{|\epsilon_n|-\sqrt{\epsilon_n^2+\Lambda^2}
+\frac{1}{2}\frac{|\epsilon_n|}{\sqrt{\epsilon_n^2+\Lambda^2}}
-\left(\epsilon_n\,\to\,\epsilon^{(0)}_n\right)\right\}
+m_\pi^2f_\pi^2\int d^3r \, \left[1-{\rm cos}F\right]\,,
\label{etot}
\end{equation}
where the $\epsilon_n$ are the eigenvalues of Dirac Hamiltonian
$h=\Vek{\alpha}\cdot\Vek{p} +\beta\, m\, \left(U_H\right)^{\gamma_5}$. Hence 
these eigenvalues are functionals of $U_H$ and thus of the chiral angle $F(r)$. The 
factor $\frac{N_C}{2}\left[1+{\rm sign}(\epsilon_{\rm v})\right]$ emerges in the first 
term because the Dirac sea carries baryon number when $\epsilon_{\rm v}<0$ and the 
bound valence level should not be occupied explicitly. The chiral angle is obtained 
by self-consistently solving this eigenvalue problem and minimizing $E_{\rm tot}[F]$.
Spin and isopsin states are generated by the collective coordinate formulation
as in Section {\it Quantization}. Substituting the chiral field from Eq.~(\ref{coll1}) 
yields the modified Dirac Hamiltonian
$h_{\vek{\Omega}}=A\left(\Vek{\alpha}\cdot\Vek{p}+\frac{\imu}{2}\Vek{\tau}\cdot\Vek{\Omega}
+\beta\, m\, \left(U_H\right)^{\gamma_5}\right)A^\dagger$. After changing the 
integration variables to $q=A\widetilde{q}$, the action, Eq.~(\ref{Aeff})
is then expanded to quadratic order in the angular velocities yielding a 
Lagrange function of same form as in Eq,~(\ref{lagrigid1}). Here the moment of inertia,
$\alpha^2$ is obtained as a regularized double sum over energy eigenvalues
$\epsilon_n$ and $\epsilon_m$. As mentioned, there are relations between the 
model parameters and pion observables. Eventually only the constituent quark
mass remains the only free parameter which, similar to the Skyrme parameter, is 
typically tuned to reproduce $\frac{3}{2\alpha^2}=M_\Delta-M_N\approx293\,{\rm MeV}$.

Rather than considering static baryon properties in chiral quark soliton models
we immediately focus on nucleon structure functions. There have been several
explorations that compute the structure functions by applying expressions for QCD 
distributions\footnote{See Chap.~18 of Ref.~\cite{ParticleDataGroup:2024cfk} for a 
discussion of QCD parton distributions and background references.} to the eigenstates 
of $h$ (or $h_{\vek{\Omega}}$), i.e., the model (constituent) 
quarks~\cite{Diakonov:1996sr,Diakonov:1997vc,Wakamatsu:1997en,Wakamatsu:1998rx,sym16111481}.
This seems as an over-interpretation of the model. Rather we only want to identify 
the symmetry currents from QCD, not the quark fields themselves. Furthermore, when 
starting from Eq.~(\ref{Aeff}) the subtle issue of how to regularize the vacuum 
contribution to the structure functions is stipulated from the onset. 

The interaction vertex for the disintegration of the nucleon is the matrix element
of the (electromagnetic) current $J_\mu(\xi)$. The cross-section contains the
squared absolute value of this matrix element and we sum/integrate over all
final states subject to energy momentum conservation. This defines the hadron
tensor for deep-inelastic electron nucleon scattering
\begin{equation}
W_{\mu \nu}(p,q;s) = \frac{1}{4\pi} \sum_X \big\langle p,s\big|J_{\mu}(0)\big|X\big\rangle
\big\langle X\big|J^\dagger_{\nu}(0)\big|p,s\big\rangle (2\pi)^4\delta^4(p+q-p_X)
=\frac{1}{4\pi} \int d^4\xi \, e^{iq\cdot \xi}\,
\big\langle p,s\big| \big[J_{\mu}(\xi),J_{\nu}^{\dagger}(0)\big]
\big| p,s \Big\rangle\,,
\label{hten0}
\end{equation}
where $p$ and $s$ the nucleon momentum and spin. The momentum of the virtual photon
is the difference of the momenta of the initial and final electrons:
$q=k-k^\prime$ with $q_0>0$ and $Q^2=-q^2>0$ since the interaction is inelastic
and space-like. The second part in Eq.~(\ref{hten0}) results from translational 
invariance. The tensor $W_{\mu \nu}$ defines Lorentz invariant form factors via the 
decomposition
\begin{align}
W_{\mu \nu}(p,q;s)
&=\left(-g_{\mu \nu} + \frac{q_{\mu} q_{\nu}}{q^2}\right) M_N W_{1} (x,Q^2)
+\left(p_{\mu} - q_{\mu}\frac{p\cdot q}{q^2}\right)
\left(p_{\nu} - q_{\nu}\frac{p\cdot q}{q^2}\right)
\frac{1}{M_N} W_{2} (x,Q^2)\cr
&\hspace{3cm}+\imu\epsilon_{\mu \nu \lambda \sigma} \frac{q^{\lambda} M_N}{p\cdot q}
\left(\left[G_1(x,Q^2)+G_2(x,Q^2)\right]s^{\sigma}
-\frac{q\cdot s}{q\cdot p} p^{\sigma} G_2(x,Q^2) \right)\,,
\label{hten1a}
\end{align}
with the kinematical variables $x=\frac{Q^2}{2M_N\nu}$ and $\nu=\frac{p\cdot q}{M_N}$.
The structure functions are defined as the limit $Q^2\,\to\,\infty$ of the above form 
factors at fixed $x$. This limit is often called the Bjorken limit and $x$ is the Bjorken
variable.

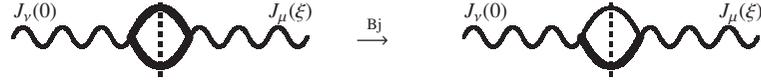
\begin{figure}[h]
\parbox[l]{3.5cm}{~}
\setlength{\unitlength}{1.3mm}
\begin{picture}(34.0,8.0)
\linethickness{1pt}
\bezier{40}(0,4)(1,6)(2,4)
\bezier{40}(2,4)(3,2)(4,4)
\bezier{40}(4,4)(5,6)(6,4)
\bezier{40}(6,4)(7,2)(8,4)
\bezier{40}(8,4)(9,6)(10,4)
\bezier{40}(10,4)(11,2)(12,4)
\put(0,6){\small $J_\nu(0)$}
\put(12,4){\circle*{0.8}}
\linethickness{2pt}
\bezier{200}(12,4)(15,10)(18,4)
\bezier{200}(12,4)(15,-2)(18,4)
\put(18,4){\circle*{0.8}}
\linethickness{1pt}
\bezier{40}(18,4)(19,6)(20,4)
\bezier{40}(20,4)(21,2)(22,4)
\bezier{40}(22,4)(23,6)(24,4)
\bezier{40}(24,4)(25,2)(26,4)
\bezier{40}(26,4)(27,6)(28,4)
\bezier{40}(28,4)(29,2)(30,4)
\linethickness{1.4pt}
\put(26,6){\small $J_\mu(\xi)$}
\multiput(15,0)(0,1){8}{\line(0,1){0.5}}
\end{picture}
\parbox[t]{1.4cm}{\vspace{-0.8cm}
${\,{\stackrel{\scriptstyle{\rm Bj}}
{\textstyle\longrightarrow\,}}}$}
\begin{picture}(32.0,8.0)
\linethickness{1pt}
\bezier{40}(0,4)(1,6)(2,4)
\bezier{40}(2,4)(3,2)(4,4)
\bezier{40}(4,4)(5,6)(6,4)
\bezier{40}(6,4)(7,2)(8,4)
\bezier{40}(8,4)(9,6)(10,4)
\bezier{40}(10,4)(11,2)(12,4)
\put(0,6){\small $J_\nu(0)$}
\put(12,4){\circle*{0.8}}
\bezier{200}(12,4)(15,10)(18,4)
\linethickness{2pt}
\bezier{200}(12,4)(15,-2)(18,4)
\put(18,4){\circle*{0.8}}
\linethickness{1pt}
\bezier{40}(18,4)(19,6)(20,4)
\bezier{40}(20,4)(21,2)(22,4)
\bezier{40}(22,4)(23,6)(24,4)
\bezier{40}(24,4)(25,2)(26,4)
\bezier{40}(26,4)(27,6)(28,4)
\bezier{40}(28,4)(29,2)(30,4)
\linethickness{1.2pt}
\put(26,6){\small $J_\mu(\xi)$}
\multiput(15,0)(0,1){8}{\line(0,1){0.5}}
\end{picture}
\caption{\label{fig:Comp}Two photon coupling to fermion loop. Thick lines are the full
fermion propagators in the soliton background. The thin line in the loop on the right 
represents a free (massless) fermion propagator, $\left(\imu\beta\dslash\right)^{-1}$.
Dashed lines denote Cutkosky cuts to extract the imaginary part, Eq.~(\ref{Comp1}).}
\end{figure}

The key ingredient for the model calculation is the relation between the hadron tensor
and the (forward virtual) Compton amplitude, $T_{\mu \nu}^{ab}$
\begin{equation}
W_{\mu \nu}(p,q;s)=\frac{1}{2\pi}\, {\rm Im}\, T_{\mu \nu}(p,q;s)
\qquad {\rm with}\qquad
T_{\mu \nu}(p,q;s) = \imu \int d^4\xi e^{\imu q\cdot \xi}\,
\big\langle p,q;s\big| T\left\{J_{\mu}(\xi) 
J_{\nu}^{\dagger}(0)\right\}\big| p,q;s \big\rangle\,.
\label{Comp1}\end{equation}
This time-ordered product in $T_{\mu \nu}$ is obtained from first principles via the 
functional derivative
\begin{equation}
T\left\{J_\mu(\xi) J_\nu(0)\right\}=
\frac{\delta^2}{\delta v^\mu(\xi)\delta v^\nu(0)}
\mathcal{A}_{\rm F}(v)\,\Bigg|_{v_\mu=0}\,,
\label{Comp2}\end{equation}
where $v_\mu$ is the photon field introduced by minimal substitution in Eq.~(\ref{Aeff})
to implement electromagnetic gauge invariance. A main simplification arising from the 
Bjorken limit in conjunction with the Pauli-Villars scheme is that one of the two propagators 
$\left(\imu\partial_t-h\right)^{-1}$ can be taken to be the one of a free, massless 
fermion, $\left(\imu\beta\dslash\right)^{-1}$. This is indicated in Fig.~\ref{fig:Comp}.
Whether this simplification holds in other regularization schemes is unclear.

\noindent
Now the stage is set to evaluate Eqs.~(\ref{Comp1}) and~(\ref{Comp2}) within this chiral
soliton model. This results in bulky expressions~\cite{Takyi:2019ahv,Takyi:2019kov} for the 
structure functions that will not even be indicated here. The numerical simulation yields
the structure functions in the nucleon rest frame (RF) at a low renormalization scale $\mu^2$.
In the rest frame the calculated structure functions have support outside the kinematically
allowed regime $x\in[0,1]$ because the soliton is localized. This is cured by a Lorentz boost
to the infinite momentum frame (IMF)~\cite{Jaffe:1980qx,Gamberg:1997qk}, which also makes
contact with the parton model picture of structure functions. The low renormalization 
scale is a new parameter in the model that sets the initial condition for the DGLAP-evolution 
program~\cite{Dokshitzer:1977sg,Gribov:1972ri,Altarelli:1977zs}\footnote{This program 
constitutes differential equations derived from perturbative QCD to relate structure
functions at different energy scales. Alternatively this program can be used to construct
parton distributions at low energy scales (something like the nucleon wave-function)
from data at high energies~\cite{Reya:1979zk}.}.
\begin{figure}[t]
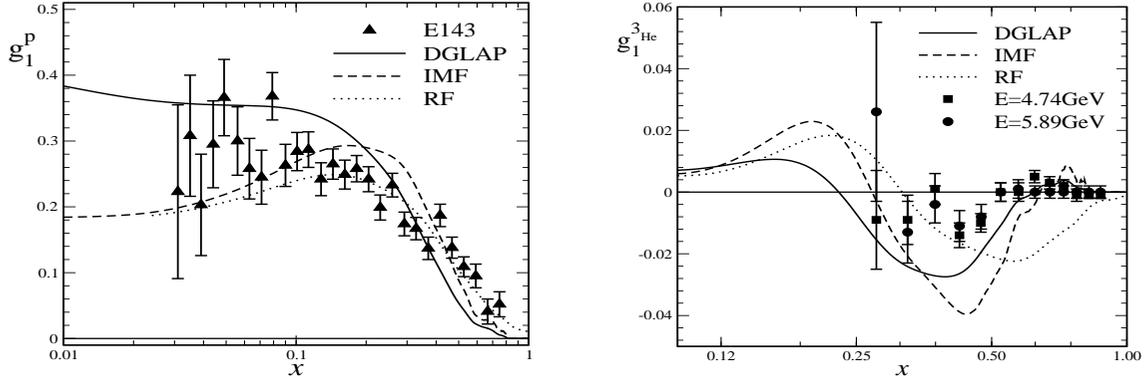
 
~\bigskip

\centerline{
\includegraphics[width=7cm,height=5.0cm]{E143_g1_400.eps}\hspace{1cm}
\includegraphics[width=7cm,height=5.0cm]{g1_helium_400.eps}
}
\caption{\label{fig:g1}Model prediction for the longitudinal polarized proton structure functions.
Left panel: $g_{1}^{p}(x)$ ; right panel: $g_{1}^{^3{\rm He}}(x)$. These functions are "DGLAP'' evolved 
from $\mu^{2}=\mathrm{0.4\,GeV^{2}}$ to $Q^{2}=\mathrm{3\,GeV^{2}}$ after being projected from the RF to 
the IMF. Data are from Refs.~\cite{Abe:1994cp,Abe:1998wq} for the proton and from Ref.~\cite{Flay:2016wie} 
for helium. In the latter case $E$ refers to the energy of the incident electron.}
\end{figure}
As an example we compare the model prediction for the axial structure function $g_1$ to data in 
Fig.~\ref{fig:g1}. Proton data are directly available, while neutron data are accessible via scattering 
electrons off ${^3{\rm He}}$. Though there are discrepancies, we nevertheless observe that chiral soliton 
models reproduce the main characteristics of the data and that these models indeed have their say on 
nucleon structure functions.

\section{Further applications}\label{Appl}

There are many other applications of chiral solitons in hadron 
physics and beyond. We list some of them with going into much detail.

\subsection{Flavor symmetry breaking and soliton extension}
\label{solextension}

According to Eq.~(\ref{asympF}) the asymptotic behavior of the flavor rotating 
soliton, Eq.~(\ref{rotsol}) is governed by the pion mass. However, the 
kaon mass should enter for the strangeness degrees of freedom so that a hedgehog
pointing into kaon direction would be narrower. There are two approaches 
that address this problem. In the slow-rotator approach
the strangeness changing angle $\nu$ defined after Eq.~(\ref{Guadagnini})
is considered a constant parameter in the classical field equation that arises 
from the Euler-Lagrange equations for $\mathcal{L}_{\rm SK}+\mathcal{L}_{\rm sb}$;
recall that $\mathcal{L}_{\rm sb}$  contains $1-D_{88}(A)=\frac{3}{2}\sin^2\nu$.
In turn the chiral angle depends parametrically on $\nu$ and so do the classical
mass and the moments of inertia. This parameter dependence is included in the 
energy eigenvalue problem formulated as differential equations with respect 
to $\nu$~\cite{Schwesinger:1992mi}. In the other, the breathing mode approach, 
the extension of the soliton is quantized as a collective coordinate. Here the 
starting point is the field parameterization\footnote{This was first explored
in the two flavor Skyrme model, Ref.~\cite{Hajduk:1984as}.}
\begin{equation}
U\left(t,\Vek{x}\right)=A(t)U_{\rm H}\left(s(t)\Vek{x}\right)A^\dagger(t)\,.
\label{breath}\end{equation}
In Section {\it Extension to SU3} we have discussed the leading order treatment~\cite{Guadagnini:1983uv} 
of flavor symmetry breaking when quantizing the collective rotations $A(t)$  but then immediately 
turned to the exact diagonalization~\cite{Yabu:1987hm}. The results of the latter can essentially 
reproduced in perturbation theory when expanding up to third order~\cite{Park:1989wz}. In 
doing so, the spin $\frac{1}{2}$ baryons are no longer pure octet states but acquire admixtures 
from spin $\frac{1}{2}$ states with the same flavor quantum numbers in higher dimensional 
$SU(3)$ representations like the anti-decuplet and the $27^{\rm th}$-plet. With the 
{\it ansatz} of Eq.~(\ref{breath}) we get an attractive potential for the scaling variable $s$
that is proportional to $D_{88}(A)$, i.e., the strength of the attraction depends 
on the considered element in a particular $SU(3)$ representation. When quantizing $s(t)$ 
the eigenstates are thus combinations of elements of different $SU(3)$ representations 
and radial excitations. Though the nucleon still has its dominant contribution from the 
radial ground state in the octet, the Roper (1440) resonance has major contributions from 
the first radial excitation in the octet and the nucleon radial ground states in the 
anti-decuplet (called $N^\prime$ in Fig.~\ref{fig:anti10}) and the 
$27^{\rm th}$-plet~\cite{Schechter:1990ce}. In either case, there are substantial 
improvements for the predictions of baryon properties in the two approaches. As an 
example we briefly discuss the magnetic moments of the spin $\frac{1}{2}$ baryons. 
From Tab.~(\ref{bsmagmom}) we see that the rigid rotator results closely follow the 
$U$-spin relations: $\mu_p=\mu_{\Sigma^+}$ ($1:0.87$), $\mu_n=\mu_{\Xi^0}=2\mu_\Lambda$ 
($-0.68:-0.45:-0.44$) and $\mu_{\Sigma^-}=\mu_{\Xi^-}$ ($-0.42:-0.25$). The model results
substantially deviate from the experimental data in parenthesis (in units of $\mu_p$). 
The slow-rotator improves these relations to  ($1:0.85$), ($-0.83:-0.54:-0.50$) and 
($-0.40:-0.20$) and so does the breathing mode approach: ($1:0.78$), ($-0.89:-0.41:-0.39$) 
and ($-0.47:-0.14$); which even overshoots the empirical ratios. Including a scalar field 
as motivated by the QCD scale anomaly~\cite{Gomm:1985ut} has a mitigating effect on this
overshooting as it does for the mass differences~\cite{Schechter:1991ki}. Either of the two
approaches corroborates that the extension of the soliton should reflect flavor symmetry
breaking.

\subsection{Pentaquarks}\label{pentaquarks} 
\begin{wrapfigure}{r}{6.0cm}
\begin{center}
~\vskip-4mm
\centerline{
\setlength{\unitlength}{1.1mm}
\begin{picture}(40,30)
\put(20,0){\vector(0,1){27}}
\put(22,25){$Y$}
\put(0,10){\vector(1,0){40}}
\put(37,12){$I_3$}
\multiput(5,5)(10,0){4}{\circle*{1}}
\put(0,6){$\Xi_5$}
\multiput(10,10)(10,0){3}{\circle*{1}}
\put(5,11){$\Sigma^\prime$}
\multiput(15,15)(10,0){2}{\circle*{1}}
\put(10,16){$N^\prime$}
\put(20,20){\circle*{1}}
\put(15,21){$\Theta^+$}
\end{picture}}
~\vskip-4mm
\caption{\label{fig:anti10}The flavor content of the 
anti-decuplet. The quantum numbers are isospin projection ($I_3$) and
hypercharge ($Y$) which is the sum of baryon- and strangeness number.
Note that the strange quark has strangeness negative one.} 
\end{center}
~\vskip-12mm
\end{wrapfigure}
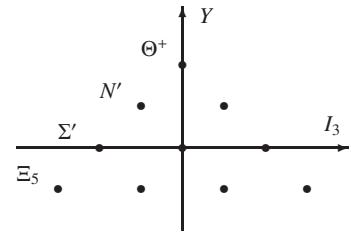

Chiral soliton models had another revival around the beginning of the millennium in the
context of the (in)famous pentaquark, $\Theta^{+}$. The relevance of higher dimensional $SU(3)$ 
representations has already been mentioned in the context of flavor symmetry breaking.
The lowest dimensional such representation is the anti-decuplet                               
whose particle content is displayed in Fig.~\ref{fig:anti10}.

Equally interesting is the fact that these representations contain exotic baryon states with 
quantum numbers that cannot be described as three quark composites. The most prominent 
exotic baryon is the $\Theta^{+}$ whose quantum numbers can only be represented by a quark 
composition of two up, two down and an anti-strange quark. The relevance of these, eventually 
low-mass, exotic states in chiral soliton models was recognized quite 
early~\cite{Biedenharn:1984su,Praszalowicz:1987em,Walliser:1992vx}. Yet, it only 
attracted appreciable recognition when it was conjectured that in particular the $\Theta^{+}$ 
was long-lived~\cite{Diakonov:1997mm}\footnote{It should be mentioned that this estimate contained 
an arithmetic error~\cite{Jaffe:2004qj}.} and when there were experimental indications 
that this was indeed the case~\cite{LEPS:2003wug}. In the meantime refined 
experiments have suggested otherwise unless the width of the $\Theta^{+}$ is two orders 
of magnitude smaller than a typical one of a hadron resonance. The current situation has 
been recently sketched in Ref.~\cite{Praszalowicz:2024mji}. Yet, there are reservations 
on the width estimate of Refs.~\cite{Diakonov:1997mm,Ellis:2004uz}. The approach is based 
on determining a Yukawa coupling between exotic and ground state baryons as well as 
pseudoscalar mesons from the model axial vector current, Eq.~(\ref{axialcurr1}). First, 
the emergence of a Yukawa coupling, i.e., a term linear in the fluctuations, in 
soliton models is an artifact (shortcoming?) of not knowing the exact time-dependent solution 
but rather approximating it by Eq.~(\ref{rotsol}). Second, this axial vector current detour employs 
the Goldberger-Treiman relation~\cite{Goldberger:1958vp}, which assumes that the Yukawa coupling, 
as a function of momentum transfer, does not (significantly) vary between zero and the meson mass. 
This may not be the case for $m_K=495\,{\rm MeV}$. Furthermore this relation utilizes PCAC which, 
as mentioned before does not hold for the current in Eq.~(\ref{axialcurr1}). To avoid these issues, 
Ref.~\cite{Walliser:2005pi} considers kaon-nucleon scattering similar to the approach in Section 
{\it Pion-nucleon scattering} but beyond the adiabatic approximation and extracts the resonance 
parameters from scattering data (as experiment does). In that approach the $\Theta^{+}$ 
is predicted to be about $700\,{\rm MeV}$ heavier than the nucleon (which is not low) 
with a width of something like $40\,{\rm MeV}$ which is only slightly smaller than a 
typical hadron width. 

\subsection{Heavy flavor symmetry}\label{HFS}

For mesons containing a single charm or bottom quark the pseudoscalar and vector 
states are almost degenerate. Hence chiral symmetry does not apply to hadrons with 
charm or bottom quarks: The dynamics of the heavy quark component is governed 
by the heavy spin-flavor symmetry~\cite{Neubert:1993mb,Manohar:2000dt,Blaschke:2004xi}
while the light flavor content follows the laws of chiral symmetry.

To incorporate heavy flavor symmetry, two-component arrays $P$ and $Q_\nu$ are introduced 
that parameterize the pseudoscalar (mass $M$) and vector mesons (mass $M^\ast$) with a single 
heavy quark. The two components stand for the light quark, i.e., up or down content. 
This can be extended by a third component for the strange quark. These (almost degenerate) 
mesons are combined in the heavy meson multiplet
\begin{equation}
\mathcal{H}=\frac{1}{2}\left(1+\gamma_\mu v^\mu\right)
\left(i\gamma_5\widetilde{P}+\gamma^\alpha\widetilde{Q}_\alpha\right)
\qquad {\rm with}\qquad
\widetilde{P}={\rm e}^{iM V\cdot x} P 
\qquad {\rm and} \qquad 
\widetilde{Q_\alpha}={\rm e}^{iM^\ast V\cdot x}Q_\alpha\,.
\label{heavymultiplet}
\end{equation}
The transformations $P\,\to\,\widetilde{P}$ and $Q_\alpha\,\to\,\widetilde{Q}_\alpha$
with four velocity $V_\mu$ ($V^2=1$) define the reference frame for the 
heavy mesons such that the time derivative produces their kinetic energies.
A model Lagrangian with the above postulated symmetries and as few as possible 
derivatives acting on the light pseudoscalar meson fields reads~\cite{Gupta:1993kd}
\begin{equation}
\widetilde{\mathcal{L}}_{\rm H}=
\imu M V^\mu {\rm Tr}\left\{\mathcal{H}\left(\partial_\mu-iv_\mu\right)
\bar{\mathcal{H}}\right\}
-d\,{\rm Tr}\left\{\mathcal{H}\gamma_\mu\gamma_5 p^\mu\bar{\mathcal{H}}\right\}
+ \ldots 
\qquad{\rm with}\qquad
p_\mu=\frac{\imu}{2}\left(\xi\partial_\mu\xi^\dagger-\xi^\dagger\partial_\mu\xi\right)
\quad{\rm and}\quad
v_\mu=\frac{\imu}{2}\left(\xi\partial_\mu\xi^\dagger+\xi^\dagger\partial_\mu\xi\right)\,,
\label{Lheavy1}\end{equation}
where the ellipsis indicate subleading pieces in $1/M$ and eventual interactions
with light vector mesons, {\it cf.\@} Section {\it Vector mesons}. The single new 
parameter can be extracted from the semi-leptonic $D\to K$ transition:
$d\approx0.53$~\cite{Gupta:1993kd,Jain:1994rb}\footnote{These articles contain 
numerous references to related approaches.}. Similar to the bound state approach in 
Section {\it Extension to SU3}, substituting the hedgehog into $p_\mu$ and $v_\mu$ 
generates an attractive potential for the heavy mesons leading to a bound state whose 
extension is $\mathcal{O}\left(\frac{1}{M}\right)$. In the limit $M\,\to\,\infty$ 
it is peaked at $r=0$ and the binding energy is given by the soliton profile(s) at 
the center: $E_B=\frac{3}{2}dF^\prime(0)-\frac{3\sqrt{2}c}{gm_V}G^{\prime\prime}(0)
+\frac{\alpha}{2}\omega(0)$ \cite{Gupta:1993kd}. The coupling constant $c$ can be 
estimated from the semi-leptonic decay $D\to K^*$~\cite{Jain:1994rb} to be $c\approx1.60$.
Unfortunately there is no direct empirical information about the parameter 
$\alpha$. Assuming light vector meson dominance for the electromagnetic
form factors of the heavy mesons suggests $\alpha\approx1$. 

A more realistic treatment~\cite{Schechter:1995vr} starts with a Lorentz-invariant 
Lagrangian and demands that it approaches $\widetilde{\mathcal{L}}_{\rm H}$ when 
$M=M^\ast\,\to\,\infty$:
\begin{equation}
\mathcal{L}_H=\left(D^\mu P\right)^\dagger D_\mu P
-\frac{1}{2}\left(Q^{\mu\nu}\right)^\dagger Q_{\mu\nu}
-M^2P^\dagger P+M^{*2}Q^\dagger_\mu Q^\mu
+2iMd\left[P^\dagger p_\mu Q^\mu-Q^\dagger_\mu p^\mu P\right]
-\frac{d}{2}\,\epsilon^{\alpha\beta\mu\nu}
\left[Q_{\nu\alpha}^\dagger p_\mu Q_\beta+
Q_\beta^\dagger p_\mu Q_{\nu\alpha}\right]+\ldots\,,
\label{Lheavy2}
\end{equation}
with $D_\mu=\partial_\mu-\imu v_\mu$ and $Q_{\mu\nu}=D_\mu Q_\nu-D_\nu Q_\mu$. Again, terms 
with light vector mesons are not written out. The condition to assume Eq.~(\ref{Lheavy1}) 
dictated the relative coefficients in the square brackets. Both $S$- and $P$-wave
(as defined by the angular momentum of the pseudoscalar component) bound states have 
been determined numerically. The heavy limit binding energy, $E_B$, in which the lowest
energy bound states in these channels are degenerate, is approached only very slowly,
even for $M=M^\ast=50\,{\rm GeV}$ deviations of 10\% ($P$-wave) or 20\% ($S$-wave)  are 
observed. Certainly, this limit does not apply for the charm sector. It is therefore 
mandatory to use the full model, Eq.~(\ref{Lheavy2}) for a realistic description for 
baryons with a heavy quark. For two light flavors the Skyrme model soliton predicts 
their masses on the light side when compared to experiment, while the vector meson 
model yields fair agreement when $\alpha=0,\ldots,0.3$~\cite{Harada:1997we}. When 
extending to three light flavors with exact diagonalization of the rigid rotator as in 
Section {\it Extension to SU3}, substituting the Skyrme model soliton into $p_\mu$ and 
$v_\mu$ slightly overestimates the flavor symmetry breaking effects within a heavy baryon 
multiplet in the $P$-wave but predicts too small mass differences in the $S$-wave 
channel~\cite{Blanckenberg:2015dsa}; though only a limited number of experimental 
data are available for the latter.

\subsection{Beyond unit baryon number}\label{Bne1} 

So far we have only considered solitons describing hadrons with baryon number one.
According to Eq.~(\ref{Bnumber}) the most suggestive procedure to construct solitons 
with baryon $B\ge2$ is to take $F(0)=n\pi$. For $n=2$ the resulting classical energy
is about three times as large as that of the $B=1$ hedgehog and thus exceeds the 
bound derived from Eq.~(\ref{Bogobound1}) considerably. A smaller energy solution
should exist. Indeed, doubling the velocity in the azimuthal angle, i.e.,
taking Eq.~(\ref{hedgehog1}) and replacing~~$\Vek{\tau}\cdot\hat{\Vek{x}}=
\begin{pmatrix}\cos\theta & {\rm e}^{-\imu\varphi} \cr
{\rm e}^{\imu\varphi}& -\cos\theta\end{pmatrix} \,\longrightarrow\,
\begin{pmatrix}\cos\theta & {\rm e}^{-2\imu\varphi} \cr
{\rm e}^{2\imu\varphi}& -\cos\theta\end{pmatrix}$~~with a radially symmetric chiral angle 
and $F(0)=\pi$ reduces that factor $3$ to about $2.14$~\cite{Weigel:1986zc}. Allowing furthermore 
the chiral angle to additionally depend on the polar angle $\theta$ the numeric (lattice)
simulation finally produces bound configurations not only for $B=2$ but also for 
$B=3,4,5$~\cite{Verbaarschot:1987au}. For $B=2$ the contour lines of equal mass density are 
(approximately) circles in planes containing the $z$-axis and that are centered at $z=0$ and a 
finite distance, $\rho_0$ away from that axis. These circles form tori which are frequently 
called {\it donuts}. More recently these numeric simulations have been pushed to even produce 
bound configurations with baryon number larger than one hundred~\cite{Feist:2012ps}.
Whether or not these solutions should be identified with existing nuclei is 
not clear: the above argument that the eventually large $\mathcal{O}(N_C^0)$ quantum 
corrections can be ignored does not hold anymore, since they might vary substantially with 
baryon number \cite{Graham:2025bwh}.

Another way to consider two nucleon systems is the so-called product {\it ansatz}
\begin{equation}
U_2(\Vek{x},t)=A_1(t)U_H(\Vek{x}_1)A_1^\dagger(t)A_2(t)U_H(\Vek{x}_2)A_2^\dagger(t)\,,
\label{Uproduct}
\end{equation}
where $U_H(\Vek{x})$ is from Eq.~(\ref{hedgehog1}) and represents a single baryon. The spatial 
separation is parameterized by $\Vek{x}_i=\Vek{x}\pm\Vek{R}/2$ and $A_i(t)\in SU(N_f)$ are the
respective spin-flavor collective coordinates. From this {\it ansatz} the nucleon-nucleon 
potential can be extracted~\cite{VinhMau:1984sc,Jackson:1985bn,Yabu:1985vx} by exploring the 
energy difference $E[U_2]-2E[U_H]$ as a function of $|\Vek{R}|$. The model reproduces the 
long-range one pion exchange attraction as expected from the asymptotic behavior, Eq.~(\ref{asympF}). 
To account for the intermediate-range attraction (which is essential for binding nucleons to nuclei) 
the Skyrme model must contain $\mathcal{L}_6$ ~\cite{Riska:1989fw}\footnote{The 
omission of a dominant contribution to $\mathcal{L}_6$ in Ref.~\cite{Riska:1989fw} doubts that 
conclusion \cite{Abada:1996ux}. D.~Harland is thanked for bringing Ref.~\cite{Abada:1996ux} to 
the author's attention.} or be augmented by scalar fields~\cite{Riska:1989fw,Abada:1996jg}; on 
the other hand vector meson fields generate (short range) repulsion~\cite{Kalafatis:1992vv}.

For three flavors another $B=2$ configuration was found from the 
{\it ansatz}~\cite{Balachandran:1983dj}
\begin{equation}
U_{\rm HDB}=\ID_{3\times3}\,{\rm e}^{i\psi}
+\imu\Vek{\Lambda}\cdot\hat{\Vek{x}}\,{\rm e}^{-\imu\psi/2}\,\sin\chi
+\left(\Vek{\Lambda}\cdot\hat{\Vek{x}}\right)^2
\left[{\rm e}^{-\imu\psi/2}\cos\chi-{\rm e}^{\imu\psi}\right]\,,
\label{hdibar}
\end{equation}
where $\psi$ and $\chi$ are radial functions 
and $\Vek{\Lambda}=\left(\lambda_7,-\lambda_5,\lambda_2\right)$.
This configuration has strangeness $S=-2$ and is thus a model for the
H-dibaryon, explored earlier in the bag model~\cite{Jaffe:1976yi}. This 
baryon can only decay into two $\Lambda$-hyperons. Yet, for the physical
value of $f_\pi$ it has a binding energy as large as about $1\,{\rm GeV}$.
Further estimates of quantum corrections, however, suggest that this 
binding energy could be much smaller~\cite{Scholtz:1993jg} ; though 
the approximation underlying this estimate has been questioned in 
Ref.~\cite{Graham:2025bwh}.

A conceptually different way to look at large baryon numbers is to 
compactify the target space $\mathbb{R}^3$ to a sphere $\mathbb{S}^3(L)$,
where $L$ is the radius of the sphere~\cite{Jackson:1988bd,Jackson:1987sy}. 
A unit baryon number solution in the compactified space then has global baryon number 
density $\propto L^{-3}$.  The sphere is embedded in Minkowski space by 
introducing a second polar angle $0\le\mu\le\pi$, such that $x\in\mathbb{R}^4$ 
is parameterized as 
$x^\alpha=\frac{L}{ef_{\pi}}\left(\cos\mu,\hat{\Vek{r}}\sin\mu\right)$.
The chiral angle $F(r)$ in Eq.~(\ref{hedgehog1}) is replaced by $f(\mu)$.
The field equations for the Skyrme model (without pion mass term) have the
identity map $f_{\rm I}(\mu)=\mu$ as a unit baryon number solution.
Its energy, $E[f_{\rm I}]=3\frac{\pi^2}{e}f_\pi \left(L+\frac{1}{L}\right)$
saturates the bound, mentioned after Eq.~(\ref{Bogobound1}), at its minimum, 
$L=1$. Hence for this radius the identity map is the true solution; but as
$L\,\to\,\infty$ the hedgehog from Fig.~\ref{fig:Xangle} must be the true solution. 
It turns out that $f_{\rm I}$ is stable for $L<\sqrt{2}$ but unstable for $L>\sqrt{2}$. 
Since ${\rm tr}U=2\cos\left[f(\mu)\right]$, the VEV $\langle{\rm tr}U\rangle$
vanishes for the identity map, i.e., for small $L$ and high densities.
Numerical studies~\cite{Jackson:1987sy} show that $\langle{\rm tr}U\rangle$,
which can be viewed as the quark condensate ({\it cf.\@} the discussion after 
Eq.~(\ref{Aeff})), is a continuous but not a differentiable function of $L$, 
evidencing that the phase transition is second order. This suggests to relate 
the observed phase transition to the chiral phase transition. Yet, the 
predicted critical density of about $0.19{\rm fm}^{-3}$~\cite{Jackson:1987sy} is 
lower than the QCD-related one which is expected to occur at several times the 
nuclear density which by itself is already $0.15{\rm fm}^{-3}$.

\section{Conclusion}\label{concl}

In this article we have reviewed the basic concepts of chiral solitons. We have discussed
their motivation from the chiral symmetry of strong interactions and large-$N_C$ QCD. 
We explored the Skyrme model for baryons in some detail. Especially we employed that 
model to explain the main concepts of and calculational techniques for chiral solitons; 
both for two and three light, or almost light, flavors. We also had a glance at refinements 
of the model, which improve on the agreement with empirical data. Though these model predictions 
are not extremely precise, it is in total fascinating to observe that the single and maybe 
simple concept of chiral solitons is able to capture almost all baryon physics, from the very 
small (structure functions), via the obvious (spectrum and static properties) to the very 
large (nuclei and dense matter).

This article has focused on chiral soliton models in the context of particle and nuclear 
physics. Yet, there are applications outside this realm. For example, the two (space) 
dimensional Skyrmions, called baby~\cite{Piette:1994mh} or 
magnetic~\cite{Schollwock:2004aa,Nagasoa:2013} Skyrmions, are highly relevant in 
condensed matter physics where they describe spin textures with topological 
charges~\cite{PhysRevB.103.L060404}.

\begin{ack}[Acknowledgments]%
The author has gained his insight into the topic of chiral soliton models for baryons 
by collaborations with many colleagues: R.\@ Alkofer, L.\@ Gamberg, G.\@ Holzwarth, 
Ulf-G.\@ Mei{\ss}ner, H.\@ Reinhardt, J.\@ Schechter, B.\@ Schwesinger, and H.\@ Walliser.
Their input is  gratefully acknowledged. The author is supported in part 
by the NRF (South Africa) under grant~150672.

\end{ack}

%%%%%%%%%%%%%%%%%%%%%%%%%%%%%%%%%%%%%%%%%%%%
%% Optional: A list of references to other relevant works/articles/websites which are not cited in the text but that would further enhance a readers understanding of this topic
\seealso{Chiral Perturbation Theory, Large $N_C$ Hadron Physics}

%%%%%%%%%%%%%%%%%%%%%%%%%%%%%%%%%%%%%%%%%
%% Mandatory: Bibliography using bibtex 
\bibliographystyle{Numbered-Style} %% for Numbered Reference Style

\begin{thebibliography*}{100}
\providecommand{\bibtype}[1]{}
\providecommand{\url}[1]{{\tt #1}}
\providecommand{\urlprefix}{URL }
\expandafter\ifx\csname urlstyle\endcsname\relax
  \providecommand{\doi}[1]{doi:\discretionary{}{}{}#1}\else
  \providecommand{\doi}{doi:\discretionary{}{}{}\begingroup
  \urlstyle{rm}\Url}\fi
\providecommand{\bibinfo}[2]{#2}
\providecommand{\eprint}[2][]{\url{#2}}
\makeatletter\def\@biblabel#1{\bibinfo{label}{[#1]}}\makeatother

\bibtype{Article}%
\bibitem{Gell-Mann:1964nj}
\bibinfo{author}{{M. Gell-Mann}}, \bibinfo{title}{{A Schematic Model of Baryons
  and Mesons}}, \bibinfo{journal}{Phys. Lett.} \bibinfo{volume}{8}
  (\bibinfo{year}{1964}) \bibinfo{pages}{214}.

\bibtype{Book}%
\bibitem{Clo79}
\bibinfo{author}{{F. E. Close}}, \bibinfo{title}{{An Intoduction to Quarks and
  Partons}}, \bibinfo{publisher}{{Academic Press, London}}
  \bibinfo{year}{1979}.

\bibtype{Article}%
\bibitem{Fritzsch:1973pi}
\bibinfo{author}{{H. Fritzsch, M. Gell-Mann, and H. Leutwyler}},
  \bibinfo{title}{{Advantages of the Color Octet Gluon Picture}},
  \bibinfo{journal}{Phys. Lett.} \bibinfo{volume}{B47} (\bibinfo{year}{1973})
  \bibinfo{pages}{365}.

\bibtype{Book}%
\bibitem{Mu87}
\bibinfo{author}{{T. Muta}}, \bibinfo{title}{{Foundations of Quantum
  Chromodynamics}}, \bibinfo{publisher}{{World Scientific, Singapore}}
  \bibinfo{year}{1987}.

\bibtype{Article}%
\bibitem{Gross:1973id}
\bibinfo{author}{{D. J. Gross, and F. Wilczek}}, \bibinfo{title}{{Ultraviolet
  Behavior of Nonabelian Gauge Theories}}, \bibinfo{journal}{Phys. Rev. Lett.}
  \bibinfo{volume}{30} (\bibinfo{year}{1973}) \bibinfo{pages}{1343}.

\bibtype{Article}%
\bibitem{Politzer:1973fx}
\bibinfo{author}{{H. D. Politzer}}, \bibinfo{title}{{Reliable Perturbative
  Results for Strong Interactions?}}, \bibinfo{journal}{Phys. Rev. Lett.}
  \bibinfo{volume}{30} (\bibinfo{year}{1973}) \bibinfo{pages}{1346}.

\bibtype{Book}%
\bibitem{Ro92}
\bibinfo{author}{{H. J. Rothe}}, \bibinfo{title}{{Lattice Gauge Theories -- An
  Introduction}}, \bibinfo{publisher}{{World Scientific, Singapore}}
  \bibinfo{year}{1992}.

\bibtype{Misc}%
\bibitem{Meissner:2024ona}
\bibinfo{author}{Ulf-G. Meißner}, \bibinfo{title}{Chiral Perturbation Theory}
  \bibinfo{year}{2024}, \eprint{2410.21912}.

\bibtype{Article}%
\bibitem{ParticleDataGroup:2024cfk}
\bibinfo{author}{{S. Navas {\it et al.}}}, \bibinfo{title}{{Review of particle
  physics}}, \bibinfo{journal}{Phys. Rev.} \bibinfo{volume}{D110}
  (\bibinfo{number}{3}) (\bibinfo{year}{2024}) \bibinfo{pages}{030001}.

\bibtype{Article}%
\bibitem{Ad69}
\bibinfo{author}{{S. L. Adler}}, \bibinfo{title}{{Axial Vector Vertex in Spinor
  Electrodynamics}}, \bibinfo{journal}{Phys. Rev.} \bibinfo{volume}{177}
  (\bibinfo{year}{1969}) \bibinfo{pages}{2426}.

\bibtype{Article}%
\bibitem{Be69}
\bibinfo{author}{{J. S. Bell and R. Jackiw}}, \bibinfo{title}{A PCAC Puzzle:
  $\pi^0\to\gamma\gamma$ in the Sigma Model}, \bibinfo{journal}{Nuovo Cim.}
  \bibinfo{volume}{A60} (\bibinfo{year}{1969}) \bibinfo{pages}{47}.

\bibtype{Article}%
\bibitem{tHooft:1973alw}
\bibinfo{author}{{G. 't Hooft}}, \bibinfo{title}{{A Planar Diagram Theory for
  Strong Interactions}}, \bibinfo{journal}{Nucl. Phys.} \bibinfo{volume}{B72}
  (\bibinfo{year}{1974}) \bibinfo{pages}{461}.

\bibtype{Article}%
\bibitem{Witten:1979kh}
\bibinfo{author}{{E. Witten}}, \bibinfo{title}{{Baryons in the 1/N Expansion}},
  \bibinfo{journal}{Nucl. Phys.} \bibinfo{volume}{B160} (\bibinfo{year}{1979})
  \bibinfo{pages}{57}.

\bibtype{Book}%
\bibitem{Ra82}
\bibinfo{author}{{R. Rajaraman}}, \bibinfo{title}{Solitons and Instantons},
  \bibinfo{publisher}{North Holland}, \bibinfo{address}{Amsterdam}
  \bibinfo{year}{1982}.

\bibtype{Article}%
\bibitem{Holzwarth:1985rb}
\bibinfo{author}{{G. Holzwarth and B. Schwesinger}}, \bibinfo{title}{{Baryons
  in the Skyrme Model}}, \bibinfo{journal}{Rept. Prog. Phys.}
  \bibinfo{volume}{49} (\bibinfo{year}{1986}) \bibinfo{pages}{825}.

\bibtype{Article}%
\bibitem{Zahed:1986qz}
\bibinfo{author}{{I. Zahed and G. E. Brown}}, \bibinfo{title}{{The Skyrme
  Model}}, \bibinfo{journal}{Phys. Rept.} \bibinfo{volume}{142}
  (\bibinfo{year}{1986}) \bibinfo{pages}{1}.

\bibtype{Article}%
\bibitem{Meissner:1987ge}
\bibinfo{author}{{Ulf-G. Mei{\ss}ner}}, \bibinfo{title}{Low-Energy Hadron
  Physics from Effective Chiral Lagrangians with Vector Mesons},
  \bibinfo{journal}{Phys. Rept.} \bibinfo{volume}{161} (\bibinfo{year}{1988})
  \bibinfo{pages}{213}.

\bibtype{Article}%
\bibitem{Schwesinger:1988af}
\bibinfo{author}{{B. Schwesinger, H. Weigel, G. Holzwarth, and A. Hayashi}},
  \bibinfo{title}{{The Skyrme Soliton in Pion, Vector and Scalar Meson Fields:
  $\pi N$ Scattering and Photoproduction}}, \bibinfo{journal}{Phys. Rept.}
  \bibinfo{volume}{173} (\bibinfo{year}{1989}) \bibinfo{pages}{173}.

\bibtype{Article}%
\bibitem{Meier:1996ng}
\bibinfo{author}{{F. Meier and H. Walliser}}, \bibinfo{title}{Quantum
  Corrections to Baryon Properties in Chiral Soliton Models},
  \bibinfo{journal}{Phys. Rept.} \bibinfo{volume}{289} (\bibinfo{year}{1997})
  \bibinfo{pages}{383}.

\bibtype{Book}%
\bibitem{Makhankov:1993ti}
\bibinfo{author}{{V. G. Makhankov, Y. P. Rybakov, and V. I. Sanyuk}},
  \bibinfo{title}{{The Skyrme model: Fundamentals, Methods, Applications}},
  \bibinfo{publisher}{{Springer-Verlag, Berlin}} \bibinfo{year}{1993}.

\bibtype{Book}%
\bibitem{Weigel:2008zz}
\bibinfo{author}{{H. Weigel}}, \bibinfo{title}{{Chiral Soliton Models for
  Baryons}}, \bibinfo{comment}{vol.} \bibinfo{volume}{743, Lecture Notes
  Phys.}, \bibinfo{publisher}{Springer-Verlag, Berlin} \bibinfo{year}{2008}.

\bibtype{Book}%
\bibitem{BrownRho:2010}
\bibinfo{author}{{G. E. Brown and M. Rho}}, \bibinfo{title}{The Multifaceted
  Skyrmion}, \bibinfo{publisher}{World Scientific, Singapore}
  \bibinfo{year}{2010}.

\bibtype{Article}%
\bibitem{Derrick:1964ww}
\bibinfo{author}{{G. H. Derrick}}, \bibinfo{title}{Comments on Nonlinear Wave
  Equations as Models for Elementary Particles}, \bibinfo{journal}{J. Math.
  Phys.} \bibinfo{volume}{5} (\bibinfo{year}{1964}) \bibinfo{pages}{1252}.

\bibtype{Article}%
\bibitem{Skyrme:1961vq}
\bibinfo{author}{{T. H. R. Skyrme}}, \bibinfo{title}{A Nonlinear Field Theory},
  \bibinfo{journal}{Proc. Roy. Soc. Lond.} \bibinfo{volume}{A260}
  (\bibinfo{year}{1961}) \bibinfo{pages}{127}.

\bibtype{Article}%
\bibitem{Skyrme:1962vh}
\bibinfo{author}{{T. H. R. Skyrme}}, \bibinfo{title}{A Unified Field Theory of
  Mesons and Baryons}, \bibinfo{journal}{Nucl. Phys.} \bibinfo{volume}{31}
  (\bibinfo{year}{1962}) \bibinfo{pages}{556}.

\bibtype{Book}%
\bibitem{Pa46}
\bibinfo{author}{{W. Pauli}}, \bibinfo{title}{{Meson Theory of Nuclear
  Forces}}, \bibinfo{publisher}{Interscience Publishes, Inc., New York}
  \bibinfo{year}{1946}.

\bibtype{Article}%
\bibitem{Witten:1983tw}
\bibinfo{author}{{E. Witten}}, \bibinfo{title}{{Global Aspects of Current
  Algebra}}, \bibinfo{journal}{Nucl. Phys.} \bibinfo{volume}{B223}
  (\bibinfo{year}{1983}) \bibinfo{pages}{422}.

\bibtype{Article}%
\bibitem{Witten:1983tx}
\bibinfo{author}{{E. Witten}}, \bibinfo{title}{{Current Algebra, Baryons, and
  Quark Confinement}}, \bibinfo{journal}{Nucl. Phys.} \bibinfo{volume}{B223}
  (\bibinfo{year}{1983}) \bibinfo{pages}{433}.

\bibtype{Article}%
\bibitem{Kaymakcalan:1983qq}
\bibinfo{author}{{\"O Kaymakcalan, S. Rajeev, and J. Schechter}},
  \bibinfo{title}{{Nonabelian Anomaly and Vector Meson Decays}},
  \bibinfo{journal}{Phys. Rev.} \bibinfo{volume}{D30} (\bibinfo{year}{1984})
  \bibinfo{pages}{594}.

\bibtype{Article}%
\bibitem{Bogomolny:1975de}
\bibinfo{author}{{E. B. Bogomolny}}, \bibinfo{title}{Stability of Classical
  Solutions}, \bibinfo{journal}{Sov. J. Nucl. Phys.} \bibinfo{volume}{24}
  (\bibinfo{year}{1976}) \bibinfo{pages}{449}.

\bibtype{Article}%
\bibitem{'tHooft:1974qc}
\bibinfo{author}{{G. 't Hooft}}, \bibinfo{title}{Magnetic Monopoles in Unified
  Gauge Theories}, \bibinfo{journal}{Nucl. Phys.} \bibinfo{volume}{B79}
  (\bibinfo{year}{1974}) \bibinfo{pages}{276}.

\bibtype{Article}%
\bibitem{Polyakov:1974ek}
\bibinfo{author}{{A. M. Polyakov}}, \bibinfo{title}{{Particle Spectrum in
  Quantum Field Theory}}, \bibinfo{journal}{JETP Lett.} \bibinfo{volume}{20}
  (\bibinfo{year}{1974}) \bibinfo{pages}{194}.

\bibtype{Article}%
\bibitem{Adkins:1983ya}
\bibinfo{author}{{G. S. Adkins, C. R. Nappi, and E. Witten}},
  \bibinfo{title}{{Static Properties of Nucleons in the Skyrme Model}},
  \bibinfo{journal}{Nucl. Phys.} \bibinfo{volume}{B228} (\bibinfo{year}{1983})
  \bibinfo{pages}{552}.

\bibtype{Book}%
\bibitem{Var88}
\bibinfo{author}{{D. A. Varshalovich, and Moskalev, and V. K. Khersonskii}},
  \bibinfo{title}{{Quantum Theory of Angular Momentum}},
  \bibinfo{publisher}{{World Scientific, Singapore}} \bibinfo{year}{1988}.

\bibtype{Article}%
\bibitem{Adkins:1983hy}
\bibinfo{author}{{G. S. Adkins and C. R. Nappi}}, \bibinfo{title}{{The Skyrme
  Model with Pion Masses}}, \bibinfo{journal}{Nucl. Phys.}
  \bibinfo{volume}{B233} (\bibinfo{year}{1984}) \bibinfo{pages}{109}.

\bibtype{Article}%
\bibitem{Langacker:1973hh}
\bibinfo{author}{{P. Langacker and H. Pagels}}, \bibinfo{title}{{Chiral
  Perturbation Theory}}, \bibinfo{journal}{Phys. Rev.} \bibinfo{volume}{D8}
  (\bibinfo{year}{1973}) \bibinfo{pages}{4595}.

\bibtype{Article}%
\bibitem{Kaymakcalan:1984bz}
\bibinfo{author}{{\"O Kaymakcalan and J. Schechter}}, \bibinfo{title}{{Chiral
  Lagrangian of Pseudoscalars and Vectors}}, \bibinfo{journal}{Phys. Rev.}
  \bibinfo{volume}{D31} (\bibinfo{year}{1985}) \bibinfo{pages}{1109}.

\bibtype{Article}%
\bibitem{Brodsky:1988ip}
\bibinfo{author}{{S. J. Brodsky, J. R. Ellis, and M. Karliner}},
  \bibinfo{title}{{Chiral Symmetry and the Spin of the Proton}},
  \bibinfo{journal}{Phys. Lett.} \bibinfo{volume}{B206} (\bibinfo{year}{1988})
  \bibinfo{pages}{309}.

\bibtype{Article}%
\bibitem{Braaten:1986md}
\bibinfo{author}{{E. Braaten, S.-M. Tse, and C. Willcox}},
  \bibinfo{title}{{Electroweak Form-Factors of the Skyrmion}},
  \bibinfo{journal}{Phys. Rev.} \bibinfo{volume}{D34} (\bibinfo{year}{1986})
  \bibinfo{pages}{1482}.

\bibtype{Article}%
\bibitem{Jackson:1985yz}
\bibinfo{author}{{A. Jackson, A. D. Jackson, A. S. Goldhaber, G. E. Brown, and
  L. C. Castillejo}}, \bibinfo{title}{{A Modified Skyrmion}},
  \bibinfo{journal}{Phys. Lett.} \bibinfo{volume}{B154} (\bibinfo{year}{1985})
  \bibinfo{pages}{101}.

\bibtype{Article}%
\bibitem{Hayashi:1984bc}
\bibinfo{author}{{A. Hayashi, G. Eckart, G. Holzwarth, and H. Walliser}},
  \bibinfo{title}{{Pion Nucleon Scattering Phase Shifts in the Skyrme Model}},
  \bibinfo{journal}{Phys. Lett.} \bibinfo{volume}{B147} (\bibinfo{year}{1984})
  \bibinfo{pages}{5}.

\bibtype{Article}%
\bibitem{Walliser:1984wn}
\bibinfo{author}{{H. Walliser and G. Eckart}}, \bibinfo{title}{{Baryon
  Resonances as Fluctuations of the Skyrme Soliton}}, \bibinfo{journal}{Nucl.
  Phys.} \bibinfo{volume}{A429} (\bibinfo{year}{1984}) \bibinfo{pages}{514}.

\bibtype{Article}%
\bibitem{Mattis:1986wc}
\bibinfo{author}{{M. P. Mattis}}, \bibinfo{title}{Skyrmions and Vector Mesons},
  \bibinfo{journal}{Phys. Rev. Lett.} \bibinfo{volume}{56}
  (\bibinfo{year}{1986}) \bibinfo{pages}{1103}.

\bibtype{Article}%
\bibitem{Holzwarth:1990eh}
\bibinfo{author}{{G. Holzwarth, G. Pari, and B. K. Jennings}},
  \bibinfo{title}{{Low-Energy Pion-Nucleon P-Wave Scattering in the Skyrme
  Model}}, \bibinfo{journal}{Nucl. Phys.} \bibinfo{volume}{A515}
  (\bibinfo{year}{1990}) \bibinfo{pages}{665}.

\bibtype{Article}%
\bibitem{Guadagnini:1983uv}
\bibinfo{author}{{E. Guadagnini}}, \bibinfo{title}{Baryons as Solitons and Mass
  Formulae}, \bibinfo{journal}{Nucl. Phys.} \bibinfo{volume}{B236}
  (\bibinfo{year}{1984}) \bibinfo{pages}{35}.

\bibtype{Article}%
\bibitem{Mazur:1984yf}
\bibinfo{author}{{P. O. Mazur, M. A. Nowak, and M. Prasza{\l}owicz}},
  \bibinfo{title}{{SU(3) Extension of the Skyrme Model}},
  \bibinfo{journal}{Phys. Lett.} \bibinfo{volume}{B147} (\bibinfo{year}{1984})
  \bibinfo{pages}{137}.

\bibtype{Article}%
\bibitem{deSwart:1963gc}
\bibinfo{author}{{J. J. de Swart}}, \bibinfo{title}{The Octet Model and its
  Clebsch-Gordan Coefficients}, \bibinfo{journal}{Rev. Mod. Phys.}
  \bibinfo{volume}{35} (\bibinfo{year}{1963}) \bibinfo{pages}{916}.

\bibtype{Article}%
\bibitem{Park:1989wz}
\bibinfo{author}{{N. W, Park, J. Schechter, and H. Weigel}},
  \bibinfo{title}{{Higher Order Perturbation Theory for the SU(3) Skyrme
  Model}}, \bibinfo{journal}{Phys. Lett.} \bibinfo{volume}{B224}
  (\bibinfo{year}{1989}) \bibinfo{pages}{171}.

\bibtype{Article}%
\bibitem{Yabu:1987hm}
\bibinfo{author}{{H. Yabu and K. Ando}}, \bibinfo{title}{{A New Approach to the
  SU(3) Skyrme Model}}, \bibinfo{journal}{Nucl. Phys.} \bibinfo{volume}{B301}
  (\bibinfo{year}{1988}) \bibinfo{pages}{601}.

\bibtype{Article}%
\bibitem{Weigel:2007yx}
\bibinfo{author}{{H. Weigel}}, \bibinfo{title}{{Axial Current Matrix Elements
  and Pentaquark Decay Widths in Chiral Soliton Models}},
  \bibinfo{journal}{Phys. Rev. D} \bibinfo{volume}{75} (\bibinfo{year}{2007})
  \bibinfo{pages}{114018}.

\bibtype{Article}%
\bibitem{Weigel:1989iw}
\bibinfo{author}{{H. Weigel, J. Schechter, N. W, Park, and Ulf-G.
  Mei{\ss}ner}}, \bibinfo{title}{{Kaon Excitation in the SU(3) Skyrme Model}},
  \bibinfo{journal}{Phys. Rev.} \bibinfo{volume}{D42} (\bibinfo{year}{1990})
  \bibinfo{pages}{3177}.

\bibtype{Book}%
\bibitem{Ge64}
\bibinfo{author}{M. Gell-Mann}, \bibinfo{author}{Y. Ne{'e}man},
  \bibinfo{title}{The Eightfold Way}, \bibinfo{publisher}{Benjamin, New York}
  \bibinfo{year}{1964}.

\bibtype{Book}%
\bibitem{Garcia:1985xz}
\bibinfo{author}{{A. Garcia and P. Kielanowski}}, \bibinfo{title}{{The Beta
  Decay of Hyperons}}, \bibinfo{comment}{vol.} \bibinfo{volume}{222, Lecture
  Notes Phys.}, \bibinfo{publisher}{{Springer, Berlin, Heidelberg}}
  \bibinfo{year}{1985}.

\bibtype{Article}%
\bibitem{Callan:1985hy}
\bibinfo{author}{{C. G. Callan Jr. and I. R. Klebanov}}, \bibinfo{title}{{Bound
  State Approach to Strangeness in the Skyrme Model}}, \bibinfo{journal}{Nucl.
  Phys.} \bibinfo{volume}{B262} (\bibinfo{year}{1985}) \bibinfo{pages}{365}.

\bibtype{Article}%
\bibitem{Callan:1987xt}
\bibinfo{author}{{C. G. Callan Jr., K. Hornbostel, and I. R. Klebanov}},
  \bibinfo{title}{{Baryon Masses in the Bound State Approach to Strangeness in
  the Skyrme Model}}, \bibinfo{journal}{Phys. Lett.} \bibinfo{volume}{B202}
  (\bibinfo{year}{1988}) \bibinfo{pages}{269}.

\bibtype{Article}%
\bibitem{Kunz:1989zc}
\bibinfo{author}{{J. Kunz and P. J. Mulders}}, \bibinfo{title}{{Magnetic
  Moments of Hyperons in the Bound State Approach to the Skyrme Model}},
  \bibinfo{journal}{Phys. Lett.} \bibinfo{volume}{B231} (\bibinfo{year}{1989})
  \bibinfo{pages}{335}.

\bibtype{Article}%
\bibitem{Schat:1994gm}
\bibinfo{author}{{C. L. Schat, N. N. Scoccola, and C. Gobbi}},
  \bibinfo{title}{{Lambda (1405) in the Bound State Soliton Model}},
  \bibinfo{journal}{Nucl. Phys.} \bibinfo{volume}{A585} (\bibinfo{year}{1995})
  \bibinfo{pages}{627}.

\bibtype{Article}%
\bibitem{Jain:1987sz}
\bibinfo{author}{{P. Jain, R. Johnson, Ulf-G. Mei{\ss}ner, N. W. Park, and J.
  Schechter}}, \bibinfo{title}{{Realistic Pseudoscalar Vector Chiral Lagrangian
  and its Soliton Excitations}}, \bibinfo{journal}{Phys. Rev.}
  \bibinfo{volume}{D37} (\bibinfo{year}{1988}) \bibinfo{pages}{3252}.

\bibtype{Article}%
\bibitem{Bando:1987br}
\bibinfo{author}{{M. Bando, T. Kugo, and K. Yamawaki}},
  \bibinfo{title}{Nonlinear Realization and Hidden Local Symmetries},
  \bibinfo{journal}{Phys. Rept.} \bibinfo{volume}{164} (\bibinfo{year}{1988})
  \bibinfo{pages}{217}.

\bibtype{Article}%
\bibitem{Meissner:1986js}
\bibinfo{author}{{Ulf-G. Mei{\ss}ner, N. Kaiser, and W. Weise}},
  \bibinfo{title}{{Nucleons as Skyrme Solitons with Vector Mesons:
  Electromagnetic and Axial Properties}}, \bibinfo{journal}{Nucl. Phys.}
  \bibinfo{volume}{A466} (\bibinfo{year}{1987}) \bibinfo{pages}{685}.

\bibtype{Article}%
\bibitem{Meissner:1988iv}
\bibinfo{author}{{Ulf-G. Mei{\ss}ner, N. Kaiser, H. Weigel, and J. Schechter}},
  \bibinfo{title}{{Relatitic Pseudoscalar - Vector Lagrangian. 2. Static and
  Dynamical Baryon Properties}}, \bibinfo{journal}{Phys. Rev.}
  \bibinfo{volume}{D39} (\bibinfo{year}{1989}) \bibinfo{pages}{1956}.

\bibtype{Article}%
\bibitem{Johnson:1990kr}
\bibinfo{author}{{R. Johnson, N. W. Park, J. Schechter, V. Soni, and H.
  Weigel}}, \bibinfo{title}{{Singlet Axial Current and the 'Proton Spin'
  Question}}, \bibinfo{journal}{Phys. Rev.} \bibinfo{volume}{D42}
  (\bibinfo{year}{1990}) \bibinfo{pages}{2998}.

\bibtype{Inproceedings}%
\bibitem{Ellis:1995de}
\bibinfo{author}{{J. R. Ellis and M. Karliner}}, \bibinfo{title}{{The Strange
  Spin of the Nucleon}}, in: \bibinfo{booktitle}{{Ettore Majorana International
  School of Nucleon Structure: 1st Course: The Spin Structure of the Nucleon}}
  \bibinfo{year}{1995}, p. \bibinfo{pages}{300}.

\bibtype{Article}%
\bibitem{Jain:1989kn}
\bibinfo{author}{{P. Jain, R. Johnson, N. W. Park, J. Schechter, and H.
  Weigel}}, \bibinfo{title}{{The Neutron - Proton Mass Splitting Puzzle in
  Skyrme and Chiral Quark Models}}, \bibinfo{journal}{Phys. Rev.}
  \bibinfo{volume}{D40} (\bibinfo{year}{1989}) \bibinfo{pages}{855}.

\bibtype{Article}%
\bibitem{Park:1991fb}
\bibinfo{author}{{N. W. Park and H. Weigel}}, \bibinfo{title}{{Static
  Properties of Baryons from an SU(3) Pseudoscalar Vector Meson Lagrangian}},
  \bibinfo{journal}{Nucl. Phys.} \bibinfo{volume}{A541} (\bibinfo{year}{1992})
  \bibinfo{pages}{453}.

\bibtype{Article}%
\bibitem{Alkofer:1994ph}
\bibinfo{author}{{R. Alkofer, H. Reinhardt, and H. Weigel}},
  \bibinfo{title}{{Baryons as Chiral Solitons in the Nambu-Jona-Lasinio
  Model}}, \bibinfo{journal}{Phys. Rept.} \bibinfo{volume}{265}
  (\bibinfo{year}{1996}) \bibinfo{pages}{139}.

\bibtype{Article}%
\bibitem{Christov:1995vm}
\bibinfo{author}{{C. V. Christov, A. Blotz, H.-C. Kim, P. Pobylitsa, T. Watabe,
  T. Mei{\ss}ner, E. Ruiz Arriola, and K. Goeke}}, \bibinfo{title}{{Baryons as
  Nontopological Chiral Solitons}}, \bibinfo{journal}{Prog. Part. Nucl. Phys.}
  \bibinfo{volume}{37} (\bibinfo{year}{1996}) \bibinfo{pages}{91}.

\bibtype{Article}%
\bibitem{Christov:1993ny}
\bibinfo{author}{{C. V. Christov, A. Blotz, K. Goeke, P. Pobylitsa, V. Petrov,
  W. Wakamatsu, and T. Watabe}}, \bibinfo{title}{{$1/N_C$ Rotational
  Corrections to $g_A$ and Isovector Magnetic Moment of the Nucleon}},
  \bibinfo{journal}{Phys. Lett.} \bibinfo{volume}{B325} (\bibinfo{year}{1994})
  \bibinfo{pages}{467}.

\bibtype{Article}%
\bibitem{Alkofer:1993pv}
\bibinfo{author}{{R. Alkofer and H. Weigel}}, \bibinfo{title}{{$1/N_C$
  Corrections to $g_A$ in the Light of PCAC}}, \bibinfo{journal}{Phys. Lett.}
  \bibinfo{volume}{B319} (\bibinfo{year}{1993}) \bibinfo{pages}{1}.

\bibtype{Article}%
\bibitem{Friedberg:1976eg}
\bibinfo{author}{{R. Friedberg, and T. D. Lee}}, \bibinfo{title}{{Fermion Field
  Nontopological Solitons. 1.}}, \bibinfo{journal}{Phys. Rev.}
  \bibinfo{volume}{D15} (\bibinfo{year}{1977}) \bibinfo{pages}{1694}.

\bibtype{Article}%
\bibitem{Kahana:1984dx}
\bibinfo{author}{{S. Kahana, G. Ripka, and V. Soni}}, \bibinfo{title}{{Soliton
  with Valence Quarks in the Chiral Invariant Sigma Model}},
  \bibinfo{journal}{Nucl. Phys.} \bibinfo{volume}{A415} (\bibinfo{year}{1984})
  \bibinfo{pages}{351}.

\bibtype{Article}%
\bibitem{Jain:1988ix}
\bibinfo{author}{{P. Jain, R. Johnson, and J. Schechter}},
  \bibinfo{title}{{Aspects of the Chiral Quark Model}}, \bibinfo{journal}{Phys.
  Rev.} \bibinfo{volume}{D38} (\bibinfo{year}{1988}) \bibinfo{pages}{1571}.

\bibtype{Book}%
\bibitem{Alkofer:1995mv}
\bibinfo{author}{{R. Alkofer and H. Reinhardt}}, \bibinfo{title}{{Chiral Quark
  Dynamics}}, \bibinfo{comment}{vol.} \bibinfo{volume}{33, Lecture Notes
  Phys.}, \bibinfo{publisher}{Springer-Verlag, Berlin} \bibinfo{year}{1995}.

\bibtype{Article}%
\bibitem{Schwinger:1951nm}
\bibinfo{author}{{J. S. Schwinger}}, \bibinfo{title}{{On Gauge Invariance and
  Vacuum Polarization}}, \bibinfo{journal}{Phys. Rev.} \bibinfo{volume}{82}
  (\bibinfo{year}{1951}) \bibinfo{pages}{664}.

\bibtype{Article}%
\bibitem{Diakonov:1996sr}
\bibinfo{author}{{D. Diakonov, V. Petrov, P. Pobylitsa, M. V. Polyakov, and C.
  Weiss}}, \bibinfo{title}{{Nucleon Parton Distributions at Low Normalization
  Point in the Large $N_C$ Limit}}, \bibinfo{journal}{Nucl. Phys.}
  \bibinfo{volume}{B480} (\bibinfo{year}{1996}) \bibinfo{pages}{341}.

\bibtype{Article}%
\bibitem{Diakonov:1997vc}
\bibinfo{author}{{D. Diakonov, V. Petrov, P. Pobylitsa, M. V. Polyakov, and C.
  Weiss}}, \bibinfo{title}{{Unpolarized and Polarized Quark Distributions in
  the Large $N_C$ Limit}}, \bibinfo{journal}{Phys. Rev.} \bibinfo{volume}{D56}
  (\bibinfo{year}{1997}) \bibinfo{pages}{4069}.

\bibtype{Article}%
\bibitem{Wakamatsu:1997en}
\bibinfo{author}{{M. Wakamatsu and T. Kubota}}, \bibinfo{title}{{Chiral
  Symmetry and the Nucleon Structure Functions}}, \bibinfo{journal}{Phys. Rev.}
  \bibinfo{volume}{D57} (\bibinfo{year}{1998}) \bibinfo{pages}{5755}.

\bibtype{Article}%
\bibitem{Wakamatsu:1998rx}
\bibinfo{author}{{M. Wakamatsu and T. Kubota}}, \bibinfo{title}{{Chiral
  Symmetry and the Nucleon Spin Structure Functions}}, \bibinfo{journal}{Phys.
  Rev.} \bibinfo{volume}{D60} (\bibinfo{year}{1999}) \bibinfo{pages}{034020}.

\bibtype{Article}%
\bibitem{sym16111481}
\bibinfo{author}{{M. Wakamatsu}}, \bibinfo{title}{{Extraordinary Nature of the
  Nucleon Scalar Charge and Its Densities as a Signal of Nontrivial Vacuum
  Structure of QCD}}, \bibinfo{journal}{Symmetry} \bibinfo{volume}{16}
  (\bibinfo{year}{2024}).

\bibtype{Article}%
\bibitem{Takyi:2019ahv}
\bibinfo{author}{{I. Takyi and H. Weigel}}, \bibinfo{title}{{Nucleon Structure
  Functions from the NJL-Model Chiral Soliton}}, \bibinfo{journal}{Eur. Phys.
  J.} \bibinfo{volume}{A55} (\bibinfo{year}{2019}) \bibinfo{pages}{128}.

\bibtype{Phdthesis}%
\bibitem{Takyi:2019kov}
\bibinfo{author}{{I. Takyi}}, \bibinfo{title}{{Structure Functions of the
  Nucleon in a Soliton Model}}, \bibinfo{comment}{Ph.D. thesis},
  \bibinfo{school}{Stellenbosch University} \bibinfo{year}{2019}.

\bibtype{Article}%
\bibitem{Jaffe:1980qx}
\bibinfo{author}{{R. L. Jaffe}}, \bibinfo{title}{{Operators in a Translation
  Invariant Two-dimensional Bag Model}}, \bibinfo{journal}{Annals Phys.}
  \bibinfo{volume}{132} (\bibinfo{year}{1981}) \bibinfo{pages}{32}.

\bibtype{Article}%
\bibitem{Gamberg:1997qk}
\bibinfo{author}{{L. P. Gamberg, and H. Reinhardt, and H. Weigel}},
  \bibinfo{title}{{Nucleon Structure Functions from a Chiral Soliton in the
  Infinite Momentum Frame}}, \bibinfo{journal}{Int. J. Mod. Phys.}
  \bibinfo{volume}{A13} (\bibinfo{year}{1998}) \bibinfo{pages}{5519}.

\bibtype{Article}%
\bibitem{Dokshitzer:1977sg}
\bibinfo{author}{{L. Y. Dokshitzer}}, \bibinfo{title}{{Calculation of the
  Structure Functions for Deep Inelastic Scattering and e+ e- Annihilation by
  Perturbation Theory in Quantum Chromodynamics.}}, \bibinfo{journal}{Sov.
  Phys. JETP} \bibinfo{volume}{46} (\bibinfo{year}{1977}) \bibinfo{pages}{641},
  \bibinfo{note}{[Zh. Eksp. Teor. Fiz.73,1216(1977)]}.

\bibtype{Article}%
\bibitem{Gribov:1972ri}
\bibinfo{author}{{V. N. Gribov, V. N. and L. N. Lipatov}},
  \bibinfo{title}{{Deep Inelastic $ep$ Scattering in Perturbation Theory}},
  \bibinfo{journal}{Sov. J. Nucl. Phys.} \bibinfo{volume}{15}
  (\bibinfo{year}{1972}) \bibinfo{pages}{438}, \bibinfo{note}{[Yad.
  Fiz.15,781(1972)]}.

\bibtype{Article}%
\bibitem{Altarelli:1977zs}
\bibinfo{author}{{G. Altarelli and G. Parisi}}, \bibinfo{title}{{Asymptotic
  Freedom in Parton Language}}, \bibinfo{journal}{Nucl. Phys.}
  \bibinfo{volume}{B126} (\bibinfo{year}{1977}) \bibinfo{pages}{298}.

\bibtype{Article}%
\bibitem{Reya:1979zk}
\bibinfo{author}{{E. Reya}}, \bibinfo{title}{{Perturbative Quantum
  Chromodynamics}}, \bibinfo{journal}{Phys. Rept.} \bibinfo{volume}{69}
  (\bibinfo{year}{1981}) \bibinfo{pages}{195}.

\bibtype{Article}%
\bibitem{Abe:1994cp}
\bibinfo{author}{{K. Abe, {\it et al.}}} (\bibinfo{collaboration}{E143}),
  \bibinfo{title}{{Precision Measurement of the Proton Spin Structure Function
  $g_1(p)$}}, \bibinfo{journal}{Phys. Rev. Lett.} \bibinfo{volume}{74}
  (\bibinfo{year}{1995}) \bibinfo{pages}{346}.

\bibtype{Article}%
\bibitem{Abe:1998wq}
\bibinfo{author}{{K. Abe, {\it et al.}}} (\bibinfo{collaboration}{E143}),
  \bibinfo{title}{{Measurements of the Proton and Deuteron Spin Structure
  Functions $g_1$ and $g_2$}}, \bibinfo{journal}{Phys. Rev.}
  \bibinfo{volume}{D58} (\bibinfo{year}{1998}) \bibinfo{pages}{112003}.

\bibtype{Article}%
\bibitem{Flay:2016wie}
\bibinfo{author}{{D. Flay, {\it et al.}}} (\bibinfo{collaboration}{Jefferson
  Lab Hall A}), \bibinfo{title}{{Measurements of $d_{2}^{n}$ and $A_{1}^{n}$:
  Probing the Neutron Spin Structure}}, \bibinfo{journal}{Phys. Rev.}
  \bibinfo{volume}{D94} (\bibinfo{number}{5}) (\bibinfo{year}{2016})
  \bibinfo{pages}{052003}.

\bibtype{Article}%
\bibitem{Schwesinger:1992mi}
\bibinfo{author}{{B. Schwesinger and H. Weigel}}, \bibinfo{title}{{SU(3)
  Symmetry Breaking for Masses, Magnetic Moments and Sizes of Baryons}},
  \bibinfo{journal}{Nucl. Phys.} \bibinfo{volume}{A540} (\bibinfo{year}{1992})
  \bibinfo{pages}{461}.

\bibtype{Article}%
\bibitem{Hajduk:1984as}
\bibinfo{author}{{C. Hajduk and B. Schwesinger}}, \bibinfo{title}{{Static
  Deformations and Rotational Excitations of Baryons in the Skyrme Model}},
  \bibinfo{journal}{Phys. Lett.} \bibinfo{volume}{B145} (\bibinfo{year}{1984})
  \bibinfo{pages}{171}.

\bibtype{Article}%
\bibitem{Schechter:1990ce}
\bibinfo{author}{{J. Schechter and H. Weigel}}, \bibinfo{title}{{The Breathing
  Mode in the SU(3) Skyrme Model}}, \bibinfo{journal}{Phys. Rev.}
  \bibinfo{volume}{D44} (\bibinfo{year}{1991}) \bibinfo{pages}{2916}.

\bibtype{Article}%
\bibitem{Gomm:1985ut}
\bibinfo{author}{{R. Gomm and P. Jain, R. Johnson, and J. Schechter}},
  \bibinfo{title}{Scale Anomaly and the Scalars}, \bibinfo{journal}{Phys. Rev.}
  \bibinfo{volume}{D33} (\bibinfo{year}{1986}) \bibinfo{pages}{801}.

\bibtype{Article}%
\bibitem{Schechter:1991ki}
\bibinfo{author}{{J. Schechter and H. Weigel}}, \bibinfo{title}{Breathing Mode
  Quantization in an Extended SU(3) Skyrme Model}, \bibinfo{journal}{Phys.
  Lett.} \bibinfo{volume}{B261} (\bibinfo{year}{1991}) \bibinfo{pages}{235}.

\bibtype{Misc}%
\bibitem{Biedenharn:1984su}
\bibinfo{author}{{L. C. Biedenharn and Y. Dothan}}, \bibinfo{title}{Monopolar
  Harmonics in SU(3)-f as Eigenstates of the Skyrme-Witten Model for Baryons},
  \bibinfo{note}{print-84-1039 (DUKE)}.

\bibtype{Inproceedings}%
\bibitem{Praszalowicz:1987em}
\bibinfo{author}{{M. Prasza{\l}owicz}}, \bibinfo{title}{{SU(3) Skyrmion}}, in:
  \bibinfo{booktitle}{{Workshop on Skyrmions and Anomalies}}
  \bibinfo{year}{1987}, p. \bibinfo{pages}{112}.

\bibtype{Article}%
\bibitem{Walliser:1992vx}
\bibinfo{author}{{H. Walliser}}, \bibinfo{title}{{The SU(n) Skyrme Model}},
  \bibinfo{journal}{Nucl. Phys.} \bibinfo{volume}{A548} (\bibinfo{year}{1992})
  \bibinfo{pages}{649}.

\bibtype{Article}%
\bibitem{Diakonov:1997mm}
\bibinfo{author}{{D. Diakonov, V. Petrov, and M. V. Polyakov}},
  \bibinfo{title}{{Exotic Anti-Decuplet of Baryons: Prediction from Chiral
  Solitons}}, \bibinfo{journal}{Z. Phys.} \bibinfo{volume}{A359}
  (\bibinfo{year}{1997}) \bibinfo{pages}{305}.

\bibtype{Article}%
\bibitem{Jaffe:2004qj}
\bibinfo{author}{{R. L. Jaffe}}, \bibinfo{title}{{Comment on 'Exotic
  Anti-Decuplet of Baryons: Predictions from Chiral Solitons' by D. Diakonov,
  V. Petrov, and M. Polyakov}}, \bibinfo{journal}{Eur. Phys. J.}
  \bibinfo{volume}{C35} (\bibinfo{year}{2004}) \bibinfo{pages}{221}.

\bibtype{Article}%
\bibitem{LEPS:2003wug}
\bibinfo{author}{{T. Nakano, {\it et al.}}}, \bibinfo{title}{{Evidence for a
  Narrow S = +1 Baryon Resonance in Photoproduction from the Neutron}},
  \bibinfo{journal}{Phys. Rev. Lett.} \bibinfo{volume}{91}
  (\bibinfo{year}{2003}) \bibinfo{pages}{012002}.

\bibtype{Misc}%
\bibitem{Praszalowicz:2024mji}
\bibinfo{author}{{M. Prasza{\l}owicz}}, \bibinfo{title}{{Odyssey of the Elusive
  $\Theta^+$}} \bibinfo{year}{2024}, \eprint{2411.08429}.

\bibtype{Article}%
\bibitem{Ellis:2004uz}
\bibinfo{author}{{J. R. Ellis, M. Karliner, and M. Prasza{\l}owicz}},
  \bibinfo{title}{{Chiral-Soliton Predictions for Exotic Baryons}},
  \bibinfo{journal}{JHEP} \bibinfo{volume}{05} (\bibinfo{year}{2004})
  \bibinfo{pages}{002}.

\bibtype{Article}%
\bibitem{Goldberger:1958vp}
\bibinfo{author}{{M. L. Goldberger and S. B. Treiman}},
  \bibinfo{title}{{Form-Factors in Beta Decay and Muon Capture}},
  \bibinfo{journal}{Phys. Rev.} \bibinfo{volume}{111} (\bibinfo{year}{1958})
  \bibinfo{pages}{354}.

\bibtype{Article}%
\bibitem{Walliser:2005pi}
\bibinfo{author}{{H. Walliser and H. Weigel}}, \bibinfo{title}{{Bound State
  versus Collective Coordinate Approaches in Chiral Soliton Models and the
  Width of the $\Theta^+$ Pentaquark}}, \bibinfo{journal}{Eur. Phys. J.}
  \bibinfo{volume}{A26} (\bibinfo{year}{2005}) \bibinfo{pages}{361}.

\bibtype{Article}%
\bibitem{Neubert:1993mb}
\bibinfo{author}{{M. Neubert}}, \bibinfo{title}{{Heavy Quark Symmetry}},
  \bibinfo{journal}{Phys. Rept.} \bibinfo{volume}{245} (\bibinfo{year}{1994})
  \bibinfo{pages}{259}.

\bibtype{Article}%
\bibitem{Manohar:2000dt}
\bibinfo{author}{{A. V. Manohar and M. B. Wise}}, \bibinfo{title}{{Heavy Quark
  Physics}}, \bibinfo{journal}{Camb. Monogr. Part. Phys. Nucl. Phys. Cosmol.}
  \bibinfo{volume}{10} (\bibinfo{year}{2000}) \bibinfo{pages}{1}.

\bibtype{Book}%
\bibitem{Blaschke:2004xi}
\bibinfo{author}{{D. Blaschke, M. A. Ivanov, and T. Mannel}},
  \bibinfo{title}{{Heavy Quark Physics}}, \bibinfo{comment}{vol.}
  \bibinfo{volume}{647, Lecture Notes Phys.},
  \bibinfo{publisher}{{Springer-Verlag, Berlin}} \bibinfo{year}{2004},
  \bibinfo{note}{proceedings: International School on Heavy Quark Physics,
  Dubna, Russia, 27 May - 5 Jun 2002}.

\bibtype{Article}%
\bibitem{Gupta:1993kd}
\bibinfo{author}{{K. S. Gupta, M. A. Momen, J. Schechter, and A. Subbaraman}},
  \bibinfo{title}{{Heavy Quark Solitons}}, \bibinfo{journal}{Phys. Rev.}
  \bibinfo{volume}{D47} (\bibinfo{year}{1993}) \bibinfo{pages}{R4835}.

\bibtype{Article}%
\bibitem{Jain:1994rb}
\bibinfo{author}{{P. Jain, M. A. Momen, and J. Schechter}},
  \bibinfo{title}{{Heavy Meson Radiative Decays and Light Vector Meson
  Dominance}}, \bibinfo{journal}{Int. J. Mod. Phys.} \bibinfo{volume}{A10}
  (\bibinfo{year}{1995}) \bibinfo{pages}{2467}.

\bibtype{Article}%
\bibitem{Schechter:1995vr}
\bibinfo{author}{{J. Schechter, A. Subbaraman, and S. Vaidya, and H. Weigel}},
  \bibinfo{title}{{Heavy Quark Solitons: Towards Realistic Masses}},
  \bibinfo{journal}{Nucl. Phys.} \bibinfo{volume}{A590} (\bibinfo{year}{1995})
  \bibinfo{pages}{655}.

\bibtype{Article}%
\bibitem{Harada:1997we}
\bibinfo{author}{{M. Harada, A. Qamar, F. Sannino, J. Schechter, and H.
  Weigel}}, \bibinfo{title}{{Hyperfine Splitting of Low-Lying Heavy Baryons}},
  \bibinfo{journal}{Nucl. Phys.} \bibinfo{volume}{A625} (\bibinfo{year}{1997})
  \bibinfo{pages}{789}.

\bibtype{Article}%
\bibitem{Blanckenberg:2015dsa}
\bibinfo{author}{{J. P. Blanckenberg and H. Weigel}}, \bibinfo{title}{{Heavy
  Baryons with Strangeness in a Soliton Model}}, \bibinfo{journal}{Phys. Lett.}
  \bibinfo{volume}{B750} (\bibinfo{year}{2015}) \bibinfo{pages}{230}.

\bibtype{Article}%
\bibitem{Weigel:1986zc}
\bibinfo{author}{{H. Weigel, B. Schwesinger, and G. Holzwarth}},
  \bibinfo{title}{{Exotic Baryon Number $B=2$ States in the $SU(2)$ Skyrme
  Model}}, \bibinfo{journal}{Phys. Lett.} \bibinfo{volume}{B168}
  (\bibinfo{year}{1986}) \bibinfo{pages}{321}.

\bibtype{Article}%
\bibitem{Verbaarschot:1987au}
\bibinfo{author}{{J. J. M. Verbaarschot}}, \bibinfo{title}{{Axial Symmetry of
  Bound Baryon Number Two Solution of the Skyrme Model}},
  \bibinfo{journal}{Phys. Lett.} \bibinfo{volume}{B195} (\bibinfo{year}{1987})
  \bibinfo{pages}{235}.

\bibtype{Article}%
\bibitem{Feist:2012ps}
\bibinfo{author}{{D. T. J. Feist, P. H. C. Lau, and N. S. Manton}},
  \bibinfo{title}{{Skyrmions up to Baryon Number 108}}, \bibinfo{journal}{Phys.
  Rev.} \bibinfo{volume}{D87} (\bibinfo{year}{2013}) \bibinfo{pages}{085034}.

\bibtype{Article}%
\bibitem{Graham:2025bwh}
\bibinfo{author}{{N. Graham and H. Weigel}}, \bibinfo{title}{{Quantum Energies
  of Solitons with Different Topological Charges}}, \bibinfo{journal}{Phys.
  Rev. D} \bibinfo{volume}{111} (\bibinfo{number}{8}) (\bibinfo{year}{2025})
  \bibinfo{pages}{085031}.

\bibtype{Article}%
\bibitem{VinhMau:1984sc}
\bibinfo{author}{{R. Vinh Mau, M. Lacombe, B. Loiseau, W. N. Cottingham, and P.
  Lisboa}}, \bibinfo{title}{{The Static Baryon-Baryon Potential in the Skyrme
  Model}}, \bibinfo{journal}{Phys. Lett.} \bibinfo{volume}{B150}
  (\bibinfo{year}{1985}) \bibinfo{pages}{259}.

\bibtype{Article}%
\bibitem{Jackson:1985bn}
\bibinfo{author}{{A. Jackson, A. D. Jackson, and V. Pasquier}},
  \bibinfo{title}{{The Skyrmion-Skyrmion Interaction}}, \bibinfo{journal}{Nucl.
  Phys.} \bibinfo{volume}{A432} (\bibinfo{year}{1985}) \bibinfo{pages}{567}.

\bibtype{Article}%
\bibitem{Yabu:1985vx}
\bibinfo{author}{{H. Yabu and K. Ando}}, \bibinfo{title}{{Static N-N and
  N-anti--N Interaction in the Skyrme Model}}, \bibinfo{journal}{Prog. Theor.
  Phys.} \bibinfo{volume}{74} (\bibinfo{year}{1985}) \bibinfo{pages}{750}.

\bibtype{Article}%
\bibitem{Riska:1989fw}
\bibinfo{author}{{D. O. Riska and B. Schwesinger}}, \bibinfo{title}{{The
  Isospin Independent Spin Orbit Interaction in the Skyrme Model}},
  \bibinfo{journal}{Phys. Lett.} \bibinfo{volume}{B229} (\bibinfo{year}{1989})
  \bibinfo{pages}{339}.

\bibtype{Article}%
\bibitem{Abada:1996ux}
\bibinfo{author}{{A. Abada}}, \bibinfo{title}{{On the Skyrme Model Prediction
  for the N-N spin Orbit Force}}, \bibinfo{journal}{J. Phys. G}
  \bibinfo{volume}{22} (\bibinfo{year}{1996}) \bibinfo{pages}{L57}.

\bibtype{Article}%
\bibitem{Abada:1996jg}
\bibinfo{author}{{A. Abada}}, \bibinfo{title}{{The Isoscalar N-N Spin Orbit
  Potential from a Skyrme Model with Scalar Mesons}}, \bibinfo{journal}{Z.
  Phys. A} \bibinfo{volume}{358} (\bibinfo{year}{1997}) \bibinfo{pages}{85}.

\bibtype{Article}%
\bibitem{Kalafatis:1992vv}
\bibinfo{author}{{D. Kalafatis and R. Vinh Mau}}, \bibinfo{title}{{Soliton
  Interactions from Low-Energy Meson Phenomenology}}, \bibinfo{journal}{Phys.
  Rev.} \bibinfo{volume}{D46} (\bibinfo{year}{1992}) \bibinfo{pages}{3903}.

\bibtype{Article}%
\bibitem{Balachandran:1983dj}
\bibinfo{author}{{A. P. Balachandran, A. Barducci, F. Lizzi, V. G. J. Rodgers,
  and A. Stern}}, \bibinfo{title}{A Doubly Strange Dibaryon in the Chiral
  Model}, \bibinfo{journal}{Phys. Rev. Lett.} \bibinfo{volume}{52}
  (\bibinfo{year}{1984}) \bibinfo{pages}{887}.

\bibtype{Article}%
\bibitem{Jaffe:1976yi}
\bibinfo{author}{{R. L. Jaffe}}, \bibinfo{title}{{Perhaps a Stable Dihyperon}},
  \bibinfo{journal}{Phys. Rev. Lett.} \bibinfo{volume}{38}
  (\bibinfo{year}{1977}) \bibinfo{pages}{195}.

\bibtype{Article}%
\bibitem{Scholtz:1993jg}
\bibinfo{author}{{F. G. Scholtz, B. Schwesinger, and H. B. Geyer}},
  \bibinfo{title}{{The Casimir Energy of Strongly Bound $B=2$ Configurations in
  the Skyrme Model}}, \bibinfo{journal}{Nucl. Phys.} \bibinfo{volume}{A561}
  (\bibinfo{year}{1993}) \bibinfo{pages}{542}.

\bibtype{Article}%
\bibitem{Jackson:1988bd}
\bibinfo{author}{{A. D. Jackson, A. Wirzba, and N. S. Manton}},
  \bibinfo{title}{{New Skyrmion Solutions on a Three Sphere}},
  \bibinfo{journal}{Nucl. Phys.} \bibinfo{volume}{A495} (\bibinfo{year}{1989})
  \bibinfo{pages}{499}.

\bibtype{Article}%
\bibitem{Jackson:1987sy}
\bibinfo{author}{{A. D. Jackson}}, \bibinfo{title}{{Skyrmions and Dense
  Matter}}, \bibinfo{journal}{Prog. Part. Nucl. Phys.} \bibinfo{volume}{20}
  (\bibinfo{year}{1988}) \bibinfo{pages}{65}.

\bibtype{Article}%
\bibitem{Piette:1994mh}
\bibinfo{author}{{B. M. A. G. Piette, and B. J. Schroers, and W. J.
  Zakrzewski}}, \bibinfo{title}{{Dynamics of Baby Skyrmions}},
  \bibinfo{journal}{Nucl. Phys.} \bibinfo{volume}{B439} (\bibinfo{year}{1995})
  \bibinfo{pages}{205}.

\bibtype{Book}%
\bibitem{Schollwock:2004aa}
\bibinfo{author}{{U. Schollw\"{o}ck, {\it et al.}}}, \bibinfo{title}{{Quantum
  Magnetism}}, \bibinfo{comment}{vol.} \bibinfo{volume}{645, Lecture Notes
  Phys.}, \bibinfo{publisher}{Springer-Verlag, Berlin} \bibinfo{year}{2004}.

\bibtype{Article}%
\bibitem{Nagasoa:2013}
\bibinfo{author}{{N. Nagaosa and Y. Tokura}}, \bibinfo{title}{{Topological
  Properties and Dynamics of Magnetic Skyrmions}}, \bibinfo{journal}{{Nature
  Nanotech}} \bibinfo{volume}{8} (\bibinfo{year}{2013}) \bibinfo{pages}{1}.

\bibtype{Article}%
\bibitem{PhysRevB.103.L060404}
\bibinfo{author}{{O. M. Sotnikov, V. V. Mazurenko, J. Colbois, F. Mila, M. I.
  Katsnelson, and E. A. Stepanov}}, \bibinfo{title}{{Probing the Topology of
  the Quantum Analog of a Classical Skyrmion}}, \bibinfo{journal}{Phys. Rev.}
  \bibinfo{volume}{B103} (\bibinfo{year}{2021}) \bibinfo{pages}{L060404}.

\end{thebibliography*}
\end{document}